\documentclass{emulateapj}

\newcommand{\sil}{\sigma_{\textrm{\tiny{lens}}}}

\newcommand{\re}{R_{\textrm{\tiny{e}}}}
\newcommand{\rl}{R_{\textrm{\tiny{lens}}}}
\newcommand{\rein}{R_{\textrm{\tiny{Ein}}}}

\newcommand{\rv}{R_{\textrm{\tiny{vir}}}}

\begin{document}

\title{Resolving the baryon-fraction profile in lensing galaxies}

\author{Dominik Leier\altaffilmark{1},  Ignacio Ferreras\altaffilmark{2}, Prasenjit Saha\altaffilmark{3}, and  Emilio~E. Falco\altaffilmark{4}}
\altaffiltext{1}{Astronomisches Rechen-Institut, Zentrum f\"ur Astronomie, Universit\"at Heidelberg, M\"onchhofstr. 12-14, D-69120 Heidelberg, Germany;  leier@ari.uni-heidelberg.de}
\altaffiltext{2}{Mullard Space Science Laboratory, University College London, Holmbury St Mary, Dorking, Surrey RH5 6NT, UK}
\altaffiltext{3}{Institute for Theoretical Physics, University of Z\"urich, Winterthurerstrasse 190, CH-8057 Z\"urich, Switzerland}
\altaffiltext{4}{Harvard-Smithsonian Center for Astrophysics, 60 Garden Street,Cambridge, MA 02138, USA}

\begin{abstract}
  The study of the distribution of baryonic matter within dark halos
  enriches our understanding of galaxy formation. We show
  the radial dependence of stellar baryon fraction curves derived for
  21 lensing galaxies from the CfA-Arizona Space Telescope LEns Survey
  by means of stellar population synthesis and pixel-based mass
  reconstruction. The sample covers a stellar mass range of $M_s\simeq
  2 \times 10^{9} - 3 \times 10^{11} M_\odot$ (solar masses) which
  corresponds to a total enclosed mass range of $M_L \simeq 7 \times 10^{9} - 3
  \times 10^{12} M_\odot$ on radial scales from $0.25 \re$ to $5 \re$
  (effective radii). By examining the $M_s$ and $M_L$ dependence on
  radial distance to the centre of each galaxy we find that there are pairs of lenses
  on small to intermediate mass scales which approach at large radii the
  same values for their enclosed total mass but exhibit very different
  stellar masses and stellar baryon fractions.
  This peculiar behaviour subsides for the most massive lensing galaxies.
  All the baryon fraction profiles show that
  the dark matter halo overtakes the stellar content between $1.5$ and
  $2.5 \re$. At $3 \re$ most of the stellar component is enclosed. We
  find evidence for a stellar baryon fraction steadily declining over
  the full mass range. Furthermore, we shed light on the Fundamental
  Plane puzzle by showing that the slope of the $M_L(<R)$-to-$M_s(<R)$
  relation approaches the mass-to-light relation of recent Fundamental
  Plane studies at large radii. We also introduce novel concentration
  indices $c=R90/R50$ for stellar and total mass profiles (i.e., the
  ratio of radii enclosing 90\% and 50\% of the stellar or total mass).
  We show that the value $c=2.6$ originally determined by light profiles
  which separates early-type galaxies from late-type galaxies
  also holds for stellar mass. In particular, less massive
  dark matter halos turn out to be influenced by the distribution of
  stellar matter on resolved scales below 10 kpc. The ongoing study of
  resolved baryon fraction profiles will make it possible to evaluate the validity of star
  formation models as well as adiabatic contraction prescriptions commonly
  used in simulations.
\end{abstract}

  
\slugcomment{Accepted for publication in the Astrophysical Journal}

\section{Introduction}
\label{sec1}

The physics driving the evolution from the collapse of gas and dark
matter halos to the formation of galaxies remains one of the open
questions in astrophysics. In general, star formation efficiency ---
viewed as the stellar to total mass fraction within the virial radius
of a halo --- is highest for galaxies similar to the Milky Way, with
an efficiency decreasing towards higher and lower masses
\citep{mos10}. The lower escape velocities in less massive galaxies
allow the gas to be ejected by stellar feedback
\citep{LAR74,DS86}. Supernova-induced winds are energetic enough to
significantly impede galaxy formation at baryonic masses below
$10^{11}$M$_\odot$ \citep{BGB07}. Such feedback regulates the star
formation efficiency, which is responsible for the mass-metallicity
relation \citep{tre04}. For more massive galaxies, an Active Galactic
Nucleus (AGN) is believed to account for the decreasing efficiency
\citep[see e.g.][]{dim05}. This feedback mechanism explains the
exponential cut-off in the luminosity function, either by the thermal
coupling of AGN outflows with gas \citep[e.g.][]{TB93,CSW06,BBM06} or
by mechanical feedback that prevents gas cooling \citep{SSD07}.

Observational estimates of the stellar baryon fraction are thus an
essential piece of the puzzle and provide important constraints on
simulations, especially at the sub-grid level that describes the
baryon physics of galaxy formation. They also help to understand the
nature of the scaling relations, such as the Fundamental Plane and its
projections. Currently, most of the studies that resolve the central
regions of galaxies on scales below 10~kpc are based on dynamical
models applied to the kinematics of stars \citep[see
e.g.][]{ca06,co09}. Similarly, lensing studies on galaxy scales are
usually based on a parametric decomposition of the stellar and dark
matter component \citep[see e.g.][]{slacs10,trott10}, with its inherent
degeneracies. Over larger scales, \citet{GUO10} and \citet{mos10}
match the stellar mass function of SDSS galaxies with the distribution
of dark matter halos from numerical simulations to find stellar baryon
fractions $f_b\sim 3-4$\% --- significantly lower than the
cosmological fraction $f_b=\Omega_b/\Omega_m=0.17$ \citep{WMAP52} ---
with a maximum for galaxies with halo masses around
$10^{12}$M$_\odot$. However, this approach is only valid for masses
enclosed within the virial radius, and cannot resolve the radial
dependence, which offers valuable information about how baryons build
galaxies. For instance, the velocity dispersion analysis of
\citet{lin06} on a sample of SDSS early-type galaxies gives within the effective radius a low
baryon fraction ($f_b\sim 8$\%) which is lower than the cosmological
value, but twice as large as determined within the virial radius,
illustrating the importance of a resolved estimate of the baryon
fraction within galaxy halos. Galaxy formation models combining the
evolution of the dark matter and gaseous components along with a set
of sub-grid prescriptions for star formation and feedback \citep[see
e.g.][]{ka93,co94,CSW06} are only indirect methods with considerable
uncertainties. Indeed, robust observational estimates of the baryon
fraction on galaxy scales are needed to properly constrain the recipes
included in these models.\\

Gravitational lensing opens a door to smaller scales over which
baryonic processes are important. For instance, one can explore
concentrations and baryon fractions giving good evidence of adiabatic
contraction, as done e.g. by \cite{ji07}. They analyze the relation
between stellar baryon fraction and concentration in adiabatic and
non-adiabatic models. \cite{MSK06} present a galaxy-galaxy weak
lensing analysis of a large sample of early and late-type
galaxies. They obtain stellar surface masses depending on radius with
a resolution down to 10~kpc. However, their approach is also based on
a halo-model to describe the relation of galaxies and dark matter.\\

The use of mass models and model-based prescriptions
  introduces hard-to-quantify deviations from real mass distributions,
  especially over the scales of a few R$_e$ that we want to
  investigate. Assuming a mass model for a lensing system excludes
  mass distributions which are not accessible in the parameter-space
  of the model and introduces the problem of model non-uniqueness.
To avoid this, free-form methods are necessary. In this paper, we use 
the {\em PixeLens\/} method of \citep{sa04,co08} to reconstruct the
surface mass density of a sample of lens galaxies. For the stellar
component --- which represents the vast majority of the baryons in the
inner regions of early-type galaxies --- the photometric data is used
to constrain a large volume of stellar population synthesis (SPS)
models \citep{fe05,fe08}. The combination of both lensing and stellar
mass in a pixel-based manner allows for a two-dimensional mapping of
the baryon fraction. Choosing a sample of moderate redshift lenses
enables us to determine the lensing profile out to a few R$_e$. The
CASTLES sample\footnote{http://www.cfa.harvard.edu/castles} fulfills
this requirement. We present in this paper an analysis of the enclosed stellar
and total mass content in a sample of 21 lensing galaxies out to a
radial distance of $\sim 1.5-2$ times the Einstein radius, i.e. up to
several $\re$.\\

In section \ref{sec:sample} we discuss briefly the lens
  sample with respect to environment, lens morphology and photometric
  properties. By means of three lens systems, arguably rather extreme,
  we illustrate the subtleties of photometric modeling and the
  authenticity of lenses. The latter point refers to unlensed double quasars which mimic a lens system with a doubly imaged quasar.
  We will show how a real lens can be distinguished from a spurious system in our analysis.
We test the reliablility of our photometry-based results by comparing
inferred stellar surface mass densities with equivalent results from
\cite{FLP09} and \cite{SMW03}.

Section \ref{sec3} presents the results of this study regarding the
radial dependence of enclosed stellar versus total lens mass. We continue in section
\ref{sec:cse} with a closer examination of the stellar and total mass
concentration and define a simple model to study the energetic
evolution of early type galaxies. The conclusion and discussion
section \ref{sec5} summarizes our findings and puts them into the
context of recent work on galaxy formation.

\section{Sample Properties}
\label{sec:sample}
In the following we compare lens samples in general with respect to
their environment. We also explain how the environment affects the
lens systems and continue with a detailed one-by-one study of the
lenses used in this analysis according to their photometric and
morphological properties. To get a clear view on lens
galaxies whose baryonic content we want to determine, we need to
correct the data for light originating from the quasar images, by means
of PSF subtraction and masking. 

The selection criteria for our lens sample are as follows.
  For the lens mass reconstruction we must have the redshifts of the source and the
  lens as well as accurate image positions.  The stellar
  population synthesis analysis demands a sufficient separation
  between lens and quasar images in order to extract uncontaminated
  photometric estimates from the lens. Furthermore, NIR imaging
  must be available.

Constraining the SPS models using photometry in several bands is desirable, although we note that our reference H-band is the F160W filter of HST/NICMOS.
For the redshift of most of the lenses, this band maps a rest-frame region that does not change significantly for the colours found in these galaxies.
We discuss in this section the available multiband data and
respective PSFs used for the modeling of the surface brightness
distribution. Finally we discuss outliers and special cases for
comparison. All information regarding lens galaxy properties, their
environment and photometry is given in Table \ref{tab1} and Table
\ref{tab2}.

\subsection{The Environment}

To describe a lens with respect to its environment,
  one has to keep in mind that the lens shear required by
  (parametric and non-parametric) lens models can be due to
  physically interacting galaxies or to line-of-sight objects.
Regarding the former, one could estimate how the
environment of the lens galaxy evolved in its recent past, whereas
any line-of-sight objects are naturally unrelated to the local region
of the lens. Nevertheless these two sources for shear are hard to
distinguish. If located in a group or cluster environment, X-ray
measurements are expected to give reliable constraints on the DM
content \citep{BT95} and thus a hint about the
direction and strength of the shear. Only a few lens environments have been studied so far for CASTLES lenses \citep[e.g.][]{fas06,mom06}.\\

The environment for a sample of 70 SLACS lenses has been studied by
\cite{tre09}; they find $17\pm 5 \%$ are in overdense
regions. For our sample of 21 CASTLES lenses we find that 7 galaxies
are located in groups and 3 in clusters. Four galaxies have one close
galaxy or possible companion with which they may interact
gravitationally. For the remaining 7, no large shear contribution is
required and no close galaxies have been found. Thus we find that
$\sim 50\%$ of our galaxies lie in overdense regions. 
The lower fraction found by \cite{tre09} is likely due to
the smaller redshift range of SLACS (up to $z \approx 0.5$) and the
property of the SDSS selection function to pick lenses whose Einstein
radius is about the fibre-radius of the SDSS spectrograph (3~arcsec).

\subsection{The Sample}

In the following we briefly discuss the lens sample. In addition to
the previous paragraph we provide information about the environment
which in fact influences both the mass model and the light profile and
yields important insights into the evolution of early-type
galaxies. After giving the full name of the lenses we use their abbreviations only. 
The method used to model the lenses is explained in detail
in section \ref{sec:lens}. Figure \ref{fig16} shows the appearance of the
free-form mass models.\\

For 9 lensing systems all three bands were used to constrain a large number of SPS models, which consequently sets constraints on the colour-to-mass relation. Another 8 lenses could be analysed in H and I band. The remaining 4 lenses had suitable data in H band only.\\

First, we describe the 9 lenses with suitable data in all three wavebands.

The four-image lens system (``quad'') \textbf{B0712+472} is one of the 
few lenses for which a \emph{TinyTim} PSF was sufficient to remove quasar images in H
band. In V and I band the quasar images could all be masked out. Lens
models found in previous studies require significant external shear, which can be attributed to
9 or more galaxies in a foreground group at $z\sim 0.3$ found by \cite{fas02}.
Their study also shows one other galaxy at the redshift of B0712 at $\sim 100''$ from the lens.

The quads \textbf{B1422+231}, \textbf{B2045+265}, \textbf{Q0047-2808},
\textbf{Q2237+030} undergo the following treatment. In both I and V
bands, \emph{TinyTim} provided a suitable PSF. In the H band an
isolated star taken from the same or a contemporaneous NICMOS image was
used for convolution and point-source fitting if needed. B1422 is in a
poor group with 5 nearby galaxies mostly south east of the lens that
cause a significant shear \citep{mom06,hb94}. The group is visible in
X-rays at $0.5-2$ keV \citep{mom06}. In recent work by
  \cite{wkm11} 12 new members were found to be part of the group.
B2045 as found by \cite{fbc99} might be influenced by a group of
galaxies west of the lens. A shear in this direction is also required
by the lens model. The lens might also be affected by a close dwarf
galaxy causing anomalous flux ratios \citep{MKF07}. Q0047 is a lens
with only a small shear required by lens models. However,
  \cite{wkm11} find evidence for a galaxy group with 9 members.  In
the case of the Einstein Cross Q2237 the bulge of a spiral galaxy is
responsible for the lensing. The system shows only a mild external
shear due to the disk of the spiral galaxy.

For spiral galaxies the contribution of dust to the photometry is usually more significant than for early-type galaxies (the latter
morphological type constitute the majority of our lensing galaxies). However, we note that in the case of Q2237, the redshift of
the lens is very low, which implies that our reference photometric band (H) maps a similar wavelength range in the {\sl rest frame},
where dust attenuation is less severe. From the estimates of \cite{Eigenbrod08} on VLT/FORS1 spectra of Q2237, we infer a
contribution from dust in the H band photometry of Q2237 at the level of 0.05 mag \citep{fe10}.

For lenses \textbf{BRI0952-0115}, \textbf{Q0142-100} and
\textbf{PG1115+080} extensive use of the iteration method described in section \ref{sec2.1} was made if the quasar images could not be masked out. The
environments of the doubly imaged quasars BRI0952 and Q0142 have been
studied by \cite{LFK00}, \cite{mom06} and \cite{ECM07} and found to
have no dominant impact on the total shear beyond a cosmological (large-scale structure) contribution
$\gamma_{\rm LSS}$ which is additionally confirmed by lens
models. BRI0952 was previously thought to reside in a region loosely
bound to a poor group with 5 members \citep{mom06}; a later study
found it is at higher redshift and thus not connected with the group
\citep{ECM07}. For Q0142 there is not much known about the close-in
group environment, although there are some galaxies near the
line-of-sight, whose redshifts are mostly unknown. \cite{SMS87}
speculate that a galaxy about $10''$ away from the lens may be a group
member. The environment of the quad PG1115 is thoroughly analyzed by
\cite{mom06}. They find 13 galaxies in a local group with elongated
group emission in X-rays according to \cite{GBC04}. The brightest 4
members of the group are located on an axis with a position angle of
$+60^\circ$ (measured North through East) of
the lens mass which accounts well for the shear required in our
lens model.

The two-image lens \textbf{HS0818+1227} requires special treatment
as we use an isolated PSF of image B1030+071 to fit the quasar image in the H
band. In the I and V bands, the quasar images are used for
fitting. Iteration as it is used for enhancing PSFs of other
lenses does not provide better model fits for the lens because of the
large separation between images and lens. The image separation
is $2.56''$. Hence the reduction
process is further simplified by masking. Since its discovery by
\cite{HR00} no further insights into the environmental properties of
the lens are available. Nevertheless \cite{HR00} found a galaxy
$5''$ north of the lens which appears to have the same redshift of
$z=0.39$, which explains the external shear required by our lens
model. A chain of galaxies at a distance of $10''$
north-east could also be associated with the lens galaxy.\\

Next, we describe the 8 lenses with suitable data in two wavebands.

For the quad \textbf{B1608+656} and the doubles \textbf{HE1104-1805} and \textbf{HE2149-2745} H and I band data could be used to isolate the lens galaxy.
According to \cite{MMM09} \textbf{MG2016+112} exhibits quadruply imaged features of the quasar jet which can be distinguished only in the radio band. We take account of the rather complex structure in the lensing part of our analysis. For all their H band images, a sufficiently isolated star with fitted background extracted from the
image of MG0414+0534 was used to remove the quasar images with an acceptable
goodness of fit.

B1608 resides in the middle of a galaxy group with 8 other group members according to \cite{fas06}.
The photometry shows an object close to the main galaxy, which constitutes a second lens galaxy.
This is confirmed by the reconstructed mass map (Fig.~\ref{fig16}) as it predicts a conspicuously elongated mass distribution towards NE.
Images also show a prominent dust lane between the two galaxies. However, we analyzed the impact of dust reddening on our results, as shown in Appendix \ref{appa} for B1608 and B1600. The uncertainty due to dust on $\log(M_s)$ is in both cases not larger than 0.3 dex.

MG2016 is known to be a giant elliptical galaxy in a cluster with 69 probable
photometrically selected members of many different galaxy types
\citep{TSH03}. Among them is a significant fraction of merging cluster
galaxies, which is direct evidence for a hierarchical formation
history \citep{DFF00}. Most of the neighbouring objects within $30''$
lie on an east-west axis and thus explain the major shear
direction. HE1104 features the second highest image separation of
$3.19''$ and a distinct lens galaxy (the median separation is
$\sim1.5''$). Furthermore the lens appears to be near the bright image
which is rather unusual and implies the presence of a group or cluster
enhancing the separation \citep{LFK00}. Parametric as well as
free-form mass models also suggest that an external shear is mandatory
to reproduce the image configuration \citep[e.g.][]{wwl98}. The
lens galaxy is unaffected by quasar light allowing for a
good fit. However, the photometric redshifts of a few neighbouring
galaxies described in \cite{fau04} indicate that such cluster galaxies
are probable companions of the lensed quasar rather than of the
lens. The double HE2149 might be a member of a cluster as inferred by
\cite{LWW98} by a large number of red non-stellar objects in R-band
images of the field around the lens. Considering recent estimates of
the lens redshift from \cite{ECM07} ($z_{\rm lens} = 0.603$) and the
environment survey from \cite{mom06} HE2149 could be in a group with 3
neighbouring objects. The morphology of the lens shows no sign
of strong external shear.

The doubly imaged quasar \textbf{SBS1520+530} is treated like the
previous doubles but with a star from the same image file in
preference to other PSFs. This lens is a member of a galaxy group with
at least 4 other members as stated in \cite{AFW08}.

For the two quads \textbf{MG0414+0534}, \textbf{RXJ0911+0551} and the
double \textbf{Q0957+561} we obtain good residual maps by means of the
iteration method.  MG0414 at $z=0.960$ is the second most distant lens
of our sample. Judging by its luminosity and colour, the lens is likely to be a passively
evolving early-type galaxy \citep{TOK99}. \cite{SM93} find an object close to image B visible only in I-band, which might contribute to the lensing effect. Our reconstructed mass map also shows increased surface density at the position of the object. RXJ0911 is located on the
outskirts of a cluster \citep{MCM01}. \emph{Chandra} observations of
the cluster suggest a complex non-spherical cluster mass distribution
at a temperature of roughly 2.3~keV. Q0957, found
by \cite{Walsh1979}, is special in several ways. First there is a doubly
imaged galaxy component in addition to the famous double quasar used
to calculate the projected mass map. Secondly the lens is a cD galaxy
located in the centre of a cluster. The nearest cluster member lies
within $10''$ East of the lens galaxy. However a
simple external shear is insufficient to describe the effect of the
environment on the image positions. Breaking the degeneracy between
the shape of the galaxy and the cluster shear takes advantage of arc
features \citep{KFI00} and X-ray data as attempted by \cite{ccf98}.\\

Finally, we describe all the lenses with suitable data in only one waveband.

\textbf{B1030+071}, \textbf{B1152+200} and \textbf{B1600+434} are
treated similarly with regard to the fitting routine, i.e. the
isolated outermost quasar image was used for subtraction and
convolution. The three doubles have comparable angular image
separations and average velocity dispersions as well as intermediate
luminosities. Observed substructures in B1030 indicate the presence of
an interacting galaxy system \citep{JXB00} although firm statements
about the environment cannot be made \citep{LFK00}. However, shear is not strongly required by our mass model. For B1152 there
is no information about the composition of the environment. Judging by
the morphology of the image-source system no strong shear is
expected. B1600 is located in a denser group with at least 6 late-type
galaxies \citep{AFA07} which cause significant shear. The absence of
X-ray emission is suggestive of a not relaxed group, a conclusion
strengthened by the elongated morphology of the group.
Furthermore the lens galaxy appears to be almost edge-on and
  exhibits a prominent dust lane. As remarked above, dust reddening
  changes the population synthesis input and leads to underestimated
  stellar content, but even for the extreme cases in our sample the
  effect of dust on inferred stellar masses cannot be larger than 0.3
  dex (see Appendix \ref{appa}).

For the doubly imaged quasar \textbf{LBQS1009-0252} the star in the
H-band image of MG0414 is used again as a PSF with sufficient quality
for the fit.  \cite{LFK00} locate the lens galaxy close to quasar
image B. They find that a dominant shear contribution of the host
galaxy of a n earby quasar ($4.6''$ northwest of the lens --- unrelated
to the lensed quasar)
is consistent with the derived major axis of the lens when modeled by a Singular Isothermal Ellipsoid.
Using a singular isothermal sphere model \cite{CKL01} determine a smaller shear.
\cite{fau04} state that there is no significant galaxy
overdensity in the field. This is in agreement with the free-form
lens models of this study, which do not require external shear for
this lens.

\subsection{Outliers And Special Cases}
\label{sec:oasc}

We now briefly describe three special cases, \textbf{B0218+357}, \textbf{B1933+503} and \textbf{RXJ0921+4529}. With the first two we want to demonstrate the impact properties like small image separations and interfering luminous structures can have on the goodness of the SPS. The third lens shows how spurious lenses, i.e. galaxies with nearby quasars which are not lensed images of the same background object, behave in this analysis.
All three lenses are excluded from our analysis.\\

For B0218 as for 10 other systems in our sample a star was used to fit the quasar images in the H-band. Since B0218 is the
system with the smallest image separation ($0.33''$) known, it is
extremely difficult to separate the lens galaxy from the images of the
background quasar. The system is an extreme case in
several aspects and a good example for showing the impact of
degeneracies between the magnitudes of overlapping objects.  B0218
unlike any other lens in the sample did not yield reasonable
Sersic profile parameters as the wings of the quasar PSFs overlap
with the lens. For an unconstrained fit the combined light from the quasar images and lens galaxy
results most likely in an overestimated magnitude of the PSFs. However after attempting to fit
the lens system only by PSFs, a Sersic profile is needed
to achieve an acceptable residual map.  Even though one
cannot obtain zero residuals by fitting only two point
sources, there are several combinations of Sersic profile magnitudes
and two PSF magnitudes that result in the same total surface
brightness profile. Bearing this in mind, we use the fitting
parameters with the best $\chi^2$, which also yields an acceptable
residual map, to carry out the SPS. The projected lens mass
map shows that external potentials induce a shear in B0218 that was
studied in \cite{LFK00}. They find 13 possibly perturbing galaxies inside
a radius of $20''$ located roughly along the axis which connects the
two quasar images.  It should be mentioned that B0218 is according to
\cite{LFK00} a late-type galaxy which causes the SPS to predict a different mass content.\\

B1933, discovered by \cite{SBJ98}, has 10 distinct images formed from
a three-component source, promising an exceptionally well-constrained
mass profile.  A star in the same H-band image was used for
convolution. There is as yet no study of the environment of the lens
but according to the mass reconstruction, no strong shear is necessary
to explain the morphology. The resolved features of the lensed
background object cannot be fitted by PSF but are taken out of the
fitting routine by using circular masks with a 5 pixel radius, a size
chosen to cover features distinguishable from background and still
show enough of the lens galaxy to allow for a reasonable fit. The
trade-off between light contamination due to minimal masking and
information loss due to aggressive masking is in any case
problematic. In the case of B1933 almost the whole inner region is
surrounded by masked regions causing the fit parameters $\re$ and $n$
to diverge. Setting a constraint on the Sersic index ($n \leq 4$) is
necessary.  Despite all attempts at modeling this lens, it remained a
persistent outlier, and hence is removed from the analysis.\\

The double RXJ0921, has the highest angular image
separation ($6.93''$) compared to any other lens in the sample. According to \cite{MFK01} it is probably a member of an X-ray cluster. 
From model fits of the host galaxy \cite{PIR06} conclude that RXJ0921 is a binary quasar rather than a gravitational lens.
Also \cite{PMM10} find quite different spectral properties in the spectra of the two components.
For now we assume the system is a lens. Since even the smaller lens-image distance is above $3''$ and the quasar images are
isolated, we obtain a high-quality fit by taking the quasar image as a PSF for both overall convolution and quasar subtraction. No
constraints are necessary. There are 16 objects within $20''$ from the lens galaxy. Only for three of them a redshift close to that of the lens could be determined. The mass model however does not require an external shear. 
In contrast to all other lenses, RXJ0921 (when treated as a lens) turns out to exhibit an unusually low stellar-mass fraction and an almost constant $M_L(<R)$
profile. The peculiar properties of RXJ0921 can be taken as further evidence against the lens hypothesis as suspected in aforementioned studies.

\section{Analysis technique} 
\label{sec2}

In this section we describe necessary steps to obtain
baryon-fraction profiles. As explained before we need to free the
lens galaxies in given multiband data from interfering light,
originating from quasar images, to obtain best-fitting surface
brightness profiles. This reduction step is shown in the following
paragraph. The output is used to constrain SPS models and to estimate
pixel-by-pixel the stellar mass via the colour-to-mass relation, as
described in section \ref{sec2.2}. Combining stellar mass estimates
and pixel-based mass reconstruction (section \ref{sec:lens}) yields
baryon-fraction profiles for the given set of lens galaxies.

\subsection{Preparing Photometry}
\label{sec2.1}
The problems arising during this procedure can be assigned to one of the
following categories: (a)~finding an appropriate PSF for convolution
and point-source reduction, or (b)~removing perturbing light sources
that negatively affect the fitting procedure.  The latter includes
masking of image regions as well as fitting of additional light
profiles to structures which clearly do not belong to the lens. Since
we use photometric data in different filters (V,I and H bands), one of
the following PSF-picking procedures has to be suitably chosen for
each band.
\begin{enumerate}
\item Find an isolated star from the same or a contemporaneous (as nearly as possible) 
image and extract it and a sufficiently large surrounding region not contaminated by light from the lens system or other sources; hereafter referred to as the \emph{star-picking method}.
\item Select the outermost image of a lens system and use it for the quasar image fitting. 
While the lens galaxy and the other quasar images are fitted with 
GALFIT \citep{galfit}, the residual image, showing only the previously chosen outermost image
without any contaminating light, can then be taken as a qualitatively refined PSF. This step can be repeated until we reach the desired level of enhancement. This procedure will be referred to as the \emph{iteration} method. In some cases, when the picked quasar image was sufficiently isolated, iteration brought no further improvement.
\item Using a synthetic PSF generated by \emph{TinyTim} \citep{tiny} yields better results for WFPC2 images rather than for NICMOS data (presumably due to the higher stability of WFPC2 PSFs).
\end{enumerate}
Methods (1) to (3) were combined with \emph{masking} of quasar images or
any other luminous structure interfering with the fit. To prevent the fit from diverging, in some cases further constraints are
necessary. 

Details of the constraints applied to each lens can be found
  in Table~\ref{tab2}. In the following, we give an overview of the
  types of constraints we have applied to the sample:
\begin{itemize}
\item fixing the sky background for the already reduced images to a
  value we determined with SExtractor, since estimating the background
  is essential to extract a meaningful profile of the lens
  \citep{HMB07}, (low signal-to-noise objects are thus neglected,
  increasing the goodness of the fit for the generally bright lens
  galaxy),
\item fixing the surface brightness profiles to previously determined
  $(x,y)$ positions,
\item constraining $\re$ and/or the Sersic index $n$, since both
  parameters are degenerate, being basically inversely related,
  i.e. constraining $\re$ to a low value causes $n$ to diverge and
  vice versa,
\item constraining the position angle and the axis ratio to a
  physically appropriate range of values, and finally
\item restricting the range of magnitudes of the point sources, e.g.,
  constraining image A to be at least $0.5$ magnitudes brighter than
  image B.
\end{itemize}

Except for the first two constraints, which are necessary for
  only a few lensing systems, we try to keep the number of degrees of
  freedom of the fit as high as possible and fix the parameters only if
  there is no alternative.
In cases where the best-fit PA in one band was found to differ
significantly from the PA in another band, it was necessary to
constrain that parameter. The same applies to the axis ratio $b/a$. If
a highly eccentric ellipse is fitted to an actual round lens galaxy
due to interfering PSF wings, the parameter space must be constrained
to exclude less likely $b/a$ values. Since both the lensing galaxies
and quasar images are in some cases too bright to be distinguishable
in H-band, but too faint in V-band, we use the shape parameters PA or
$b/a$ from I-band as a proxy for the fits in other bands. This was done
for B1422, BRI0952, HS0818 and Q2237.  This approach is legitimized by
the fact that different stellar populations visible in different bands
do not change their relative positions and orientations considerably.
All other fitting parameters apart from the boxiness, which is set to
zero throughout the process are free. The boxiness as well as all
other parameters are defined in \cite{galfit}.

\begin{figure}
\figurenum{1}
\includegraphics[scale=0.287]{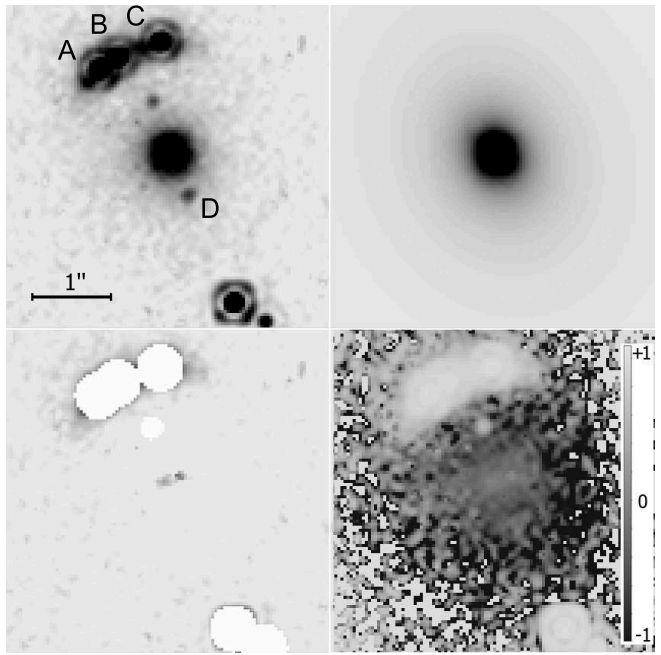}
\caption{Upper left panel: original H-band NICMOS image of the lens
  system B2045. North is left and East is down.  Best fits for the
  lens galaxy could be obtained by masking out images A, B, C, the
  stellar objects south of the lens and a ``blob'' West of the lens
  galaxy. Upper right panel: model for the lens galaxy. Lower left
  panel: Residual image. Lower right panel: Residual divided by
  original image.\label{fig:seq}}
\end{figure}

To minimize $\chi^2$ and test the stability of the fit, the fitting
procedure was repeated with slight changes to the initial
parameters. But $\chi^2$ is not the only criterion to assess the
quality of a fit. We focus on the goodness of the fit in regions most
important to our analysis, i.e. in the central region of the lens
galaxy. The ratio of the fitted image and the original image
  yields a percentage map of the lens systems showing
  pixel-by-pixel the quality of the model, as illustrated in
  Fig. \ref{fig:seq}. We note that in some cases, better $\chi^2$ fits
  were rejected in favour of the flatness of the residual in central regions $\lesssim
  1\re$ of the lens. This occurs in particular for the few lenses for
  which the short distance to a lensed quasar image makes constraints
  mandatory.

For lenses with photometric
data in more than one band the I or V-band parameters for $\re$, $n$,
$b/a$ and PA were taken as a prior to the H-band parameters if
necessary.

\subsection{Estimating stellar masses}
\label{sec2.2}
Our stellar mass estimates are based on a pixel-based comparison of
the best fits to the surface brightness of the lenses with stellar
mass-to-light ratios ($\Upsilon$) determined by population synthesis
models constrained by the available photometry. Even though GALFIT
does a parametric search to get the best fit, for this analysis we are
just interested in the 2D distribution that minimises the residuals,
regardless of the parameters themselves, i.e. we are less sensitive to
the inherent degeneracies associated with parametric fits. For each lens
we ran a grid of $32\times 32\times 32$ models, where the star
formation history is described by a decaying exponential, defined by
three free parameters --- the quantities in parentheses denote the range
explored for each one: formation epoch (defined as a redshift
$2<z_{\rm \rm FOR}<10$); exponential timescale ($-1<\log(\tau/{\rm
Gyr})<1$); and metallicity ($-1<[$m/H$]<+0.3$). Models from
\citet{bc03} are used, assuming a \citet{chab03} initial mass function
(IMF). For each choice of the three parameters, a composite population
is obtained, transformed to the redshift of the lens, and folded with
the passband response of the HST V (F555W, WFPC2), I (F814W, WFPC2)
and H (F160W, NICMOS) filters to compare with the observed
colours and to extract a mass-to-light ratio in the observer-frame H
band. The colours are corrected for Galactic extinction using the dust
maps of \citet{sc98}.

We extract the stellar mass densities from the H band image (NICMOS
F160W). Whenever model fits of the lens were available for I or V, the
colours where used on a pixel by pixel basis to constrain
$\Upsilon_H$.
Otherwise, we used integrated colours within an elliptical
  aperture defined by the half-light radius $\re$ of the H band image
  (see Table \ref{tab1}).
In general, broadband photometry alone cannot be used to constrain the
ages and metallicities of the lens galaxy. However, the stellar
masses, when estimated via ``red'' $M/L$ ratios, are less sensitive to
the age-metallicity degeneracy \citep[see e.g.][]{fe08}.
For comparison, we provide the total stellar mass-to-light
  ratio in the rest-frame V-band $M_s/L_V$ in Table \ref{tab1}.
  Colours and magnitudes are in agreement with comparable quantities
  in \cite{ru03b}.
The F160W band corresponds to a rest-frame wavelength between 0.8 and
1.2$\mu m$ (except for Q2237, which roughly samples rest-frame
H-band). Hence, for the sample considered here, the mass-to-light
ratios are not affected by the presence of young stars, an issue that
becomes important when dealing with optical or NUV indicators
\citep[see e.g.][]{ben10}. From the modeling of the old stellar
populations that these systems feature (except for lens Q2237 ---
which is a bulge --- the other lenses are early-type galaxies), an
uncertainty of $\Delta\Upsilon\lesssim 0.15$ dex is expected
\citep{ag09}.
Dust reddening, as we explain in Appendix \ref{appa}, leads to underestimated stellar mass.
 However, since no starburst galaxy is among our lenses we can safely assume that the effect of dust on $M_s$ does not exceed $20\%$.
 The number of lenses which exhibit dusty features (e.g. B1600) is, nevertheless, small.
The most significant systematic error relates to the choice of the
Initial Mass Function, especially the low-mass end, which does not
contribute to the light, but can contribute very significantly to the
total mass content. However, frequently used choices of the IMF such
as \citet{ms79}, \citet{sc86}, \citet{kr93} or \citet{chab03} have
similar distributions at the low mass end. It is only the traditional
single-power law of the \cite{salp55} IMF that gives different stellar
mass predictions. Previous detailed work on the kinematics of nearby
early-type galaxies \citep{ca06} or strong lenses \citep{fe08,fe10} shows
that the low-mass end of the Salpeter IMF is ruled out as it predicts
stellar mass surface densities higher than the dynamical or lensing
estimates. However, even using a Salpeter IMF does not strongly affect our results, as they mainly focus on the scaling
of the regions where dark matter dominates. \cite{fe10} illustrate differences between five different
population synthesis models based on different prescriptions and/or stellar
libraries. The predicted stellar masses -- measured in the H band -- agree to within $10\%$,
(at fixed IMF) especially given the ages of these lenses.

\begin{figure}
\figurenum{2}
\includegraphics[scale=0.96]{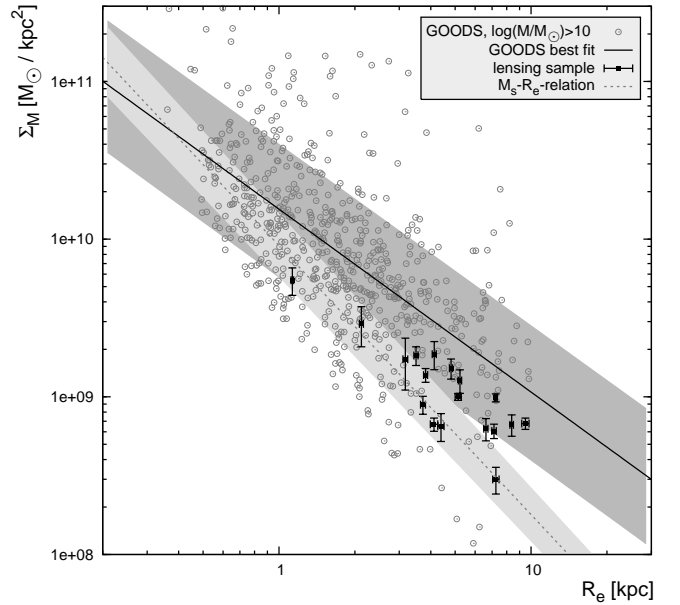}
\caption{Stellar Surface Mass density at a radius of $1\re$ versus effective
radius. Early-type galaxies with stellar mass above $10^{10} M_\odot$
(open circles) selected from the Hubble Space Telescope/Advanced
Camera for Surveys images of the Great Observatories Origins Deep
Survey (GOODS) are shown together with our lens sample (filled
squares). The dashed line denotes the stellar mass to size relation
from SDSS which accounts for early-type galaxies at $z \sim 0.1$
\citep{SMW03}. \label{fig:a1} \vspace{0.27cm}}
\end{figure}

To compare our lensing (early-type) galaxies with a
  typical field sample, we show in Fig.~\ref{fig:a1} the equivalent
  to the Kormendy relation (this time defined with respect to the
  surface stellar mass density at 1$\re$). We show as open dots the sample of
  ACS/GOODS early-type galaxies from \citep{FLP09}, and our lensing
  galaxies as squares with error bars. One can see that 14 out of 18
  lensing galaxies are located inside a $1\sigma$-band around the best
  fit of the GOODS sample.  We also provide the SDSS relation from
  \cite{SMW03} as a local ($z \sim 0.1$) reference.  The obvious
  preference of the lens sample to be at larger effective radius and
  smaller surface mass density is due to a selection bias, which is a
  combination of the lensing bias and additional requirements, such as a sufficient distinguishability from surrounding quasar
  images.

Furthermore, as we show later in this paper, the slope
of the Fundamental Plane relation $M^\eta \sim L$ can be recovered from our data.
Thus we consider the lens sample representative for early-type
galaxies in general. 

\subsection{Reconstructing the total-mass profiles}
\label{sec:lens}

For each lens, the projected total-mass distribution is reconstructed
on a circular field made up of 750 square tiles or pixels, each pixel consisting of a uniform non-negative mass distribution with a mass density of a few times the critical density. We provide mass reconstruction maps of the sample in Fig. \ref{fig16}.
The pixellated mass distribution must reproduce the observations, in
the following ways.
\begin{enumerate}
\item Multiple-image systems with the observed positions must arise as
  solutions of the lens equation.  The images are considered to be
  unresolved; for extended images, the peak of their surface
  brightness distribution is located and considered as an unresolved
  image. Note that although the mass distribution is discontinuous at
  the pixel boundaries, the lens equation is continuous.
\item For lenses with measured time delays, the model is required to
  reproduce them.  Concordance values of
  $H_0,\Omega_M,\Omega_\Lambda$ (i.e. $72$ km s$^{-1}$ Mpc$^{-1}$,0.3,0.7) are assumed.
\end{enumerate}
In addition, the mass distribution must satisfy the following prior
conditions.
\begin{enumerate}
\item The local density gradient must point no more than $45^\circ$
away from the centre of brightness. Since the central regions
of galaxies are expected to be dominated by stars, it seems safe
to assume that the mass and light peaks coincide.
\item The circular average (around the centre) of the projected
density, falls off as $R^{-1/2}$ or faster.  The three-dimensional
mass profiles of galaxies are thought to be invariably steeper than
$r^{-1.5}$, so again this appears to be a safe prior assumption.
\item No pixel is allowed to be more than twice the mean of its
neighbours, except for the central pixel, which can be arbitrarily
high to mimic central density cusps,
\item Unless the lens shows signs of asymmetry, the mass
distribution is required to be symmetric under a $180^\circ$ rotation
around the centre.
\end{enumerate}

In practice, there are infinitely many mass models that satisfy all
the above conditions, because solutions of the lens equation are
highly non-unique \citep{FGS85,sa00,LRD08}.
Accordingly, for each lens we generate an ensemble of 300 models, by a
random-walk technique in model space.  The random walk implicitly
defines a prior measure in model space, and it turns out that an
equivalent statement of the prior measure is that it must be invariant
under rescalings of units \citep{co08}.  A very useful
property of the above constraints (observational or prior) is that
they can all be formulated as linear equations or inequalities.  This
means any weighted mean of ensemble members is also an admissible
model.  Hence, we can conveniently use the ensemble mean to represent
a typical model.  However, the main results later in this paper use
the full model ensembles.\\

The lens models do not attempt to subtract off lensing mass
  outside the galaxy.  Such mass could come from the environment, 14
  of the lenses being in dense environments (see column $env$ in
  Tab.~\ref{tab1}), or it could be in an interloper along the line of
  sight.  However, given that the model-ensemble technique yields
  conservatively large error-bars on the mass maps, we expect
  that external lensing mass is unlikely to be larger than the
  estimated uncertainties.

In the following section, we will consider the circularly-averaged
enclosed mass profile $M(<R)$, see Figures~\ref{comp1} and \ref{comp2}.
The outermost radius to which the mass profiles are reconstructed is
fixed to two times the lensing radius $\rl$, which is defined as the
radial position of the outermost lensed image with respect to the
center of the lens. We choose $2\rl$ as a trade-off between
uncertainty and common radial range for the sample. The range of enclosed-mass
profiles in the ensemble, which is interpreted as the uncertainty, has
a characteristic butterfly shape. That is to say, $M(<R)$ is well
constrained in the image region, but becomes more uncertain farther in
or out. Note that the butterfly shape is less prominent or even
distorted for less symmetric lensed image configuration.
The steep limit of the butterfly shape is expected to be roughly
$M(<R)\sim R^{1.5}$, resulting from the minimal steepness of
$R^{-0.5}$ in the projected density.  The shallow limit of the
butterfly shape is given by the steepest model in the ensemble.

\section{Radial Dependence Of Stellar Versus Total Mass}
\label{sec3}

\begin{figure}
\figurenum{3}
\begin{center}
\includegraphics[scale=1.03]{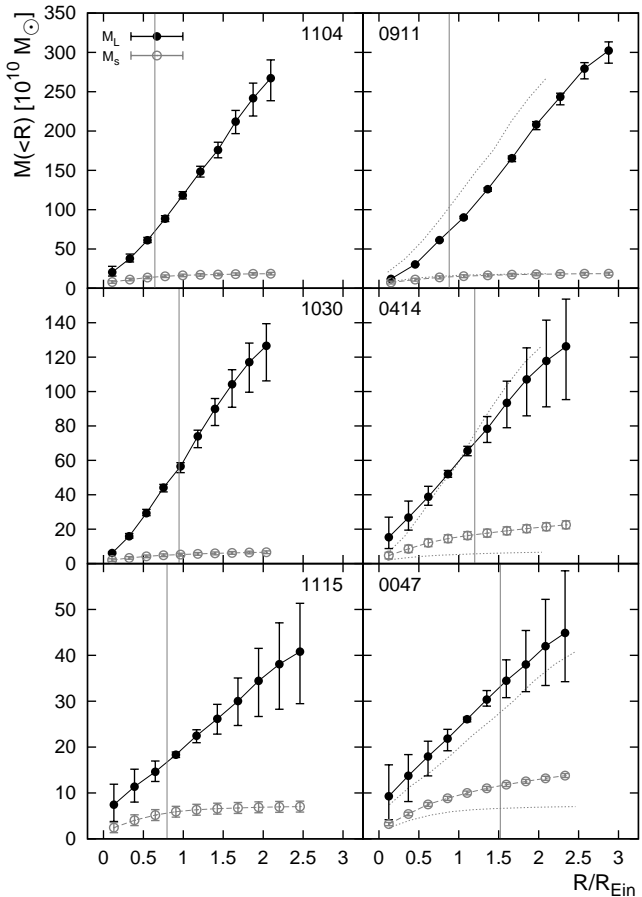}
\end{center}
\caption{Cumulative stellar mass and lens mass profiles against
  projected radius (in units of the Einstein radius) for three
  comparable pairs of lenses.  PG1115 and Q0047 (bottom row) are low
  mass, B1030 and MG0414 (middle) are intermediate mass, and HE1104
  and RXJ0911 (top) are high mass.  The gray vertical line marks
  $2\re$.  To assist comparison, each right-hand panel has the
  profiles from the corresponding left-hand panel duplicated with thin
  dotted lines.\label{comp1}}
\end{figure}

To compare the radial dependence of stellar and total mass, it is
interesting to consider pairs of lenses with matching $M_L(<R)$ or
with matching $M_s(<R)$.  To illustrate, in Fig.~\ref{comp1} we show
three pairs of galaxies with the following properties (see also Table
\ref{tab1}):

\begin{enumerate}
\item small mass, matching $M_L(<R)$ profiles, differing $M_s(<R)$,
\item intermediate mass, matching $M_L(<R)$ profiles, differing $M_s(<R)$,
\item high mass, differing $M_L(<R)$ profiles, matching $M_s(<R)$.
\end{enumerate}

The radial scale is $R/\rein$ where $\rein$ has been estimated from
the pixelated mass maps.  Error bars are $68\%$ confidence from the
population-synthesis models used for $M_s(<R)$ values. For $M_L(<R)$
we use error bars corresponding to $90 \%$ of the $M_L(<R)$ range of
the model ensemble, as described in section \ref{sec:lens}. Note that
the errors attached to $M_s(<R)$ and $M_L(<R)$ are correlated.

The matched pairs are, of course, only rough matches.
Also, the $\rein$ values are not the same for the matched pairs of galaxies.
The angular radial scale is proportional not only to the enclosed mass but also to $(d_{LS}/(d_L d_S))^{0.5}$ (corresponding to $(d_{L}d_{LS}/d_S)^{0.5}$ for the physical Einstein radius), where the $d$'s are the angular diameter distances between observer and lens (L), observer and source (S) and lens and source (LS). Latter distance ratio must be approximately equal to identify matching profiles. To enable comparison between scales we include $\rein/\re$ in Table \ref{tab1}.
With these caveats, we point out some interesting features.

Consider first the two low-mass lenses PG1115 and Q0047 (bottom
panels, Fig.~\ref{comp1}).  While the total mass within $2\rl$ is very
similar, the stellar mass of PG1115 rises only to $50 \%$ that of
Q0047.  The same qualitative behavior is seen if these two galaxies
are compared using $R/\re$ rather than $R/\rein$ as the radial scale.
Nevertheless, these two low-mass lenses have $f_s\equiv M_s/M_L
\gtrsim$ 0.17 the range of high baryon fractions. Lenses within this
range can consequently be referred to as high $f_s$ lenses.

Comparing the two intermediate mass lenses B1030 and MG0414 in the
middle panels of Fig.~\ref{comp1}, we find that their cumulative total
mass curves are very similar. However, the stellar mass of B1030 is
just $\sim 30\%$ that of MG0414, independent of the radius.  If we
consider the stellar radial scale, we find that MG0414 has $\sim 4$
times the stellar mass of B1030.  Their baryon fractions approach
values from $f_s \approx 0.05$ (B1030) to $f_s\approx 0.17$ (MG0414)
at the outermost radius to which we have estimates.  In the
intermediate mass range of our sample (roughly $5\times 10^{11}
M_\odot$ to $15\times 10^{11} M_\odot$) MG0414 has one of largest and
B1030 the lowest stellar-mass fraction. It should be noted that the
opposite behaviour, namely matching stellar profiles on both $\rl$ and
$\re$ scale with very different total mass is also possible. An
example for the latter would be a comparison between B1030 (middle
row) and PG1115 (bottom row) of Fig.~\ref{comp1} with equal
$M_s(R/R_e)$ but total mass profiles differing by a factor of $4.5$ at
$3\re$.

The two high mass lenses RXJ0911 and HE1104 have a total
stellar mass of $\simeq 2 \times 10^{11} M_\odot$.  For comparison, the
total mass profiles differ slightly for radii $\lesssim 1.5 \re$ and
$\gtrsim 2.5 \re$. But it should be noted that RXJ0911 is located at
the centre of a galaxy cluster which might lead to a lens mass
estimate slightly larger than the actual virial mass of the lens
galaxy. At $6.5 \re$, i.e., $\sim 2 \rl$ HE1104 has $12 \%$ less total
mass than RXJ0911. At $\sim 3 \re$, i.e. $\rl$ the difference is still
$6\%$. In terms of stellar-mass fraction HE1104 exhibits small values
of $f_s\approx 0.07$ and RXJ0911 of $f_s \approx 0.06$. In the high
mass regime ($>15\times 10^{10} M_\odot$) the range of possible
stellar-mass fractions appears to be small compared to low and
intermediate masses, and always close to 0.05. Those two lenses are
thus representative for low $f_s$ lenses.

\begin{figure}
\figurenum{4}
\begin{center}
\includegraphics[scale=1.03]{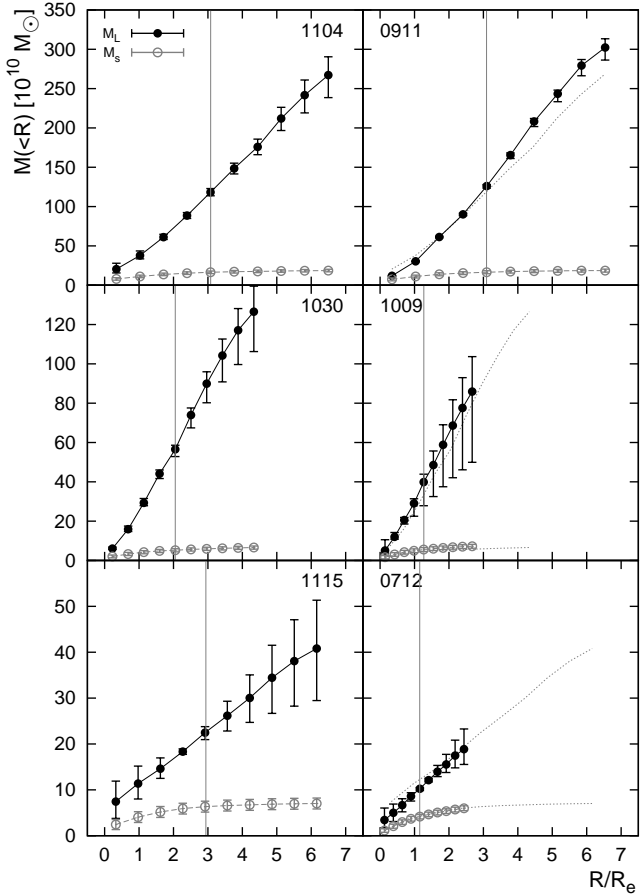}
\end{center}
\caption{As in Fig.~\ref{comp1}, but with the effective radius $\re$ as reference scale, shown for PG1115 and B0712 (bottom), B1030 and BRI1009 (middle), HE1104 and RXJ0911 (top).
Here the grey vertical line marks $\rl$. \label{comp2}}
\end{figure}

Comparing lens profiles on $\re$ scales intrinsic to the
luminous part of the galaxy, one can find many lenses with similar
stellar mass profiles, which is not surprising. After all the enclosed
mass values $M_s(<2\rl)$ cover with $\sim10^{10}$ to $\sim2\times
10^{11}$ a relatively small range in contrast to a total mass range
$M_L(<2\rl)$ of $\sim2 \times 10^{10}$ to $\sim2 \times 10^{12}$.
However, pairs of lenses with matching $M_s(<R)$ and $M_L(<R)$
profiles over the whole radial range as shown in Fig.~\ref{comp2} are
rare. Most lenses with matching $M_s(<R)$ profiles exhibit quite different $M_L(<R)$ profiles.
The above lenses HE1104 and RXJ0911 are -- apart from their data points $\gtrsim6\re$ -- matching pairs within uncertainties on $\re$ scale as they are on $\rl$ scale, a consequence of $\rl/\re$ being equal for both objects.

In Fig.~\ref{comp2} we present the left column lenses of Fig.\ref{comp1} now on baryonic scales, two of them have new counterparts with similar $M_s(<R)$ and $M_L(<R)$.
As before we present low to high mass galaxies from the bottom up.

For the two low mass lenses PG1115 and B0712, we find that at their outermost common radius $\sim 2.5\re$ their baryon fraction is $\sim 0.08$.
B1030 and BRI1009 also match well within their error bars although the mean stellar mass profile of B1030 is consistently below the one of BRI1009. The error region of its lens mass profile shows quite large error bars and thus make it easy to match. The baryon fraction at $2.7\re$ is approximately $f_s=0.08$. If we compare lenses along the vertical direction of Fig.~\ref{comp2} B0712 and BRI1009 are representative for most lenses on low to high mass scales, that is, similar $M_s(<R)$, dissimilar $M_L(<R)$ and baryon fractions. 

In summary we find on both baryonic scale $\re$ and lensing scale $\rl$:
\begin{itemize}
\item many pairs with the same enclosed total (lens) mass, but with different enclosed stellar mass, 
\item a small number of pairs (decreasing with increasing $M_L$) with the same enclosed total (lens) and stellar mass.
\end{itemize}

We can already conjecture an anti-correlation between enclosed lens mass and stellar-mass fraction, which
will be studied in detail later on. 

However, one should keep in mind that our result could be influenced by the lens environment and its history.
See also Table \ref{tab1}, column '\emph{Env}' and section \ref{sec:sample} for information on the local lens environment.
The phenomenon of same $M_L$ but different $M_s$ becomes less prominent for larger total lens masses, on both $\rl$ and $\re$ scale.
Nevertheless, global trends and interdependencies might be revealed by analysing the whole set of lenses, which is done below.

\begin{figure*}
\figurenum{5}
\includegraphics[scale=1.51]{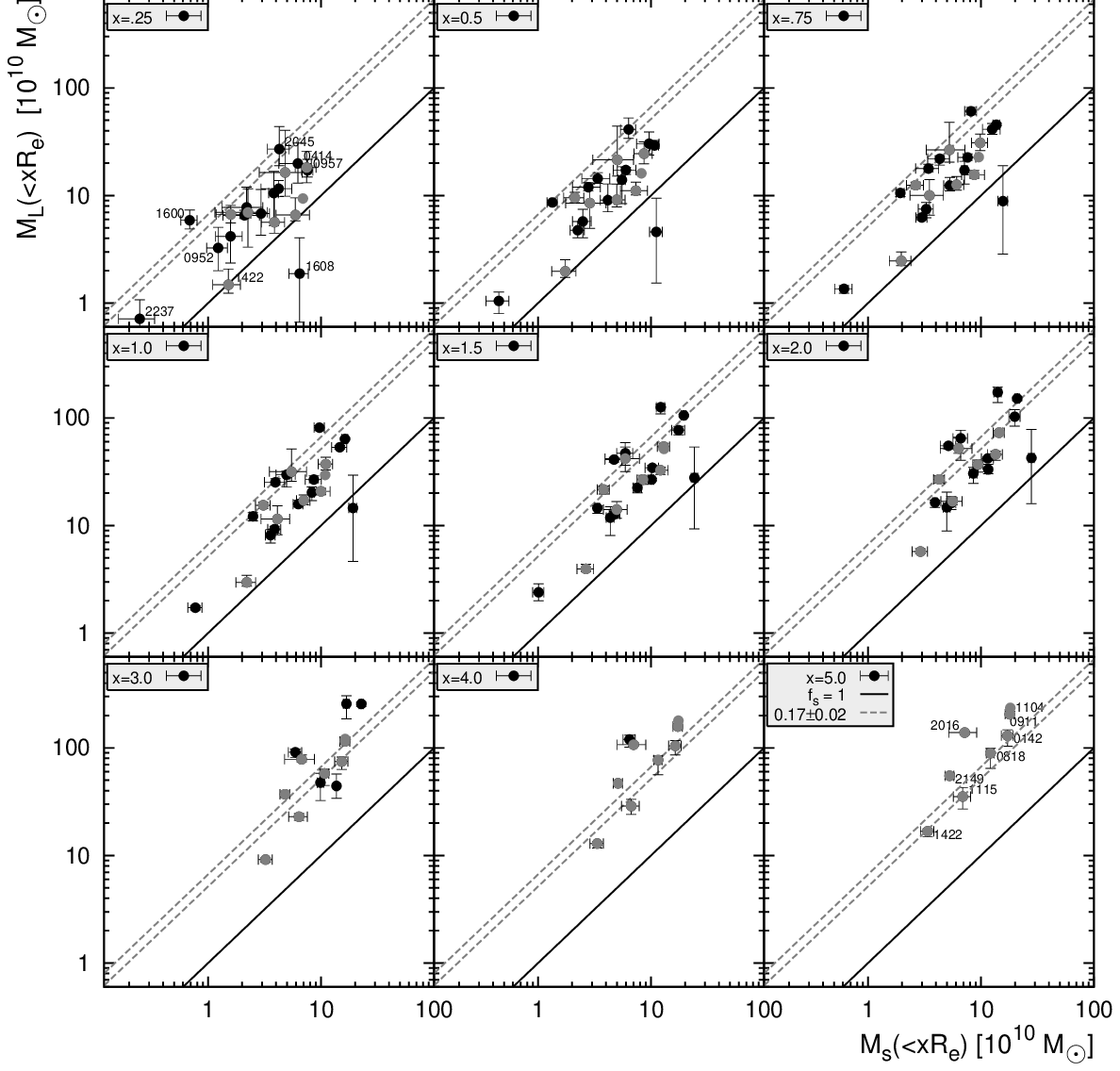}
\caption{The panels show the enclosed $M_L$ against enclosed $M_s$
  plane for a number of apertures, defined by the radial distance $x\re$
  to the centre of the lens galaxy, labelled by '$x$' in the top-left
  corner of each panel. We cover a radial distance from $0.25\times\re$
  to $5\times\re$ from upper left to lower right panels in conveniently
  chosen steps. Grey circles highlight a subset of 8 lenses which are probed out to $5\re$.
    The solid line denotes the equality of total and
  stellar mass, whereas the dashed lines represent the upper and lower
  limit of the global baryon fraction \citep{WMAP5}.\label{fig:mov1}}
\end{figure*}

Using our sample of 21 lensing objects we consider the
following relations to highlight the interdependencies in the
($M_L$,$M_s$,$R$) parameter space:
\begin{enumerate}
\item the enclosed total mass $M_L(<R)$ as a function of enclosed
  stellar mass $M_s(<R)$ at a fixed radius $R$,
\item the stellar-mass fraction as a function of radial distance,
  $f_s(R)=M_s(<R) / M_L(<R) $,
\item the stellar-mass fraction as a function of the total mass $M_L$,
\item the stellar-mass fraction as a function of redshift.
\end{enumerate}

Fig. \ref{fig:mov1} shows the first relation for a range of radial
positions from $0.25\re$ to $5\re$, parametrized by the dimensionless
quantity $x\equiv R/\re$.
For reference, we list in Table~\ref{tab1} the enclosed stellar and lens mass within $2\re$ with error bars.

The universal baryon fraction according to WMAP5, $f_b=\Omega_B
/ \Omega_M = 0.17 \pm 0.02$ \citep{WMAP5}, is included as two dashed
lines for the upper and lower bounds. The solid line denotes a stellar-mass
fraction of one, i.e. the total mass content consists of $100\%$
stellar mass. Note that the data points refer to baryonic matter in
stars and do not account for other baryonic content like gas and
dust. The gas contents of our lens sample -- mostly early type
galaxies -- is expected to be small. But for the Einstein Cross (Q2237),
which is the bulge of a spiral galaxy, and B1600, which is likely to be
a late-type galaxy viewed edge-on, one can indeed expect deviations
from the obtained $M_s$ values.

The galaxy B1608 shows an unreasonably high stellar-mass fraction for
radii $ \leq .75 \re$ (e.g. left panel in top row of Fig. \ref{fig:mov1}).
To take proper account of the light distribution we fit both the brightest galaxy and its merging companion with Sersic profiles, but we only use the information of the light profile of the brightest galaxy for the computation of stellar mass. The enclosed mass values are thus taken with respect to the center of the brightest galaxy. As a consequence of the degeneracy between the two Sersic profiles the central region of the light profile is modeled rather poorly. This causes an overestimate of the stellar content ($\lesssim 15\%$) in a region where the neighbouring galaxy, which is also responsible for light deflection, interferes with the fit. The pixels with highest total mass and highest stellar content do not match for B1608. This also causes larger deviations in the region $\lesssim 1\re$.

The late-type galaxies Q2237 and B1600 might be subject to dust reddening. In general the impact of reddening on high redshift lenses is stronger due to the bluer populations observed in H-band and the higher absorption of dust at smaller wavelength. However, on the basis of the analysis shown in Appendix \ref{appa}, we do not expect departures of more than $20\%$ towards higher $M_s$. This will shift $B1600$ closer to the bulk of lenses in Fig.~\ref{fig:mov1}.

The prominent $f_s$ curve of B1422 --- starting at twice the value of
most other lens galaxies --- might also be caused by light
contamination. This time it originates from the innermost quasar image
which lies just $0.25''$ away from the galaxy centre, an extreme among
the 21 lenses studied in this paper.

In the online material of this paper we provide a movie version of
Fig. \ref{fig:mov1} to help visualize the trend of the stellar-baryon
fraction with increasing radius. See also Appendix \ref{app}.

The lens galaxies reveal the following properties, which are qualitatively assessable already from Fig. \ref{fig:mov1}, but will be explained in detail later on:
\begin{enumerate}
\item Most lenses populate a band of $0.1<f_s<0.4$ within $5 \re$.
\item The slope of the enclosed $M_L$-to-$M_s$ relation of Fig.
  \ref{fig:mov1} within the shown radial range becomes gradually steeper for larger enclosed radii (an effect quantified in the following paragraph).
\item Between $2$ ($1.5$) and $2.5 \re$ ($2 \re$) for most lenses with
  total mass below (above) $4\times 10^{11} M_\odot$ the dark matter
  halos overtake the stellar content, that is they move primarily toward increasing total mass.
  The turning point thus depends on the halo mass. The dark matter halos of more massive galaxies
  start to dominate the matter balance at larger radii (in units of $\re$)
  than those of less massive galaxies.
\end{enumerate}

Note that by ``overtake'' we refer to the radius where $dM_L/dR
\approx dM_s/dR$ rather than to the radius where the total stellar
mass contributes $50\%$ of the total mass.
As a consequence of limited resolution this radius can only be given with larger uncertainties ($\sim 0.5\re$).\\

\begin{figure}
\figurenum{6}
\includegraphics[scale=1.03]{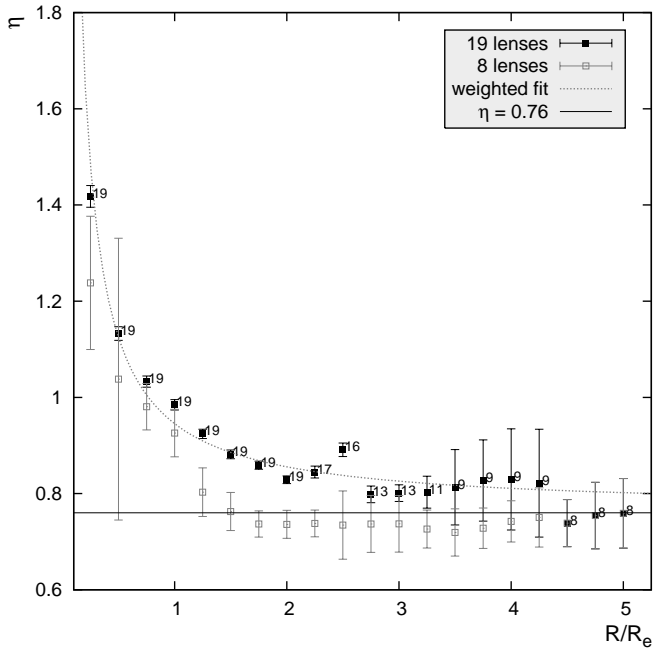}
\caption{Slope of the $M_s$-to-$M_L$ relation taken from Fig.
\ref{fig:mov1} plotted against the distance to the centre in terms of
effective radii. The median slopes are determined via a bootstrapping
fitting method with $10^4$ realizations to compute meaningful standard
errors for a sample of 19 lenses (filled squares) and a reduced 8 lens
sample (open squares). The numbers at the filled squares give the
number of lenses probed out to the respective radius. The dotted line
represents a weighted best fit of $\eta(R) \sim
1/R+$const. \label{fig:slope}}
\end{figure}

Point 2 can also be illustrated by plotting the slope $\eta$
determined from $M_s\propto M_L^\eta$ so that it represents light as a
function of mass. We find that $\eta$ asymptotically approaches
$0.75$, as one can see in Fig. \ref{fig:slope}, which is in
agreement with previous studies of the fundamental plane within error
bars \citep[e.g.,][]{gu93,jo96,DL09}. A bootstrapping method for a
large and a reduced sample is used to determine the $M_s$-to-$M_L$
relation and its standard errors respectively. Both runs are done with
$10^4$ realizations.  The 19-lens sample contains all the lenses
except for the outlier B1608 and the late-type galaxy
Q2237. Farther out in radius, the number of lenses with
profiles extending to a particular radius decreases. Because of that,
Fig. \ref{fig:slope} also shows the number of lenses used for each
fit. As a consequence of changing sample size discontinuities appear
between $2.25$ and $3.5 \re$ and at $4.5 \re$. The most extreme ones
are caused by B0712 ($2.5 \re$) and B1030 ($4.5 \re$) falling out of
the sample. The behaviour of the error bars in a bootstrap fit depends on the size of the drawn sample subset.
To get more meaningful error bars we fixed the size of the sample subset to be $~50\%$ of the available number of lenses
at each radius.
The small sample instead comprises all 8 lenses being
probed out to $5 \re$, which are highlighted in Fig.~\ref{fig:mov1} by grey filled circles.
From $M_s \propto M_L^{1.24\pm 0.14}$ at $0.25 \re$ the reciprocal slope $\eta(R)$ declines as $1/R$
and ends up at $5 \re$ with the relation $M_s \propto M_L^{0.76\pm
0.07}$.
We expect only small deviations from this slope for larger
radii since we run out of stars, and additional mass from the dark
matter halo shifts the distribution upwards, whereas possible baryonic
contributions from gas shift the whole population farther
to the right of Fig. \ref{fig:mov1}. Additionally, for the
19-lens sample a weighted best fit for $\eta(R)$ suggests that the
function approaches asymptotically a constant value of $0.77\pm
0.01$. Note that for the small sample $\eta(R)$ declines rapidly to reach the value of 0.76 already at $\sim 1.5\re$ and thereafter shows 
no significant departure from it. However, the change in slope from small to large radii
is significant for both the 19-lens and the 8-lens sample.\\

\begin{figure*}
\figurenum{7}
\includegraphics[scale=1.03]{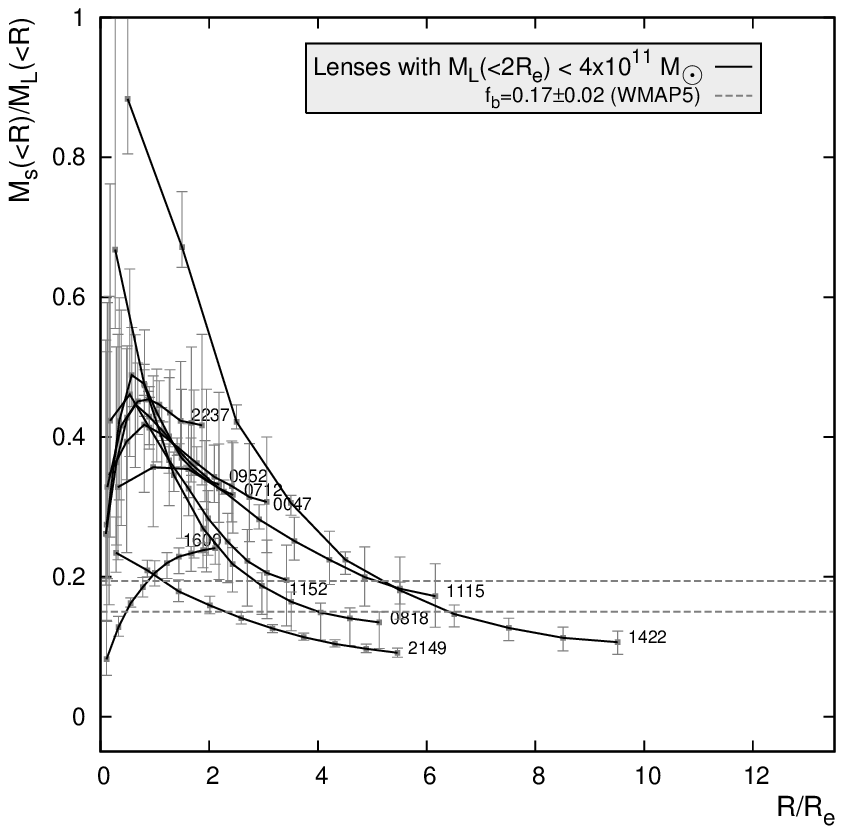}\hspace{0.67cm}\includegraphics[scale=1.03]{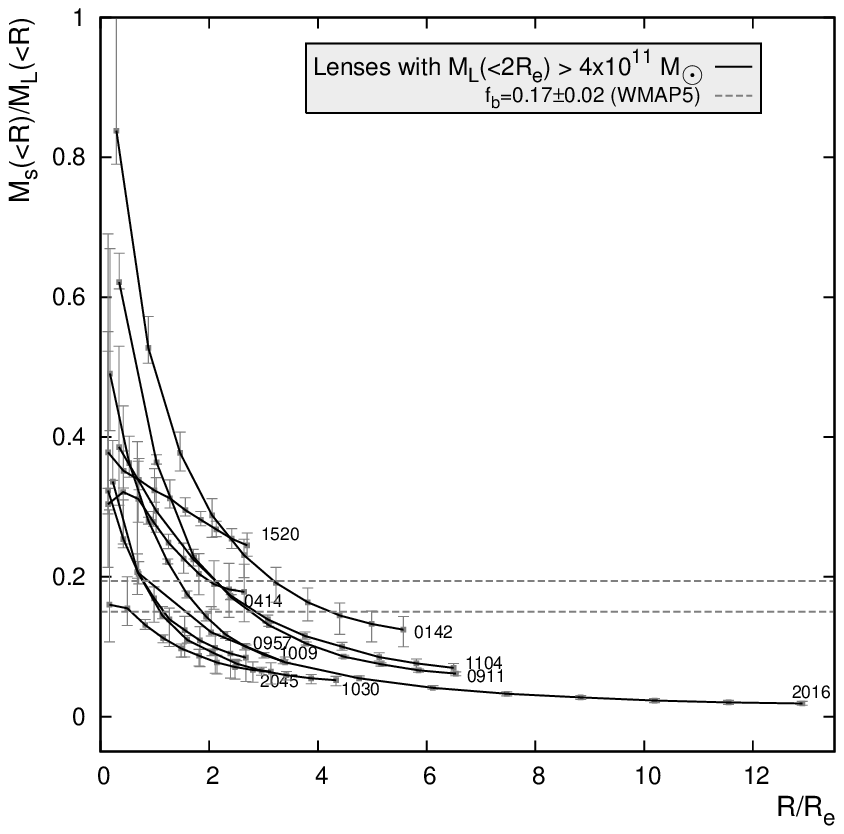}
\caption{Left panel: Stellar-mass fraction against radius in effective radii for lenses with lens mass enclosed within $2\re$ below $4\times 10^{11}$
$M_\odot$. Right panel: Similar, but for lenses with $M_L(<2\re)$ above $4\times 10^{11}$ $M_\odot$. \label{fig7-1}}
\end{figure*}

All the stellar-mass fraction curves in the left and right hand panel
of Fig.~\ref{fig7-1} turn over to a similar stellar-mass fraction
between 1.5 and 2.5 $\re$, a fact also reflected by $\eta(R)$ in Fig.~\ref{fig:slope}.
With increasing radius, the stellar-mass fractions of high mass galaxies ($M_L(<2\re) \gtrsim
4\times10^{11} M_\odot$) tend towards lower values in the majority of
cases, meaning $f_s \lesssim 0.2$. Low mass galaxies ($M_L(<2\re) \lesssim
4\times10^{11} M_\odot$) show a larger range of possible stellar-mass
fractions at high and low radii, which are in a range between $0.1$
and $0.35$ (see left hand panel of Fig.~\ref{fig7-1}). This is the
reason for the large scatter of enclosed stellar-to-total enclosed
masses at small radii in Fig. \ref{fig:mov1}.\\

Averaged over the whole lens sample we find that the stellar-mass fraction 
declines with increasing radius from its value $f_{s}(<1\re)$ enclosed in $1\re$ to only $\sim 71 \%$
of $f_{s,max}$ at $2 \re$, $\sim 55 \%$ at $3\re$,
$\sim 39 \%$ at $4 \re$ and finally $\sim 33 \%$ at $5 \re$.
Splitting the sample with respect to total mass as done before yields
a different picture: For lenses with $M_L(<2\re) \lesssim 4\times10^{11}
M_\odot$ $79 \%$ of the stellar-mass fraction at $1\re$ remains at $2
\re$, $63 \%$ at $3 \re$, $47 \%$ at $4 \re$ and finally $40 \%$ at $5
\re$. For lenses with $M_L(<2\re) \gtrsim 4\times10^{11} M_\odot$ $64 \%$ of
the stellar-mass fraction at $1\re$ is found at $2 \re$, $48 \%$ at
$3 \re$, $33 \%$ at $4 \re$ and finally $27 \%$ at $5 \re$. The uncertainties of 
stellar-mass fractions at $1\re$ for low $M_L$ lenses are only as high as $10\%$.
For larger radius and mass the $f_s$ errors decline strongly to less than $1\%$. From this
we can conclude the following.
\begin{enumerate}
\item Low mass galaxies show a shallower decline in their enclosed
  stellar-mass fraction than high mass galaxies: either their stellar
  content is less concentrated than in high mass galaxies or their
  dark matter content is more concentrated. This point becomes clearer
  in Section \ref{sec:cse}, where we calculate concentration indices
  of stellar and total mass profiles,
\item The relative stellar-mass fraction of high versus low mass galaxies is significantly offset
by a constant value within $5 \re$ from the centre, i.e.
\end{enumerate}
\begin{equation}
\frac{f_{s}(<R)}{f_{s}(<1\re)} \big|_{M_L>4E11 M_\odot} \approx
\frac{f_{s}(<R)}{f_{s}(<1\re)} \big|_{M_L<4E11 M_\odot} - 0.15
\label{eq:fsfmax}
\end{equation}
\begin{enumerate}
\item[] for $2\re<R<5\re$.
\end{enumerate}

\begin{figure*}
\figurenum{8}
\includegraphics[scale=1.195]{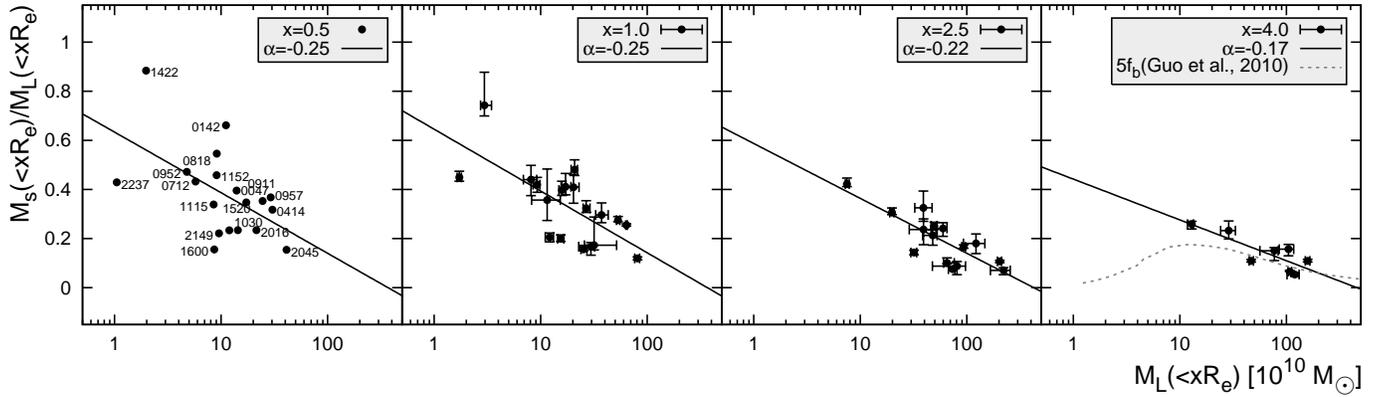}
\caption{The stellar-mass fraction determined at $0.5$,
$1.0$, $2.5$ and $4.0$ $\re$ against total mass. The best fits are
found for the sample excluding lenses with mean stellar-mass fractions
above $1$ at $\re$, which is the case only for B1608. \label{fig6-2}}
\end{figure*}

The latter phenomenon becomes more evident when plotting the
stellar-mass fraction at fixed $R/\re$ against the total mass as in
Fig.~\ref{fig6-2}.  From left to right the panels show the
$f_s$--$M_L$ relation at discrete radii of $0.5$, $1.0$, $2.5$ and
$4.0$ $\re$. It should be emphasized that the solid line fit does not
imply a physical relation extendable to the high or low mass end of
the plot. Note that the relation has a tendency to steepen gradually
towards lower radii whereas the scatter increases. Comparing this to recent results from \cite{GUO10} where the ratio of
total enclosed stellar mass and halo mass $M_{\rm halo}$ is analyzed
with an abundance matching method, we find that their stellar-mass
fraction curve shows a peak at a halo mass of around $6 \times 10^{11}
M_\odot$ and decreasing fractions towards lower and higher halo
masses. This is overplotted in the last panel of
Fig.~\ref{fig6-2}.

The effective height of the curve $f_b$ is reduced in
contrast to our results owing to the fact that there is significant dark
matter in the halo extending up to the virial radius, which is defined
as
\begin{equation}
\rv = \left( \frac{G M_{\rm halo}}{100 H^2(z)} \right)^{1/3}.
\label{eq:rvir}
\end{equation}
$\rv$ is roughly a hundred times larger than the region probed in this
study. 
We list $\rv$ values deduced from $M_s$ of this study given
  their $M_s$-to-$M_{\rm halo}$ relation in Table \ref{tab1}, which
  has additional implications on the lens environment, provided
  that the lens behaves like an SDSS-Galaxy plus simulated halo
  counterpart of respective stellar mass.
To visualize how the computed stellar-mass fractions
change between our resolution range and the virial radius we multiply
a constant factor by the stellar-mass fraction curve from \cite{GUO10}
and divide its total mass by the same factor (here we use $5$).  The
slope of the high mass end of their curve agrees with our best fit of
$M_L^{-0.16 \pm 0.04}$ within error bars. Scaling to lower radii makes
the mismatch for lower $M_L$ even more prominent. However, we conclude
that down to a certain level the $M_s$-to-$M_{\rm halo}$ from
\cite{GUO10} is scalable. In the $5\re$-to-$\rv$-range, the lower-mass
lensing galaxies need to decrease their stellar-mass fractions by a
larger amount than high mass galaxies in order to match with the
results from \cite{GUO10}.  Note that scaling our lens sample
instead towards higher total masses and lower stellar-mass fractions
yields the same result.\\

We should point out that this direct comparison of our results with
\cite{GUO10} is imperfect, since $M_{\rm halo}$ and $M_L$
are differently defined and the spatial distribution of dark matter in
a region not directly addressed in either paper is unknown. On the
other hand the steepest part of the total mass profiles is already
enclosed and the cumulative mass profiles saturate, i.e. the slope of
the $M_L$-to-$f_s$ relation is only slowly changing beyond $5
\re$. These different trends for $f_s$ at lower masses could be
indicative of an underestimated stellar-mass fraction towards smaller
halo masses or an overestimated baryonic content towards higher halo
masses. If the aforementioned study of the $f_s$-to-$M_L$ dependency
is correct, then our findings give rise to the question of what makes
the stellar-mass fraction of low mass galaxies decline less strongly
within $5 \re$ than in the range from $5\re$ up to the virial radius,
in contrast with high mass galaxies.  Expressed in terms of stellar
mass content we find a steeper decrease of stellar-mass fractions
towards larger $M_s$ than our results predict.

The above defined virial radius $\rv$ becomes smaller for lower
stellar mass content. Low $M_s$ galaxies reside in halos with larger
$f_s$ than high $M_s$ galaxies, meaning the mass in the dark matter
halo relative to $M_s$ is even larger, i.e. small galaxies have
more concentrated dark matter halos than larger ones (see also Section
\ref{sec:cse}).

\begin{figure*}
\figurenum{9}
\includegraphics[scale=1.195]{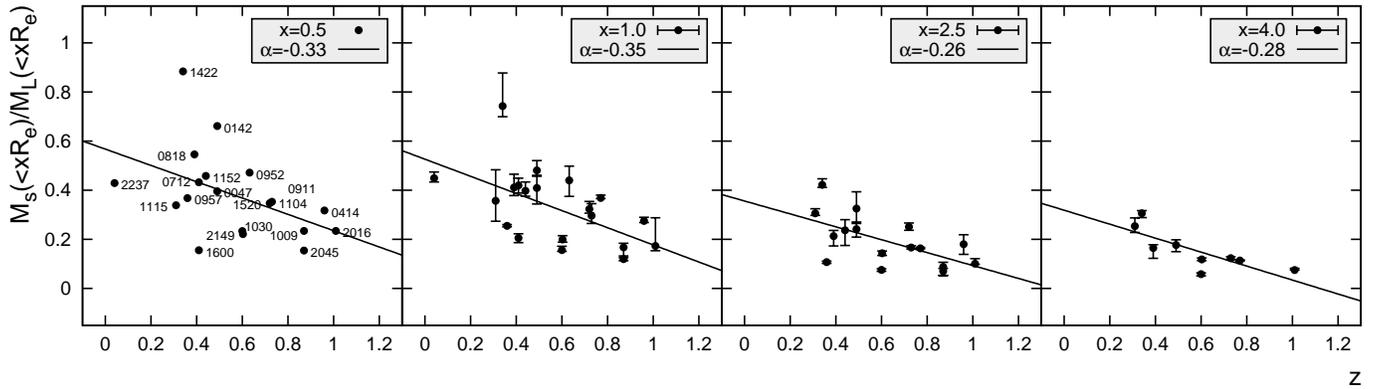}
\caption{The stellar-mass fraction determined at $0.5$, $1.0$, $2.0$ and $4.0$ $\re$ against redshift. \label{fig6}}
\end{figure*}

In order to investigate the influence of the distance/lensing-bias, we also
show the redshift dependence of the stellar-mass fraction in Fig.~\ref{fig6}.
The ordinate might be subject to several biases.
The lensing galaxies plus halo must be massive to produce an observable signature.
The galaxy should not be too faint to be seen and has to obey our selection criterion of sufficient separation from quasar image.
Fig.~\ref{fig6} shows that the correlation between stellar-mass fraction and redshift becomes more pronounced with larger radius.
However, the strongly increasing scatter below $4\re$ blurs the correlation and the slope shows no uniform trend.

\section{Diagnostics of baryon cooling} 
\label{sec:cse}

We now consider two different measures of the stellar and total-mass
profiles, with a view to gaining insight on the evolution of lens
galaxies from formation to observation redshift.

\subsection{Concentration index}

\begin{figure}
\figurenum{10}
\includegraphics[scale=0.96]{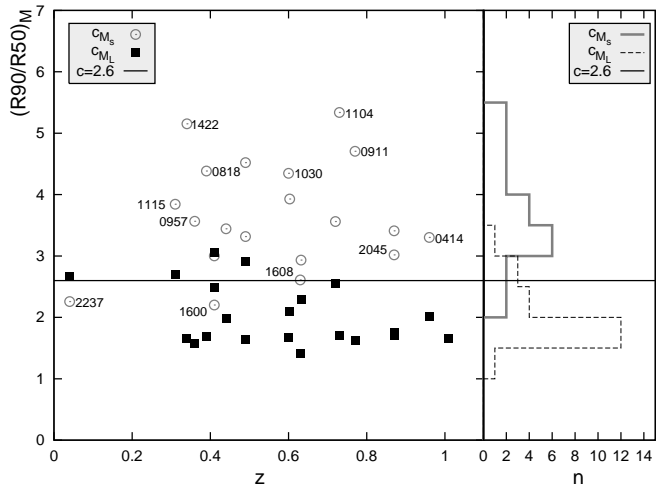}
\caption{Concentration index $c=R90/R50$ versus redshift, where $c$ is the ratio of the radii enclosing $90 \%$ and
$50 \%$ of the total stellar mass ($c_{M_s}$, open circles) and lens
mass ($c_{M_L}$, filled squares). For MG2016 R50 cannot be calculated
due to lack of data points at small radii. The solid line
indicates $c=2.6$ separating early-type ($c>2.6$) from late-type
galaxies ($c<2.6$). Error bars for index $c$ can be found in
Fig.~\ref{fig9-2}. \label{fig9}}
\end{figure}

Our spatially resolved stellar and total mass maps allow us to study
the difference in concentration of the baryon and the total mass
distribution.  We define a concentration index \citep[see
e.g.][]{ber00} as $c\equiv R90/R50$, where $R90$ and $R50$ denote the
radii enclosing 90 and 50 percent of the mass (either stellar or
lens mass). For the luminous component, a concentration index above
2.6 indicates an early-type galaxy, whereas indices below 2.6 refer to
late-type galaxies\citep[see e.g. Fig. 1 in ][]{ig05}. Previous
studies based on the surface brightness distribution use the Petrosian
radius (or a given number of Petrosian radii) to define the total
brightness. In our case, we redefine $c$ and take the respective radii
of our cumulative stellar mass and total mass profiles instead ---
100\% corresponding to enclosed masses at $2\rl$ (except for Q0957 and
HS0818 where it is $1.5 \rl$). In Fig.~\ref{fig9} we show concentration versus redshift in the left-hand panel with
no obvious correlation and the frequencies per concentration bin in
the right-hand panel. Note that defining the concentration values using $\rein$ instead of $\rl$ will change the concentration values  slightly, but even for $\rl/\rein=1.5$ we obtain changes in the lens mass concentration of less than $\sim 30\%$ and only for lenses with high concentrations.

From Fig.~\ref{fig9} the frequency distribution of $c_{M_s}$ peaks between
$3.0$ and $3.5$ which is in agreement with most concentration studies
of early-type galaxies \citep[e.g.][]{YID05,DBT10}. That is, even with
the redefined concentration quantities one can distinguish the lens
galaxies morphologically. Evidence is given by the two late-type
galaxies Q2237 and B1600 which indeed lie below 2.6. 
For the two merging galaxies in the lens B1608, the
  interfering potential (for $c_{M_L}$) or light (for $c_{M_s}$),
  causes the concentration values to be decreased, pushing $c_{M_s}$
  down to 2.6. For the same reason we obtain rather large error bars
  on the lens mass.

\begin{figure}
\figurenum{11}
\includegraphics[scale=1.06]{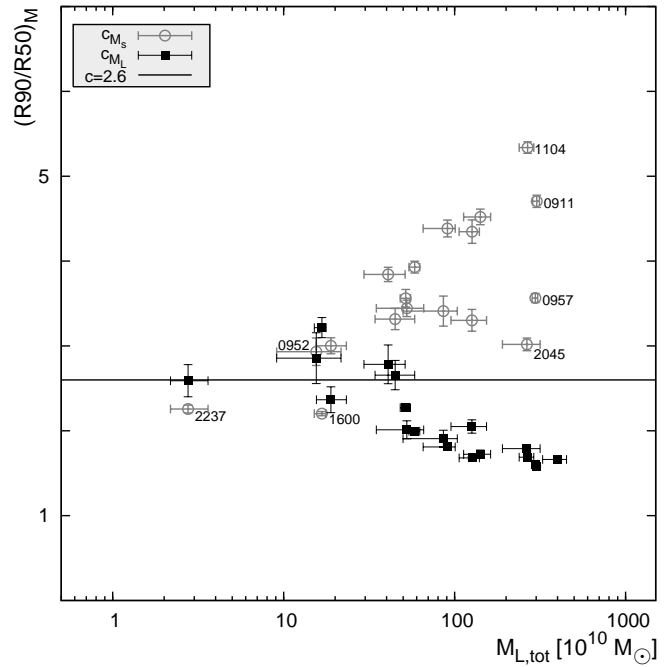}
\caption{As in Fig.~\ref{fig9} but plotted against the total mass $M_{L,tot}$ enclosed in $2\re$. The y-axis error bars represent for $c_{M_L}$ the standard errors for mean concentrations of 300 models. For $c_{M_s}$ the error bars correspond to the uncertainties originating from a $10 \%$ error in the flux per pixel. \label{fig9-2}}
\end{figure}

One could check in detail now if the strong correlation between $c$
and Hubble-type is maintained for the newly defined
$c_{M_s}$. If we define a concentration parameter by means of the total
mass profiles we expect, as our findings in Section \ref{sec3} already suggest, a totally different distribution. Most
lenses exhibit $c_{M_L}$ values in a narrow region between 1.5 and 2.
However, neither in $c_{M_s}$ nor in $c_{M_L}$ can a clear evolutionary
trend be found. Figure \ref{fig9-2} shows that the concentration parameter for
stellar mass $c_{M_s}$ has a rising trend with total
lens mass, whereas $c_{M_L}$ clearly declines with lens mass.

Note that the error bars of $c_{M_s}$ and $c_{M_L}$ are the standard
errors of the $R90/R50$ values of each model in the ensemble
multiplied by student's $t$ for a $95\%$ confidence interval. This was
done since the ensemble can be seen as being part of a normal
population. The horizontal error bars are the $M_L$ errors at the
outermost radius of the reconstructed mass profile. As we can see at low total lens masses the distributions of $M_s$
and $M_L$ are almost the same, which means that the $M_L$ profile
approaches the distribution of the baryonic matter. An interaction
between the baryonic and dark matter distribution seems to be a reasonable
explanation, since already in Section \ref{sec3} we find that the
stellar-mass fractions of less massive lenses are larger than for the more massive lenses.

\begin{figure}
\figurenum{12}
\includegraphics[scale=1.075]{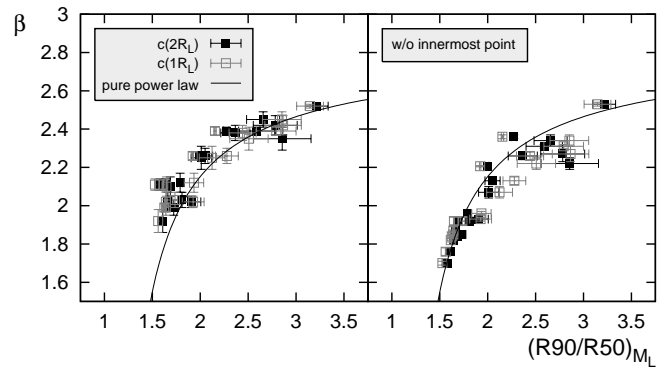}
\caption{Power law index $\beta$ of the density profile plotted against the concentration parameter $c=(R90/R50)_{M_L}$ for the total lens mass distribution.
Filled (open) squares indicate $c=(R90/R50)_{M_L}$ determined for a profile extending to $2\rl$ ($1\rl$).
The left panel uses power law fits to the whole radial range, whereas the right panel neglects the innermost data point.
The solid black line shows the pure power law case according to eq. \ref{eq:beta}. \label{fig9-3}}
\end{figure}

A possible interaction between baryons and dark matter is likely to influence the slope of the total mass distribution close to the center of galaxies.
If we assume a density following a pure power law $\rho(r) \sim r^{-\beta}$ the enclosed mass becomes $M(<R)\sim r^{3-\beta}/(3-\beta)$. Thus the concentration $c$ and
the density slope obey the relation
\begin{equation}
\beta=3-\frac{\ln{0.9/0.5}}{\ln c}.\label{eq:beta}
\end{equation}
Figure \ref{fig9-3} contrasts the relation between $\beta$ and $c$ based on a pure power law (solid line) and data for different radial extents.
The $\beta$ values represent weighted best fits to lens mass profiles with standard errors from the fit.
If the mass distribution does not follow a pure power-law $(R90/R50)$ might depend strongly on the radial extent of the lens ($2\rl$).
Therefore we compare in both panels of Fig. \ref{fig9-3} concentration values inferred from differently sized profiles, with a maximal radius of $1\rl$ and $2\rl$.
The innermost data point has rather large uncertainties and deviates in most cases strongly from the trend at larger radius.
To demonstrate its impact on the relation we contrast fits with (left panel) and without (right panel) regard of the innermost point.
According to eq. \ref{eq:beta} we find that with increasing $M_L(<2\re)$ -- i.e. decreasing concentration -- the slope $\beta$ gets shallower.
It is remarkable how extraordinarily well the weighted power law fits neglecting the innermost point reproduce the simple $\beta(c)$ model at low concentrations, but 
fail to do so at large concentrations, where the data lies below the pure power law relation.
Higher values of $\beta$ correspond to shallower $M_L$ profiles. Including the innermost point always flattens $M_L(<R)$ fits,
which explains why respective $\beta$ values, although less representative for the outer part of the profiles, are in better agreement with eq. \ref{eq:beta}.
We can conclude that
\begin{enumerate}
\item excluding the core region of the lenses, we obtain power law indices and concentration parameters ($R90/R50$) indicative of a pure power-law behaviour for small concentrations.
This notion is strengthened by only small shifts of $(R90/R50)$ going from $2$ to $1\rl$ profiles.
\item For more concentrated total mass distributions, we find evidence for a significant departure from pure power law behaviour.
This is confirmed by significant shifts of $(R90/R50)$ while reducing the extent of the lens from $2$ to $1\rl$ and an increasing $\beta$ error towards higher concentrations.
\end{enumerate}

\subsection{Energy ratio}

\begin{figure*}
\figurenum{13}
\includegraphics[scale=1.195]{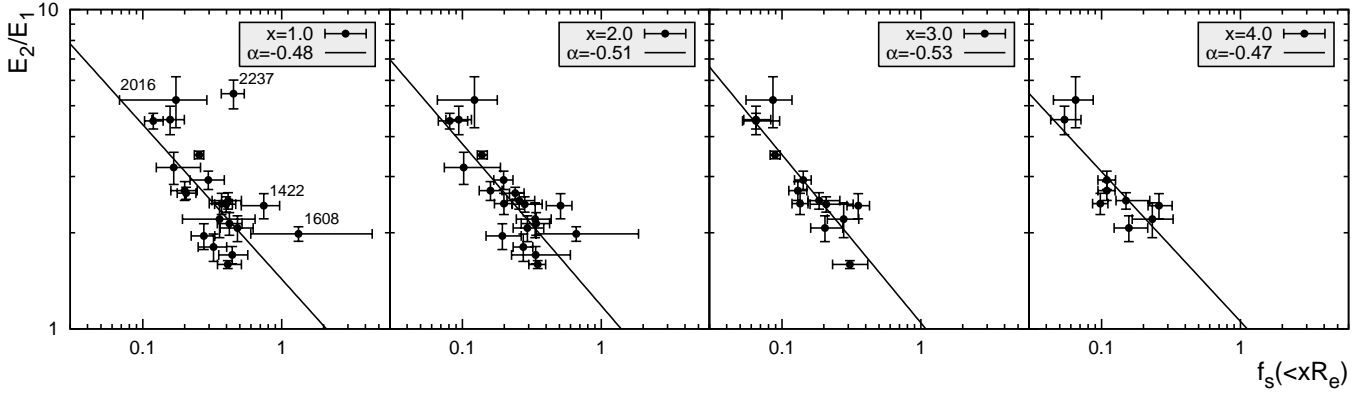}
\caption{The stellar-mass fraction determined at $1.0$, $2.0$, $3.0$ and $4.0$ $\re$ versus $E_2/E_1$. The solid line denotes the best fit of a power-law with slope $\alpha$. The fits exclude Q2237 and MG2016. \label{fig12-2}}
\end{figure*}

\begin{figure*}
\figurenum{14}
\includegraphics[scale=1.195]{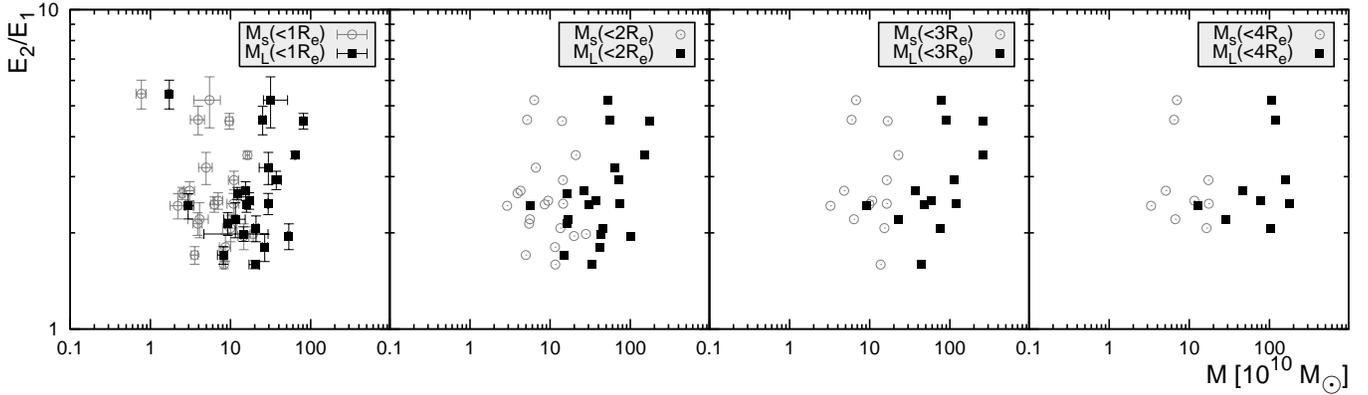}
\caption{The energy ratio versus enclosed stellar and enclosed total
mass at $1.0$, $2.0$, $3.0$ and $4.0$ $\re$. \label{fig13-2}}
\end{figure*}

By means of the stellar mass content one can approach the
subject of galaxy formation from a different viewpoint. The first question is: ``is it feasible to determine a characteristic quantity which gives us the amount of energy lost between the collapse of an initial sphere of
homogeneously distributed baryons and its later state as a lens
galaxy?''.  One could ask as well for a ratio of the radius of the pre-collapse sphere and
an observable spatial quantity, like the effective radius. Even
though this is a rough estimate, one can gain
insight in the evolution process of galaxies.

At the time of collapse $t_1$, a region decouples
from the expansion of the surrounding universe. The baryons which are
assumed to be homogeneously distributed in this sphere are for now
assumed to make up the whole stellar content of the later lens
galaxy, neglecting any kind of active evolution such as caused by mergers, ram pressure, tidal stripping, etc. The
radius $r_1$ of such a sphere at $t_1$ is
\begin{equation}
r_{1} = \left( \frac{M_s}{\frac{4}{3} \pi \Omega_b \rho_c}
\right)^{1/3} (1+z_1)^{-1}
\label{eq1}
\end{equation}
where $\Omega_b= 0.0441 \pm 0.0030$ is the baryonic energy density in
terms of critical density according to \cite{WMAP5} and $\rho_c \simeq
143.87 M_\odot/\rm kpc^3$.  The average Newtonian energy per unit mass
at $t_1$ consists only of the potential energy per unit mass, which is
\begin{equation}
E_1 = -\frac{GM_{\rm tot}}{r_1}.
\end{equation}
The total mass $M_{\rm tot}$ is defined here as
\begin{equation}
M_{\rm tot} = \Omega_m \rho_c (1+z_1)^3 \frac{4}{3} \pi r_{1}^3
\end{equation}
As the collapse goes on the baryons start to fall in to build a more
tightly bound structure. At the observation redshift, i.e. at time $t_2$,
we find mostly objects which are in virial equilibrium, that is
$E=-T$, where $T$ denotes kinetic energy. Thus we can determine the
total energy per unit mass of the galaxy at $t_2$ to be
\begin{equation}
E_2 = -T_2 =-\frac{1}{2} \sil^2,
\end{equation}
where
\begin{equation}
\sil \equiv \sqrt{GM_L(<R)/R} \label{eq:sil}
\end{equation}
is an effective velocity dispersion inferred from lensing (see Table
\ref{tab1}). It is computed at $R=\rl$ and assumed not to vary
drastically with radius.  This effective dispersion has been
shown to be an appropriate surrogate for the observed
kinematic velocity dispersion \citep{DL09}. Thus,
\begin{equation}
\frac{E_1}{E_2}  = \frac{2GM_{\rm tot}}{\sil^2 r_1} = \frac{G \Omega_m \rho_c (1+z_1)^3 \frac{8}{3} \pi r_1^2 }{ \sil^2}
\end{equation}
using Eq.~\ref{eq1},
\begin{equation}
\frac{E_2}{E_1}  \propto \frac{\sil^2}{M_s^{2/3}}.
\label{eq:6}
\end{equation}
Thus we get a quantity $E_1/E_2\propto \rl\times M_s^{2/3}/M_L$ or $\rl
M_s^{-1/3}f_s$. This is reminiscent of the Kormendy relation, except that it relates to
three-dimensional rather than projected densities. For definiteness,
we assume a formation redshift $z_1=5$, but the value only implies a
multiplicative constant.  Plotting the energy ratio against the
stellar-mass fraction, we find a strong correlation
(Fig.~\ref{fig12-2}) regardless of the enclosure radius. The slope
changes only marginally, but the scatter decreases with increasing radius.

However, $E_2/E_1$ appears to be uncorrelated with the enclosed
stellar and total mass. For different radii one
obtains Fig.~\ref{fig13-2}. The fact that $E_2/E_1$ exhibits such a
tight correlation with $f_s$ but no clear correlation to contributing
masses, can be interpreted as insensitivity of the star formation in
early-type galaxies to active evolution processes over the time span
from $z_1$ to $z_{\rm lens}$.

\section{Discussion}
\label{sec5}

A resolved, model-independent and thus non-degenerate (w.r.t. $M_s$ and $M_L$ for fixed $f_s$) estimate of stellar versus
total mass within galaxy halos is crucial to constrain current galaxy formation models and prescriptions of baryon-dark
matter interactions used therein. Besides dynamical methods to explore
scales below 10 kpc the combination of strong gravitational lensing
and population synthesis used in this paper is most promising to give
robust estimates of stellar-mass fractions.

The analysis of the radial dependence of the mass profiles of 21
CASTLES lenses presented in this paper allows us to draw the following
conclusions. The relation between basic galaxy properties, i.e. $M_L$, $M_s$ and
$\rl$ cannot simply be scaled with their mass. The scatter in this
parameter space turns out to be especially large for galaxies of
smaller size. The study of $M_s$ versus $M_L$ and of the stellar-mass
fractions ($f_s\equiv M_s/M_L$) enables us to discriminate between lens
galaxies below and above $M_L(<2\re)=4\times 10^{11} M_\odot$. The high mass class populates a lower and narrower $f_s$ regime (0.05 to 0.2)
on given scales and runs out earlier of stellar mass (i.e. at lower
enclosed radius) than low mass lenses. The latter exhibit a more
inhomogeneous behaviour with a wider range in $f_s$ (0.1 to 0.5) and
respective slopes.

We conclude that between $1.5$ and $2.5 \re$ dark matter halos start
to dominate the matter balance depending on their total mass. This
$M_L$-dependence causes high mass galaxies to gain mass primarily in the
form of dark matter already at lower radii than low mass
galaxies. Therefore the slope of the mass-to-light relation, which is
a projection of the fundamental plane --- or our equivalent
representation, $M_L^\eta \propto M_s$ --- becomes shallower with
increasing radius and asymptotically approaches a slope of $\eta =
0.76\pm0.07$. Thus the FP tilt can be recovered as a gradually growing process with radius. Equivalently, the stellar-mass fraction
shows a strong correlation to the total mass. As we contrast
$f_s(M_L)$ with a comparable curve deduced by abundance matching from
\cite{GUO10} dissimilarities for low $M_L$ galaxies become more
evident the smaller the enclosed region gets. This is likely to be a result of different halo definitions, physical properties and
processes, like baryon-dark matter interactions and adiabatic contraction which is beyond the scope of the
aforementioned study. However, the $f_s$-to-$M_{\rm halo}$ relation
scaled down to $4\re$ agrees quite well with lenses with $M_L \sim
10^{12} M_\odot$, since the biggest part of stellar matter is still enclosed.

Another important result of this study addresses the concentration of
stellar ($c_{M_s}$) and total ($c_{M_L}$) mass profiles. The
rule-of-thumb delimiter of $c=2.6$ which separates early-type galaxies
($c>2.6$) from late-types $c<2.6$) holds also for the concentration
parameter ($c_{M_s}$) defined by means of stellar mass instead of
luminosity. In the low mass regime $M_L(<2\re) \lesssim 4\times 10^{11} M_\odot$ both,
$c_{M_s}$ and $c_{M_L}$, tend to similar values around 2.6. This means
that the total mass profile is likely to be influenced by the
distribution of baryonic matter in stars. From $10^{11} M_\odot$ upwards,
$c_{M_s}$ and $c_{M_L}$ diverge, due to a stronger confinement of stars in more massive dark matter halos.
The $c_{M_L}$ values above $4\times 10^{11} M_\odot$ stay around $\sim 2$ instead.
Studying the interdependency of density slope and $c_{M_L}$ we find that the reconstructed lens
 profiles show deviations from a pure power-law mass model, which is evidence for the sensitivity to the radial trends of the dark matter distribution.

The results presented in this paper are critical to ongoing studies about
the reliability of parametric lens models and prescriptions used in
galaxy-formation models. The radially resolved stellar to dark matter
fractions should thus also serve as benchmarks for future simulations.

\section*{Acknowledgments}
We would like to thank the anonymous referee for comments that helped to improve this paper.

\appendix

\section{Animated results}
\label{app}

An animated version of Fig. \ref{fig:mov1} is provided in the
  online material of this paper.  The movie contains three panels. The
  left panel shows the enclosed $M_L$ against enclosed $M_s$ plane
  depending on the size of the aperture, defined by the radial
  distance $x\re$ to the centre of the lens galaxy. The solid black
  line denotes the equality of total and stellar mass, whereas the
  dashed lines represent the upper and lower limit of the global
  baryon fraction \citep{WMAP5}. The upper right panel shows the lens PG1115
  which is highlighted by a red label in the left panel.  The lower right panel shows stellar baryon fraction
  versus radius as in Fig. \ref{fig7-1}. The solid black line denotes
  the baryon fraction curve of PG1115.  With each time step of the
  movie the enclosure radius increases indicated by the factor x in
  the legend of the left panel and the red lines in the two right-hand
  panels. We cover a radial distance from $0.125$ $\re$ to $5$ $\re$
  in 40 time steps.

\begin{center}
\begin{table*}
\tabcolsep1.5mm
\begin{tiny}
\hspace{0.5cm}\begin{tabular}{@{} lcccrccclcccccccc @{}}
\hline\\
\textbf{Lens}  &
$\mathbf{z_L}$ &
$\mathbf{z_s}$ &
$\mathbf{\Delta\theta}$ &
$\mathbf{\Delta\theta}$ &
$\mathbf{\re}$ &
$\mathbf{\frac{\rl}{\re}}$ &
$\mathbf{\frac{\rein}{\re}}$ &
$\mathbf{M_L(<2\re)}$ &
$\mathbf{M_s(<2\re)}$ &
$\mathbf{\frac{M_s}{L_V}}$ & 
$\mathbf{\sil}$ &
$\mathbf{R_{\textrm{\tiny{vir}}}}$ &
$Env$ \\
\vspace{0.1cm} & & & [$''$] & [kpc] & [$''$] &  &  &[$10^{10}\mathbf{M_{\odot}}$]  & [$10^{10}\mathbf{M_{\odot}}$] & [$\frac{M_{\odot}}{L_{\odot}}$] & [km s$^{-1}$] & [kpc] & \\
\hline \hline\\
\vspace{0.1cm}Q0047&0.485&3.60&2.20&12.82&$0.880 \pm 0.025$&$1.45 \pm 0.04$&$1.31^{+0.02}_{-0.01}$&$\phantom{1}33.27_{-3.03}^{+3.80}$ &$11.58 \pm 0.43$&$5.04_{-0.89}^{+1.09}$&$189.2_{-4.8\phantom{0}}^{+5.7}$&538.4&G(9)\\
\vspace{0.1cm}Q0142&0.490&2.72&2.23&13.10&$0.703\pm 0.013$&$2.64 \pm 0.05$&$2.54^{+0.70}_{-0.83}$&$\phantom{1}45.81_{-5.50}^{+3.31} $&$13.45 \pm 2.05$&$2.92_{-0.69}^{+0.90}$&$245.8_{-46.8}^{+16.8}$&666.5&-\\
\vspace{0.1cm}MG0414&0.960&2.64&2.08&16.01&$0.734 \pm 0.018$&$1.85 \pm 0.05$&$1.67^{+0.09}_{-0.07}$&$102.76_{-19.04}^{+16.55}$&$19.90 \pm 2.29$&$5.62_{-1.62}^{+2.27}$&$247.0_{-12.9}^{+9.93}$&635.0&1\\
\vspace{0.1cm}B0712&0.410&1.34&1.29&6.82&$0.702 \pm 0.016$&$1.15 \pm 0.03$&$1.02^{+0.05}_{-0.02}$&$\phantom{1}16.13_{-2.04}^{+2.55}$&$\phantom{1}5.46 \pm 0.52$&$3.27_{-0.52}^{+0.62}$&$164.1_{-6.78}^{+6.73}$&295.8 & 1\\
\vspace{0.1cm}HS0818&0.390&3.21&2.56&13.15&$0.679 \pm 0.013$&$3.27 \pm 0.06$&$2.51^{+0.55}_{-0.70}$&$\phantom{1}36.88_{-5.48}^{+3.69}$&$\phantom{1}9.44 \pm 0.97$&$5.42_{-0.92}^{+1.11}$&$245.8_{-41.8}^{+16.9}$&517.8&1\\
\vspace{0.1cm}RXJ0911&0.769&2.8&3.22&23.16&$0.725 \pm 0.011$&$3.09 \pm 0.05$&$2.27^{+0.09}_{-0.14}$&$\phantom{1}73.04_{-1.36}^{+1.15}$&$14.52 \pm 1.87$&$2.99_{-0.75}^{+1.00}$&$278.6_{-6.0\phantom{1}}^{+3.9}$&590.4&C\\
\vspace{0.1cm}BRI0952&0.632&4.5&\underline{0.99}&5.27&$0.619 \pm 0.021$&$1.04 \pm 0.04$&$0.82^{+0.32}_{-0.13}$&$\phantom{1}14.76_{-5.87}^{+5.67}$&$\phantom{1}4.98 \pm 0.35$&$2.78_{-0.52}^{+0.64}$&\underline{$140.0_{-20.6}^{+15.8}$}&233.5& G(5)\\
\vspace{0.1cm}Q0957&0.356&1.41&\underline{6.17}&\underline{29.98}&\underline{$1.491 \pm 0.018$}&$3.51 \pm 0.04$&$2.39^{+0.14}_{-0.20}$&$151.29_{-6.58}^{+5.17}$&$20.92 \pm 0.96$&$2.52_{-0.50}^{+0.62}$&\underline{$374.6_{-14.0}^{+11.9}$}&950.0&C\\
\vspace{0.1cm}LBQS1009&0.880&2.74&1.54&11.56&$0.963 \pm 0.028$&$1.27 \pm 0.04$&$1.07^{+0.18}_{-0.38}$&$\phantom{1}64.73_{-24.46}^{+11.97}$&$\phantom{1}6.62 \pm 0.96$&$2.14_{-0.61}^{+0.86}$&$220.9_{-65.6}^{+17.8}$&252.1&-\\
\vspace{0.1cm}B1030&0.600&1.54&1.32&8.60&$0.675 \pm 0.019$&$2.05 \pm 0.06$&$2.11^{+0.36}_{-0.33}$&$\phantom{1}55.09_{-3.52}^{+2.11}$&$\phantom{1}5.16 \pm 0.80$&$1.12_{-0.25}^{+0.32}$&$256.8_{-17.9}^{+9.9}$&277.2& 1\\
\vspace{0.1cm}HE1104&0.730&2.32&3.19&22.54&$0.681 \pm 0.010$&$3.08 \pm 0.05$&$3.11^{+0.32}_{-0.25}$&$\phantom{1}72.80_{-3.22}^{+3.68}$&$14.43 \pm 1.62$&$3.50_{-0.80}^{+1.04}$&$302.9_{-11.0}^{+11.9}$&604.5&-\\
\vspace{0.1cm}PG1115&0.310&1.72&2.43&10.76&$0.478 \pm 0.009$&$2.94 \pm 0.06$&$2.50^{+0.22}_{-0.13}$&$\phantom{1}16.80_{-1.23}^{+1.31}$&$\phantom{1}5.61 \pm 1.19$&$3.68_{-0.92}^{+1.23}$&$191.6_{-12.9}^{+11.7}$&354.0&G(13)\\
\vspace{0.1cm}B1152&0.439&1.02&1.56&8.60&$0.691 \pm 0.013$&$1.62 \pm 0.03$&$1.43^{+0.41}_{-0.25}$&$\phantom{1}30.43_{-5.86}^{+4.23}$&$\phantom{1}8.57 \pm 0.65$&$2.84_{-0.50}^{+0.61}$ &$216.9_{-37.6}^{+22.8}$&431.4& -\\
\vspace{0.1cm}B1422&0.337&3.62&1.29&6.02&\underline{$0.241 \pm 0.003$}&$4.49 \pm 0.06$&$3.56^{+0.23}_{-0.43}$&$\phantom{10}5.72_{-0.24}^{+0.39}$&$\phantom{1}2.91 \pm 0.45$&\underline{$6.40_{-1.02}^{+1.21}$} &$162.9_{-11.9}^{+4.9}$&231.3&G(17)\\
\vspace{0.1cm}SBS1520&0.710&1.86&1.57&10.97&$0.947 \pm 0.018$&$1.28 \pm 0.02$&$0.95^{+0.08}_{-0.07}$&$\phantom{1}41.98_{-1.82}^{+1.49}$&$11.49 \pm 1.41$&$2.63_{-0.79}^{+1.14}$&$198.3_{-9.1\phantom{1}}^{+7.2}$&434.3&G(4)\\
\vspace{0.1cm}B1600&0.420&1.59&1.38&7.45&$1.015 \pm 0.007$&$1.00 \pm 0.01$&\underline{$0.75^{+0.05}_{-0.07}$}&$\phantom{1}16.38_{-1.67}^{+0.82}$&$\phantom{1}3.93 \pm 0.17$&$6.37_{-1.44}^{+1.86}$ &$154.4_{-12.1}^{+6.1}$&236.0& G(6)\\
\vspace{0.1cm}B1608&0.630&1.39&2.10&13.92&$0.839 \pm 0.047$&$1.82 \pm 0.01$&$1.55^{+0.12}_{-0.20}$&$\phantom{1}42.49_{-26.52}^{+35.50}$&\underline{$27.99 \pm 1.63$}&$2.44_{-0.45}^{+0.55}$&$266.8_{-14.9}^{+6.0}$&\underline{972.9}& G(8)\\
\vspace{0.1cm}MG2016&\underline{1.010}&3.3&3.36&26.22&$0.406 \pm 0.009$&\underline{$6.12 \pm 0.14$}&\underline{$11.1^{+1.6}_{-2.8}$}&$\phantom{1}52.05_{-5.13}^{+14.23}$&$\phantom{1}6.34 \pm 1.99$&$0.89_{-0.28}^{+0.40}$&$308.6_{-26.5}^{+9.9}$&242.6&C(69)\\
\vspace{0.1cm}B2045&0.870&1.28&1.93&14.46&$0.950 \pm 0.019$&$1.48 \pm 0.03$&$1.23^{+0.13}_{-0.15}$&\underline{$173.07_{-34.20}^{+21.02}$}&$14.05 \pm 1.03$&$2.47_{-0.50}^{+0.62}$&$338.8_{-40.2}^{+16.6}$& 517.2&- \\
\vspace{0.1cm}HE2149&0.603&2.03&1.71&10.02&$0.531 \pm 0.008$&$2.59 \pm 0.04$&$1.67^{+0.16}_{-0.23}$&$\phantom{1}26.82_{-2.19}^{+2.07}$&$\phantom{1}4.28 \pm 0.47$&\underline{$0.79_{-0.16}^{+0.20}$}&$191.2_{-9.1\phantom{1}}^{+7.1}$&242.6&-\\
\vspace{0.1cm}Q2237&\underline{0.039}&1.7&1.83&\underline{1.40}&$1.090 \pm 0.014$&\underline{$0.89 \pm 0.01 $}&$0.81^{+0.01}_{-0.01}$&\phantom{10}\underline{$2.76_{-0.58}^{+0.85}$}&\underline{$\phantom{1}1.15 \pm 0.12$}&$4.39_{-1.26}^{+1.76}$&$145.2_{-3.9\phantom{1}}^{+3.8}$&\underline{212.4}&-\\
\hline\hline
\end{tabular}
\end{tiny}
\caption{Full set of gravitational lenses used for this analysis. All quantities in the table assume $H_0 = 72$ km s$^{-1}$ Mpc$^{-1}$, $\Omega_{m}=0.3$ and $\Omega_{\Lambda}=0.7$. The underlined values show maximum and minimum. $\Delta \theta$ is the image separation. For systems with more than two images the maximal image separation between two images is given. Columns $\re$, $\rl/\re$ and $\rein/\re$ contain Petrosian radii determined in the observed $H$-band with $1\sigma$ error bars. Note that $\rein$ is computed from the critical surface density of the pixelated maps. The total and stellar masses enclosed within $2 \re$ are given in the following two columns. The stellar mass-to-light ratios in the rest-frame $V$-band are median values of all models. $\sil$ denotes the velocity dispersion at $\rl$. Column $R_{\textrm{\tiny{vir}}}$ gives the virial radius calculated using Eq. \ref{eq:rvir} and our stellar mass values in combination with the $M_s$-to-$M_{\rm halo}$ relation from Guo et al. (2010). The column labeled \emph{Env} contains environmental information. ``C'' denotes a cluster environment, ``G'' a group environment and ``1'' a lens with only one known companion. If known the number of group members is given in parentheses.  References for given values are mentioned in section \ref{sec:sample}.
Colours and magnitudes are in agreement with comparable quantities in Rusin et al. (2003).  \label{tab1}
} 
\end{table*}
\end{center}

\begin{center}
\begin{table*}
\begin{tiny}
\hspace{0.9cm}\begin{tabular}{@{} llrccclccclccclccclccc @{}}
\hline
\textbf{Lens} & \textbf{Im.} & & \multicolumn{3}{c}{\textbf{Bands}} & & \multicolumn{3}{c}{\textbf{PSF}} & & \multicolumn{3}{c}{\textbf{fitting}} & & \multicolumn{3}{c}{\textbf{masking}} & & \multicolumn{3}{c}{\textbf{constraints}} \\
              &                 & &I & V & H & & I & V & H  & & I & V & H & & I & V & H & & I & V & H\\
 \hline \hline
Q0047&   4  &\vline& \checkmark & \checkmark &\checkmark  &\vline& tt & tt & A(1030)  &\vline& - & - & -       &\vline& - & 4P & -         &\vline&- &-&- \\
Q0142&   2  &\vline& \checkmark & \checkmark &\checkmark  &\vline& it & tt & it      &\vline& 2P & 1P & 2P  &\vline& - & 1P & -         &\vline& -&$n$&- \\
MG0414&  4  &\vline& \checkmark & &\checkmark             &\vline& tt & & it          &\vline& - & & 4P       &\vline& 4P & & -           &\vline& -&&m \\
B0712&   4  &\vline& \checkmark & \checkmark & \checkmark &\vline& tt & tt & tt        &\vline& - & - & 1P&\vline& 4P & 4P & 3P   &\vline& $\re$&$\re$& - \\
HS0818&  2  &\vline& \checkmark &\checkmark&\checkmark    &\vline& A & A & A(1030)    &\vline& 1P & 1P & 2P &\vline& B & B & -          &\vline& -&PA&- \\
RXJ0911& 4  &\vline& \checkmark & &\checkmark             &\vline& it & & it         &\vline& 4P & & 4P1S     &\vline& - & & -         &\vline& $n$,\underline{S}&&\underline{$xy$}\\
BRI0952& 2  &\vline& \checkmark &\checkmark&\checkmark    &\vline& it & it & it     &\vline& 2P & 2P & 2P   &\vline& - & - & -          &\vline& $b/a$&-&$b/a$ \\
Q0957&   2+2&\vline& \checkmark & &\checkmark             &\vline& it & & it         &\vline& 1P & & 2P      &\vline& A & & -            &\vline& -&&-  \\
LBQS1009&  2  &\vline& & & \checkmark                       &\vline& & &*(0414)        &\vline& & & 2P          &\vline& & & -              &\vline& &&$n$ \\
B1030&   2  &\vline& & & \checkmark                       &\vline& & & A               &\vline&  & & 1P1S    &\vline& & & 1P         &\vline& &&$n$\\
HE1104&  2  &\vline& \checkmark &&\checkmark              &\vline& A & & *(0414)     &\vline& 1P & & 2P      &\vline& - & & -            &\vline& $b/a$,PA&&\underline{$xy$}  \\
PG1115&  4  &\vline& \checkmark & \checkmark &\checkmark  &\vline& tt & it & it      &\vline& - & - & 4P     &\vline& 4P & 4P & -        &\vline& -&$n$&$\re$ \\
B1152&   2  &\vline& & & \checkmark                       &\vline& & & A               &\vline& & & 1P1S       &\vline& & & 1P         &\vline& &&-\\
B1422&   4  &\vline& \checkmark&\checkmark&\checkmark     &\vline& tt & tt & *(0414) &\vline& 1P & 1P & 4P  &\vline& 3P & 3P &-   &\vline& $\re$&$\re,b/a$&$b/a$\\
SBS1520& 2  &\vline& \checkmark & &\checkmark             &\vline& it & & *(1520)  &\vline& 2P & & 2P        &\vline& 1* & & 1*          &\vline& $\re$&& \underline{S} \\
B1600&   2  &\vline& & & \checkmark                       &\vline& & & A        &\vline& & & 2P          &\vline& & & -        &\vline& && -\\
B1608&   4  &\vline& \checkmark&  &\checkmark &\vline& tt &  &*(0414)  &\vline& 1P1S &  & 1P1S &\vline& 3P & & 3P &\vline& $\re$,\underline{S}& &$n,\re$,\underline{S},\underline{$xy$}\\
MG2016&  2+4  &\vline& \checkmark & &\checkmark             &\vline& tt & & *(0414)    &\vline& - & & 1P1S         &\vline& 2P & & 1P2*      &\vline& $b/a$&&- \\
B2045&   4  &\vline& \checkmark &\checkmark&\checkmark    &\vline& tt & tt & *(2045)  &\vline& - & - & 1P     &\vline& 3P1b & 3P & 3P1b2* &\vline& -&-&$n$ \\
HE2149&  2  &\vline& \checkmark &&\checkmark              &\vline& it & & *(0414)   &\vline& 2P & & 2P       &\vline& - & & -            &\vline& - &&- \\
Q2237&   4  &\vline& \checkmark & \checkmark &\checkmark  &\vline& tt & tt & *(1654) &\vline& - & - & 4P      &\vline& 4P & 4P & -        &\vline& \underline{S} &PA,\underline{S}&\underline{S} \\
\hline\hline
\end{tabular}
\end{tiny}
\caption{List of lens systems and how they were treated to obtain surface brightness profiles of the lens galaxies. From left to right the Lens-ID, the number of lensed images, the bands for which fitting was feasible, the PSF picking method, fitted and masked objects and necessary constraints are given. The column \emph{PSF} includes the abbreviations \emph{tt} for a PSF created with \emph{TinyTim}, \emph{A} for the outermost and thus fairly isolated quasar image, \emph{*(0414)} for an isolated star close to lens MG0414, \emph{it} for the iteration method applied to the most isolated image and \emph{A(1030)} for an isolated quasar image taken from lens B1030 used for subtracting quasar images and for the convolution of the whole image. In column \emph{fitting} we summarize the number of objects, not significantly contributing to the lens mass, which are fitted with previously picked PSFs (\emph{P}) and Sersic profiles (\emph{S}). In column \emph{masking} \emph{xP} refers to a number of $x$ masked out point sources, mostly quasar images, \emph{b} denotes resolved but indistinct objects which are not necessarily connected to the lens mass and henceforth excluded from each fit. Point sources like foreground stars indicated by * are also masked out. \emph{A},\emph{B} refer to quasar images which could be masked out given their separation from the lens galaxy. The last column states the type of constraint used if necessary to prevent each fit from diverging. The constrained parameters are: effective radius $\re$, Sersic index $n$, axis ratio $b/a$, magnitude \emph{mag}, position angle \emph{PA}, sky background \emph{S} and position of the Sersic profile $xy$. Parameters fixed at a certain value are underlined.\label{tab2}}
\end{table*}
\end{center}

\begin{center}\begin{figure*}
\figurenum{15}
\vspace{-0.01cm}
\hspace{1cm}\begin{minipage}[l]{0.15\textwidth}
\small
\begin{tiny}
\begin{verbatim}
object B0712+472
redshifts 0.41 1.34
shear -45
quad 
-0.013 -0.804
 0.795 -0.156 0
 0.747 -0.292 0
-0.391  0.307 0
\end{verbatim}
\end{tiny}\end{minipage}
\begin{minipage}[r]{0.15\textwidth}
\includegraphics[width=49pt, bb = 0 35 320 320]{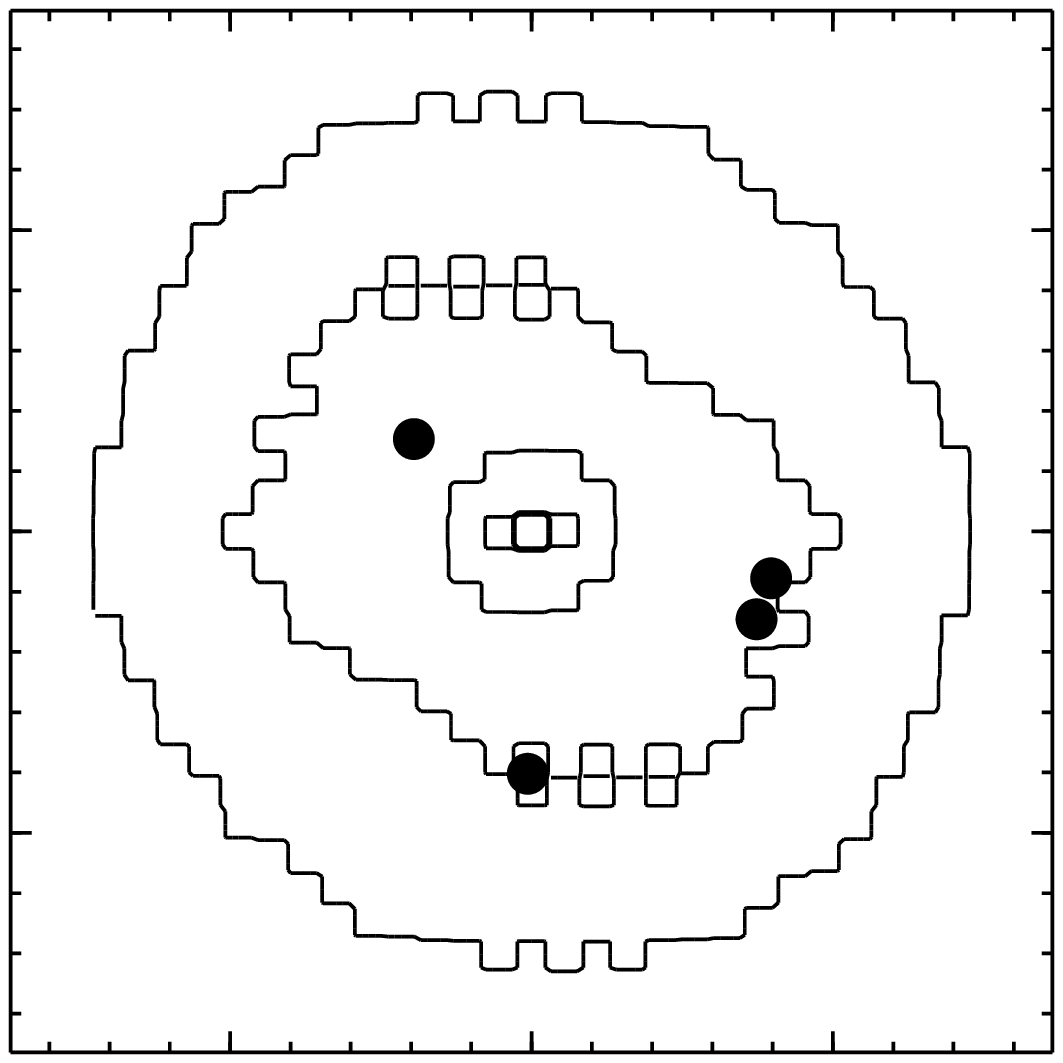}\\
\end{minipage}\begin{minipage}[r]{0.15\textwidth}
\small
\begin{tiny}
\begin{verbatim}
object 1030+071
redshifts 0.600 1.540
shear +45
double
 0.846  1.097
-0.085 -0.156 0
\end{verbatim}
\end{tiny}\end{minipage}
\begin{minipage}[r]{0.15\textwidth}
\includegraphics[width=49pt, bb = 0 36 320 320]{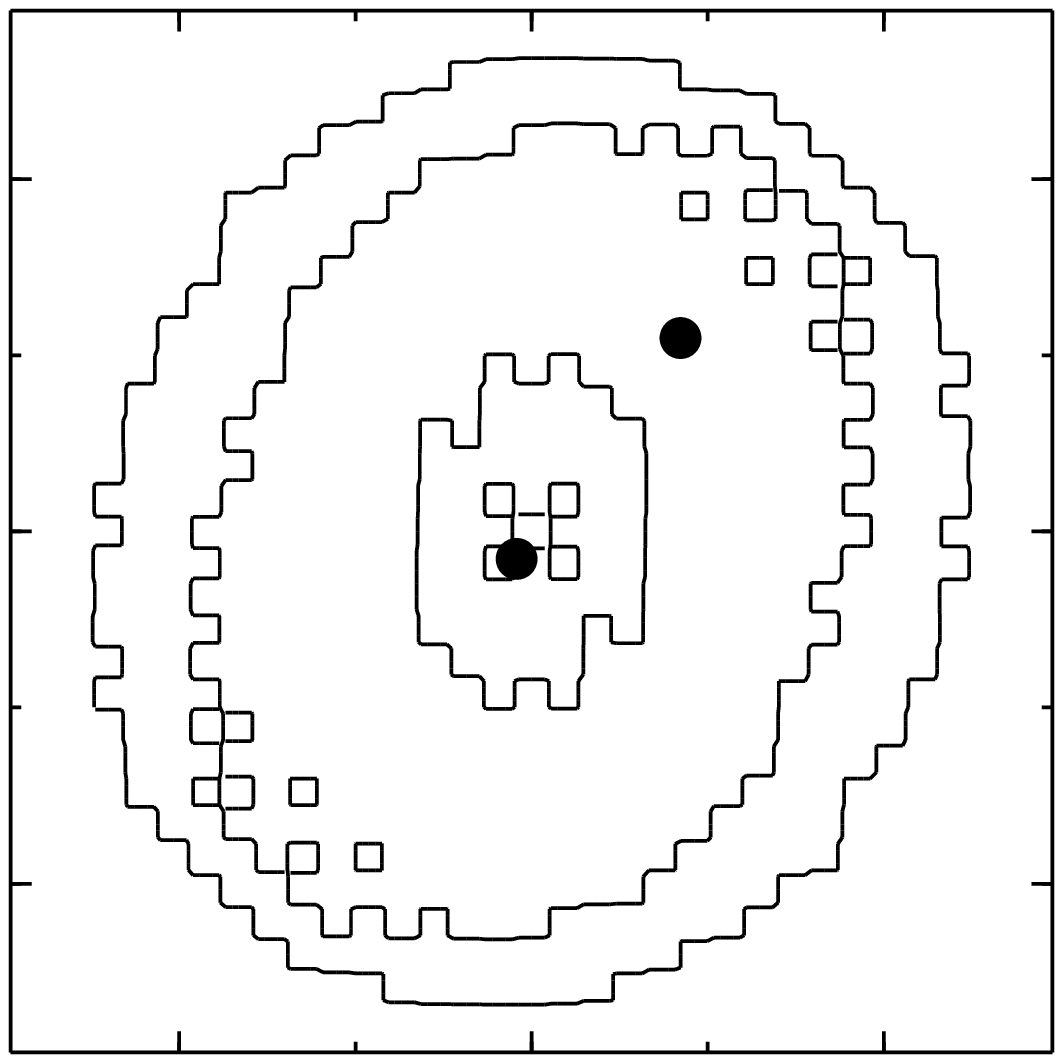}\\
\end{minipage}\begin{minipage}[l]{0.15\textwidth}
\small
\begin{tiny}
\begin{verbatim}
object B1152+200 
redshifts 0.439 1.019
shear +30 
double  
 0.549  0.978  
-0.387 -0.268 0
\end{verbatim}
\end{tiny}\end{minipage}
\begin{minipage}[r]{0.15\textwidth}
\includegraphics[width=49pt, bb = 0 35 320 320]{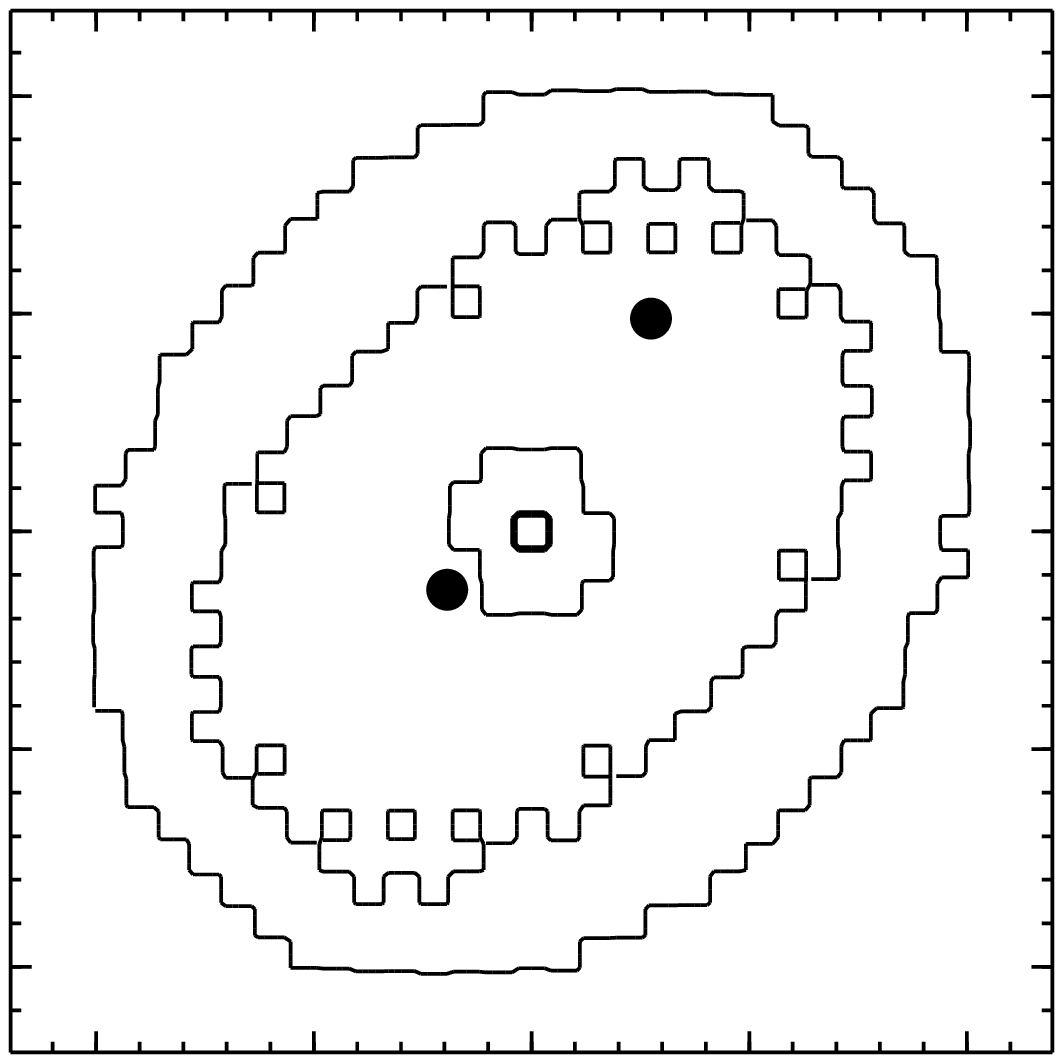}\\
\end{minipage}
\vspace{+0.1cm}

\hspace{1cm}\begin{minipage}[l]{0.15\textwidth}
\small
\begin{tiny}
\begin{verbatim}
object B1422+231
redshifts 0.337 3.620
shear +45
quad
 1.079 -0.095
 0.357  0.973 0
 0.742  0.656 0
-0.205 -0.147 0
\end{verbatim}
\end{tiny}\end{minipage}
\begin{minipage}[r]{0.15\textwidth}
\includegraphics[width=49pt, bb = 0 35 320 320]{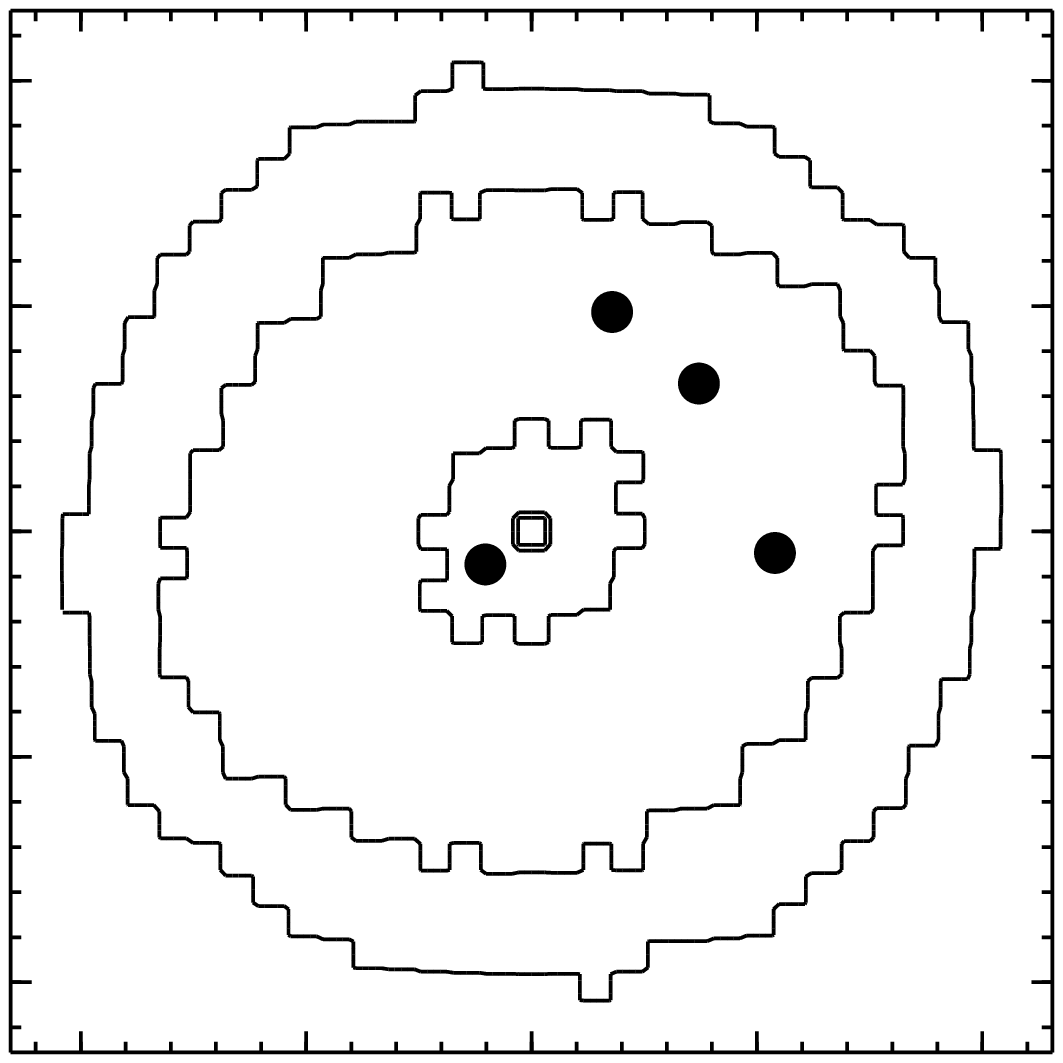}\\
\end{minipage}\begin{minipage}[l]{0.15\textwidth}
\small
\begin{tiny}
\begin{verbatim}
object B1600+434
redshifts 0.42 1.59
shear +60
double
 0.610  0.814
-0.110 -0.369 51
\end{verbatim}
\end{tiny}\end{minipage}
\begin{minipage}[r]{0.15\textwidth}
\includegraphics[width=49pt, bb = 0 36 320 320]{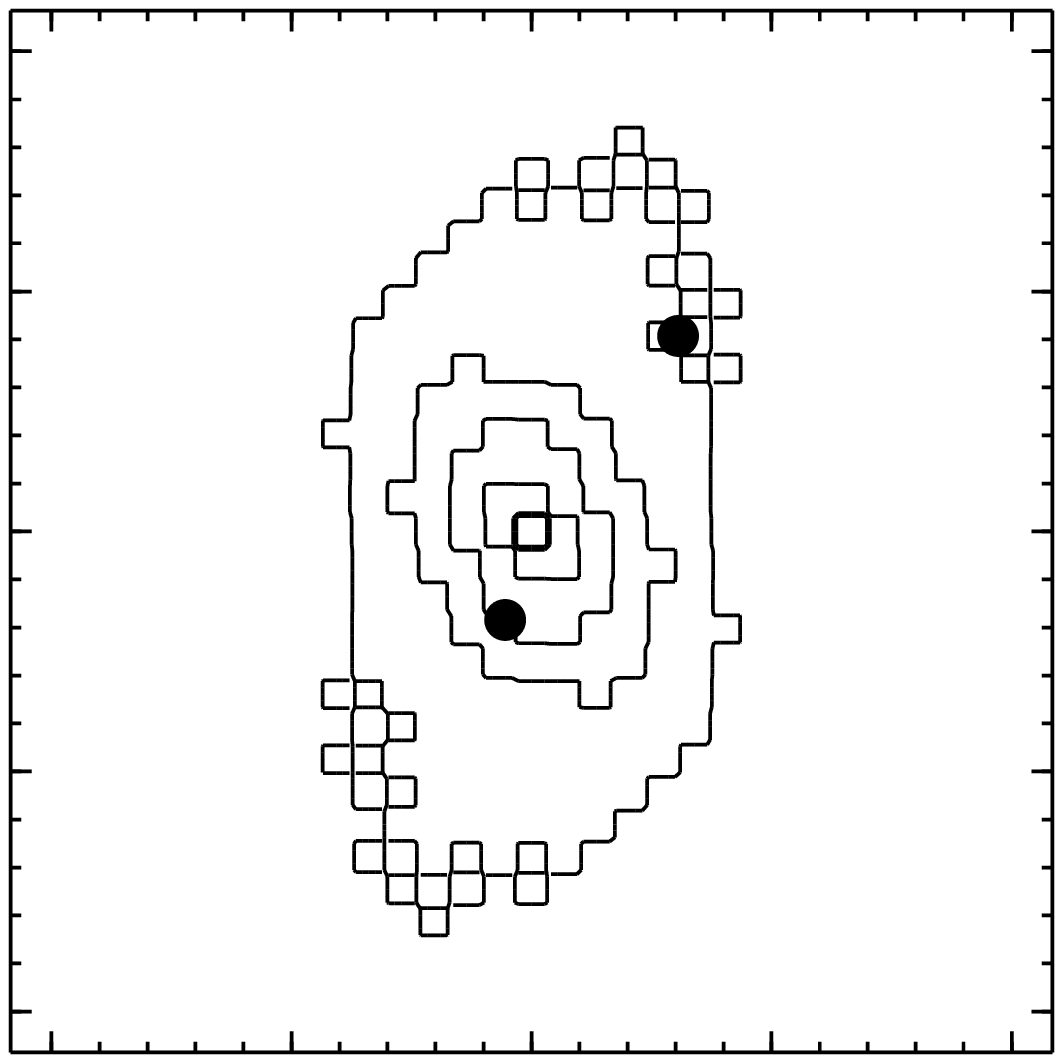}\\
\end{minipage}\begin{minipage}[l]{0.15\textwidth}
\small
\begin{tiny}
\begin{verbatim}
object B1608+656 
redshifts 0.630 1.394
shear -30 
quad 
-1.300 -0.800 
-0.560  1.160  31 
-1.310  0.700   5 
 0.570 -0.080  40 
\end{verbatim}
\end{tiny}\end{minipage}
\begin{minipage}[r]{0.15\textwidth}
\includegraphics[width=49pt, bb = 0 35 320 320]{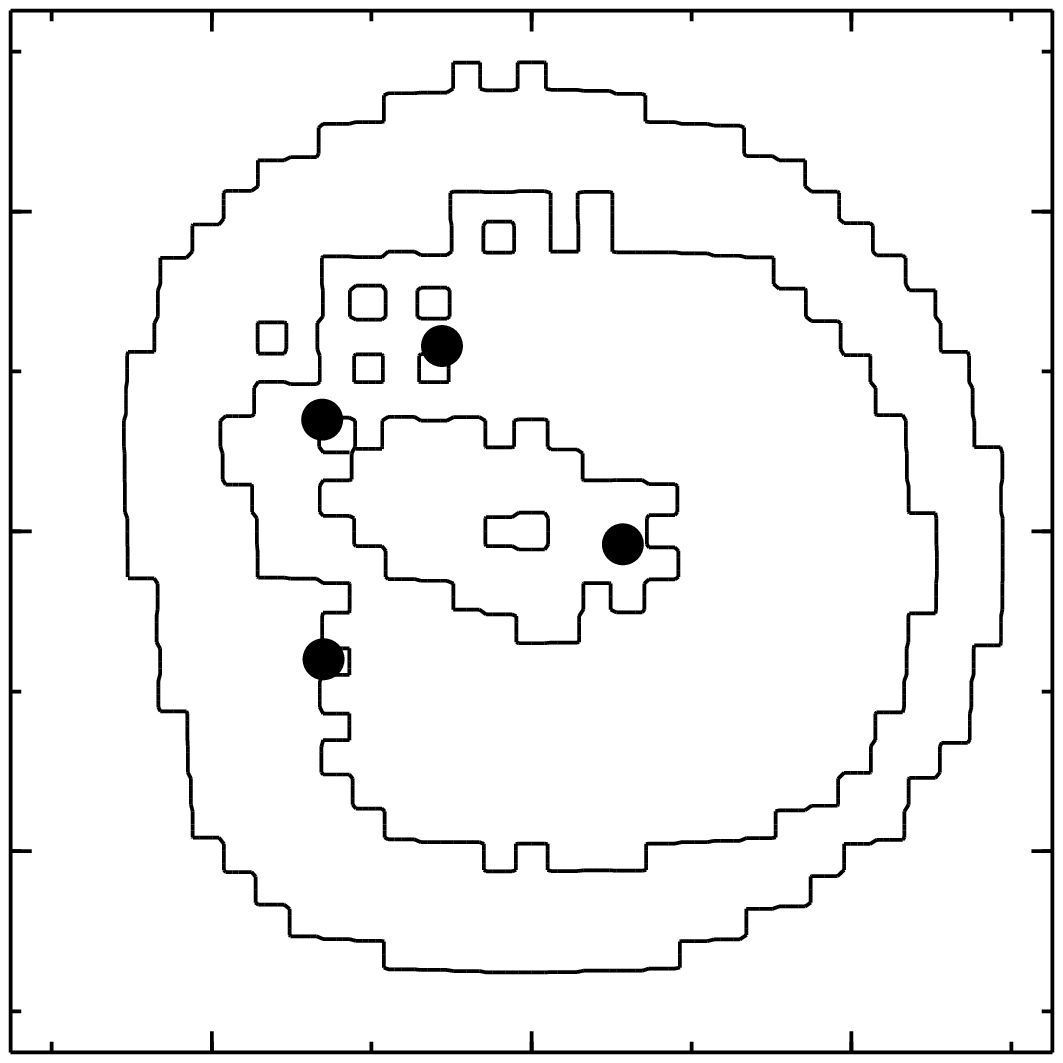}\\
\end{minipage}
\vspace{+0.1cm}

\hspace{1cm}\begin{minipage}[l]{0.15\textwidth}
\small
\begin{tiny}
\begin{verbatim}
object B2045+265
redshifts 0.870 1.280
shear +30
quad 
 1.121  0.824
 1.409  0.035 0
 1.255  0.576 0 
-0.507 -0.183 0
\end{verbatim}
\end{tiny}\end{minipage}
\begin{minipage}[r]{0.15\textwidth}
\includegraphics[width=49pt, bb = 0 35 320 320]{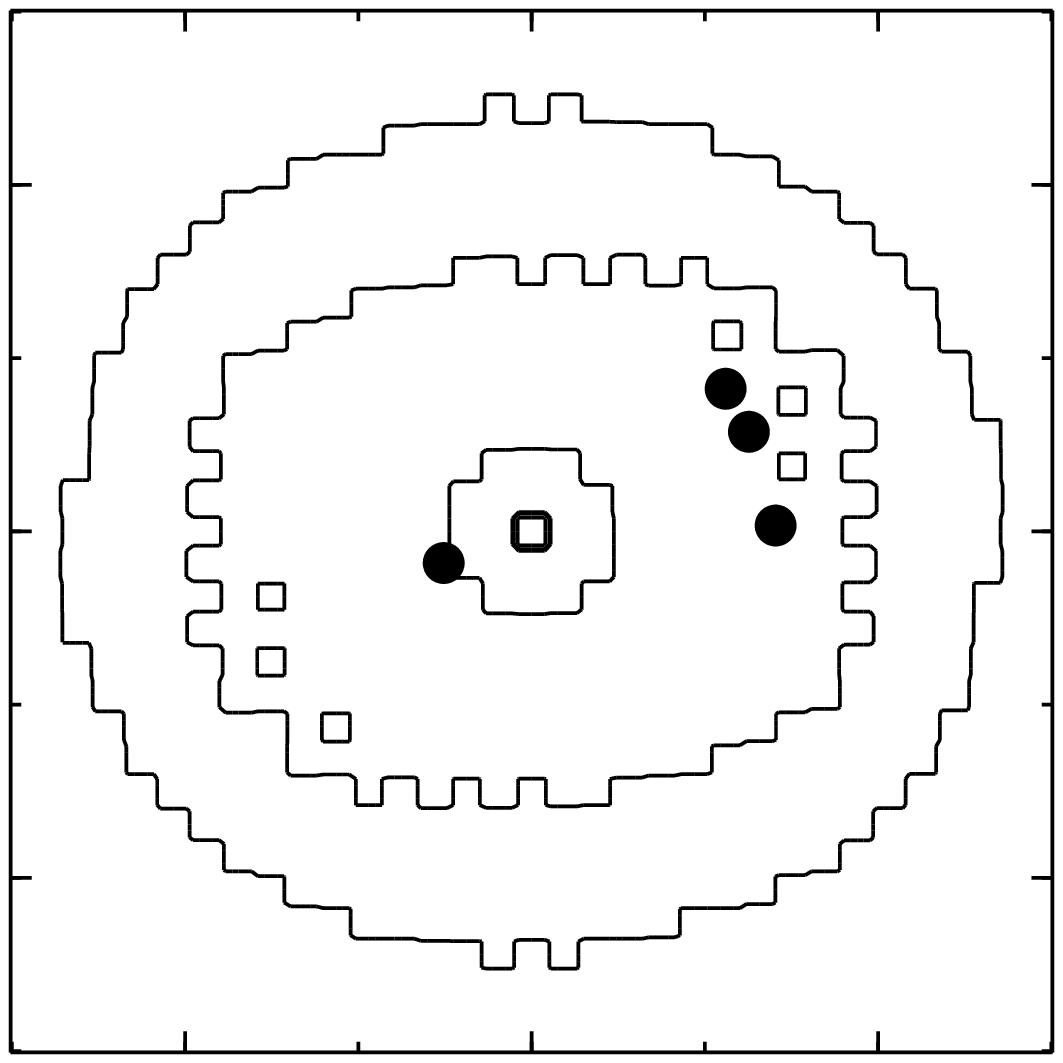}\\
\end{minipage}\begin{minipage}[l]{0.15\textwidth}
\small
\begin{tiny}
\begin{verbatim}
object BRI0952-012
redshifts 0.632 4.500
shear -45
double
-0.396  0.506
 0.300 -0.203 0
\end{verbatim}
\end{tiny}\end{minipage}
\begin{minipage}[r]{0.15\textwidth}
\includegraphics[width=49pt, bb = 0 36 320 320]{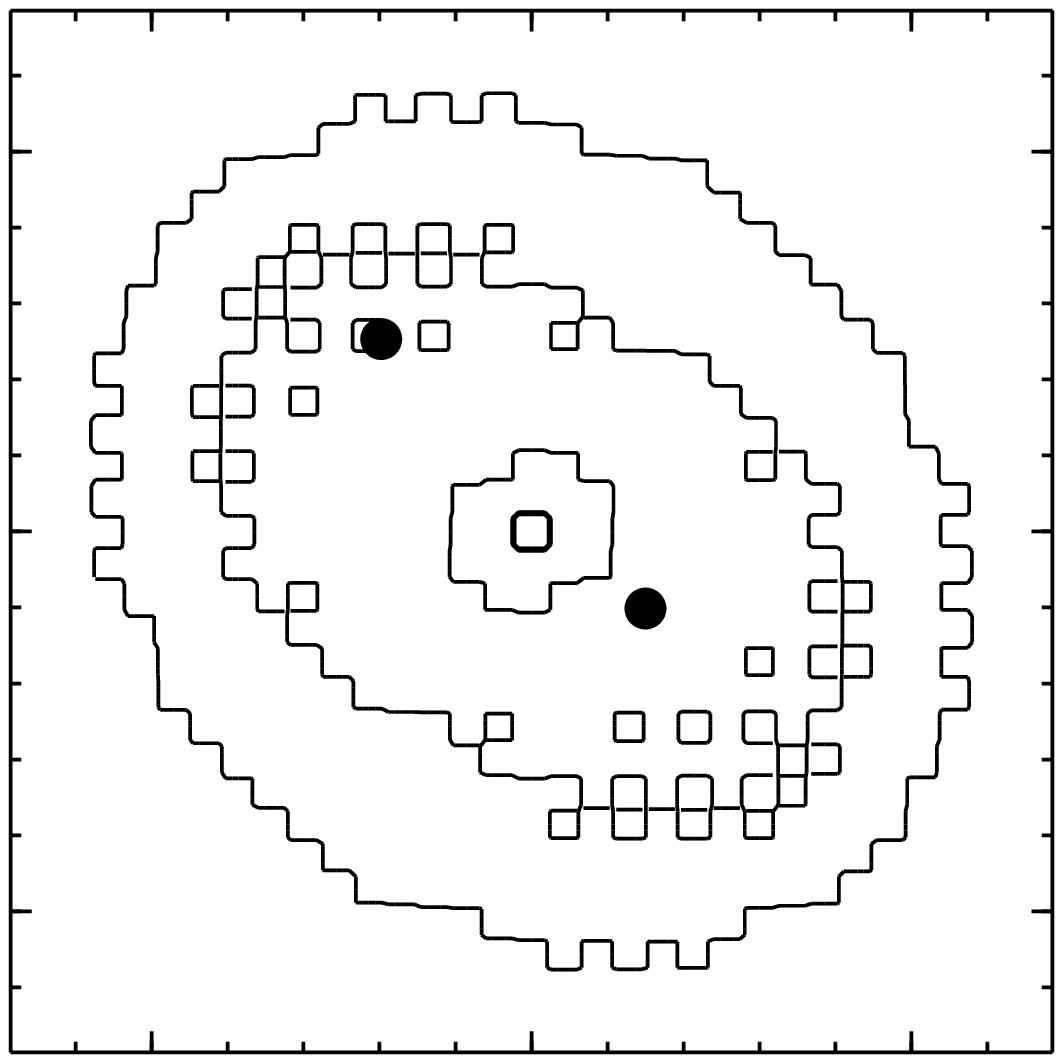}\\
\end{minipage}\begin{minipage}[l]{0.15\textwidth}
\small
\begin{tiny}
\begin{verbatim}
object HE1104-181
redshifts 0.730 2.320
shear +30
double
-1.927 -0.822
 0.974  0.510 161
\end{verbatim}
\end{tiny}\end{minipage}
\begin{minipage}[r]{0.15\textwidth}
\includegraphics[width=49pt, bb = 0 35 320 320]{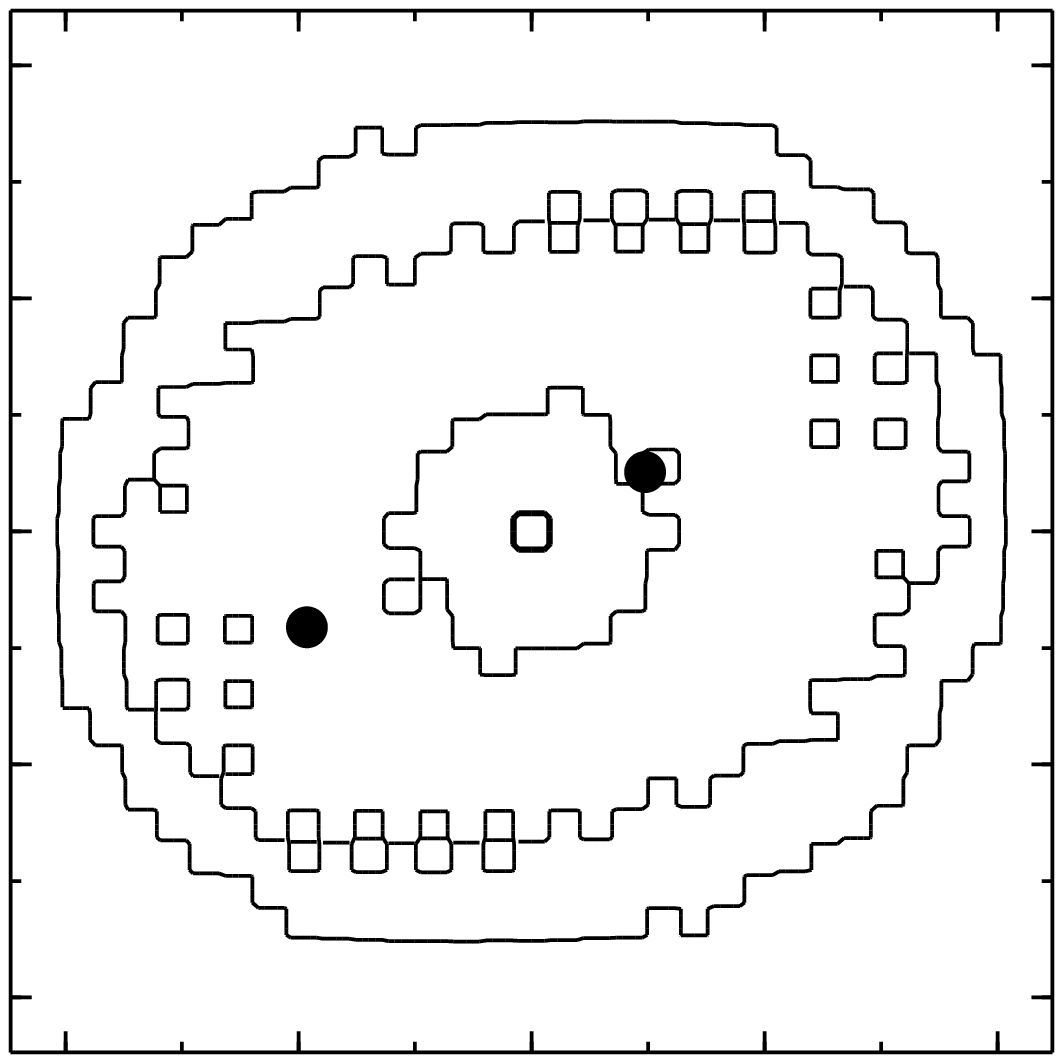}\\
\end{minipage}
\vspace{+0.1cm}

\hspace{1cm}\begin{minipage}[l]{0.15\textwidth}
\small
\begin{tiny}
\begin{verbatim}
object HE2149-275
redshifts 0.603  2.03
shear +45
double
 0.736 -1.161
-0.173  0.284 103
\end{verbatim}
\end{tiny}\end{minipage}
\begin{minipage}[r]{0.15\textwidth}
\includegraphics[width=49pt, bb = 0 35 320 320]{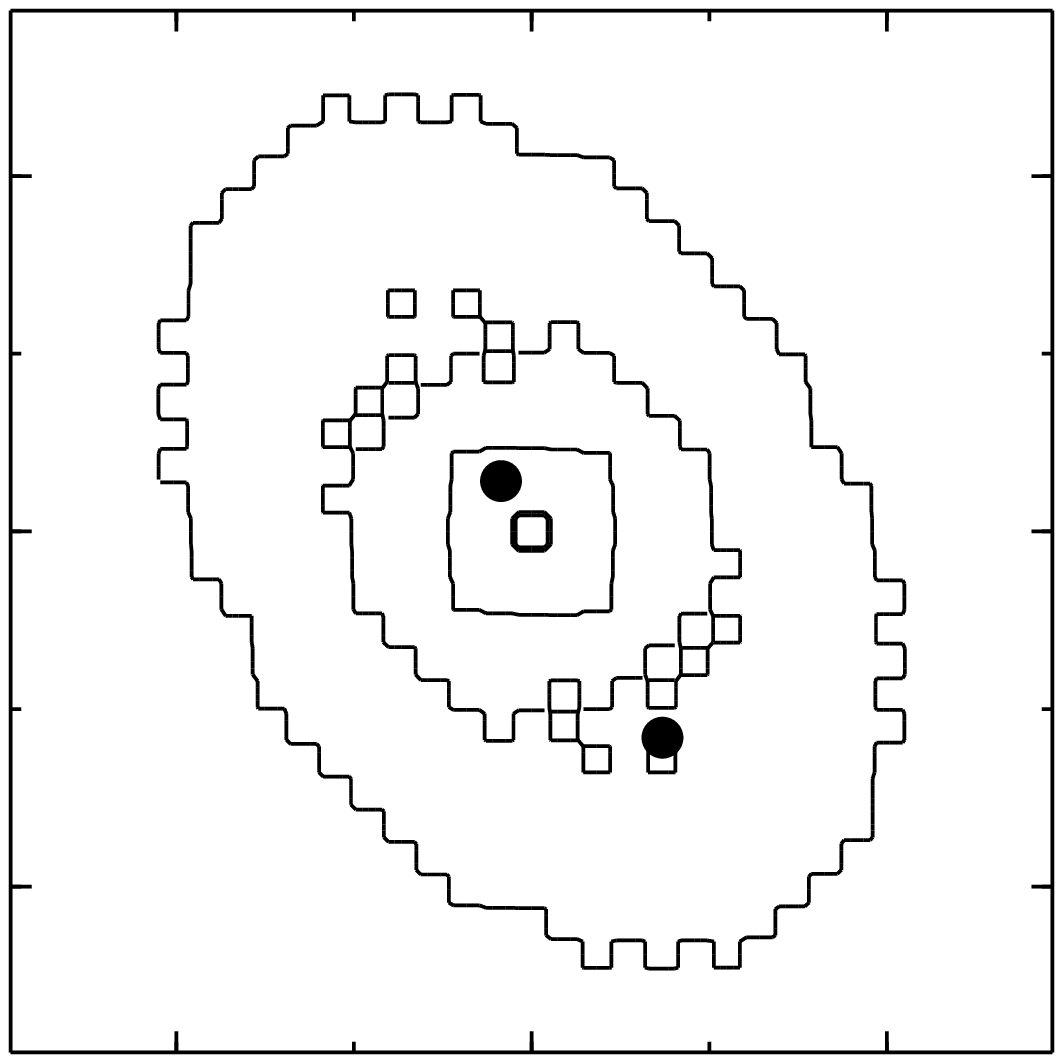}\\
\end{minipage}\begin{minipage}[l]{0.15\textwidth}
\small
\begin{tiny}
\begin{verbatim}
object HS0818+123
redshifts 0.390 3.115
ptmass -2 2 2 10
double
-1.657 -1.475
-0.152  0.592 0
\end{verbatim}
\end{tiny}\end{minipage}
\begin{minipage}[r]{0.15\textwidth}
\includegraphics[width=49pt, bb = 0 36 320 320]{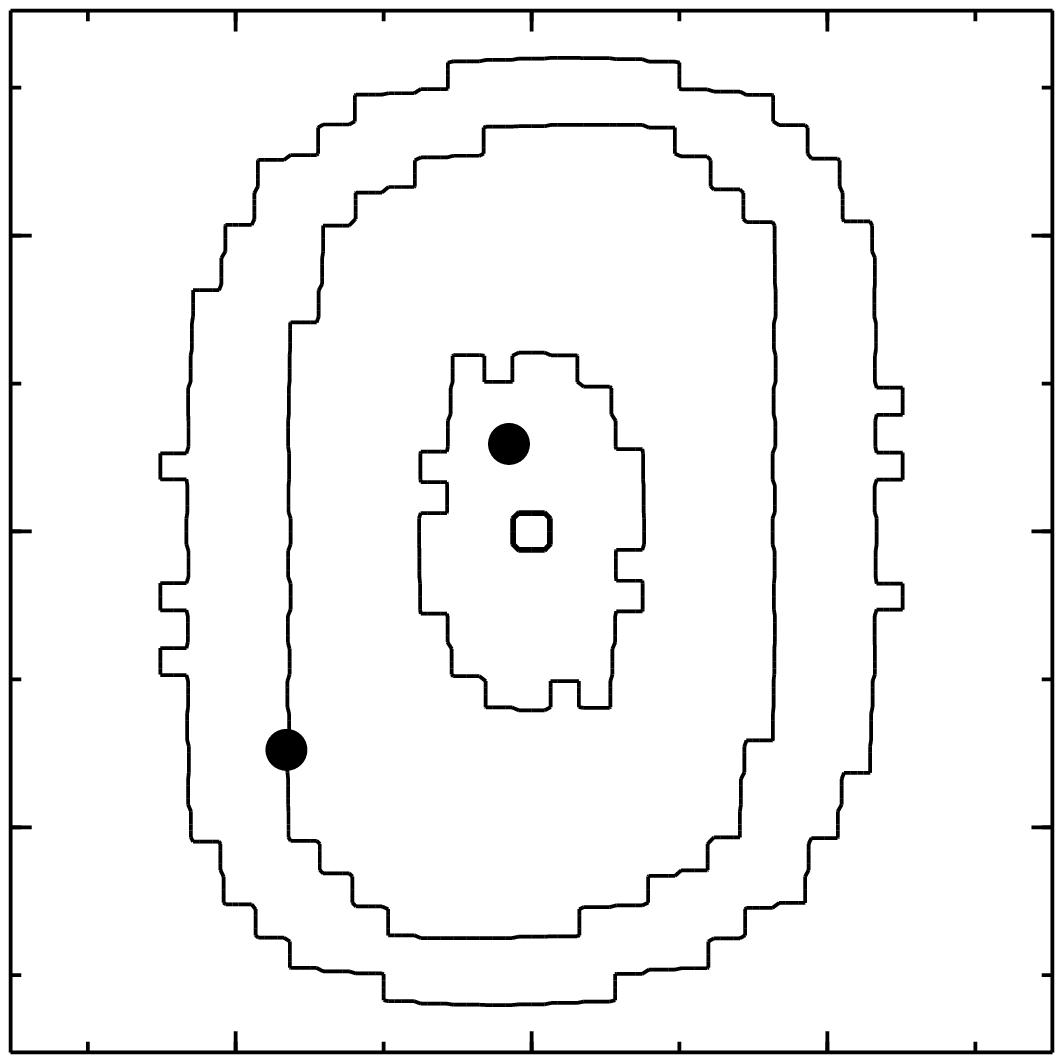}\\
\end{minipage}\begin{minipage}[l]{0.15\textwidth}
\small
\begin{tiny}
\begin{verbatim}
object LBQS1009-025
redshifts 0.880 2.740
shear +90
double
-0.537  1.097
 0.133 -0.286 0
\end{verbatim}
\end{tiny}\end{minipage}
\begin{minipage}[r]{0.15\textwidth}
\includegraphics[width=49pt, bb = 0 35 320 320]{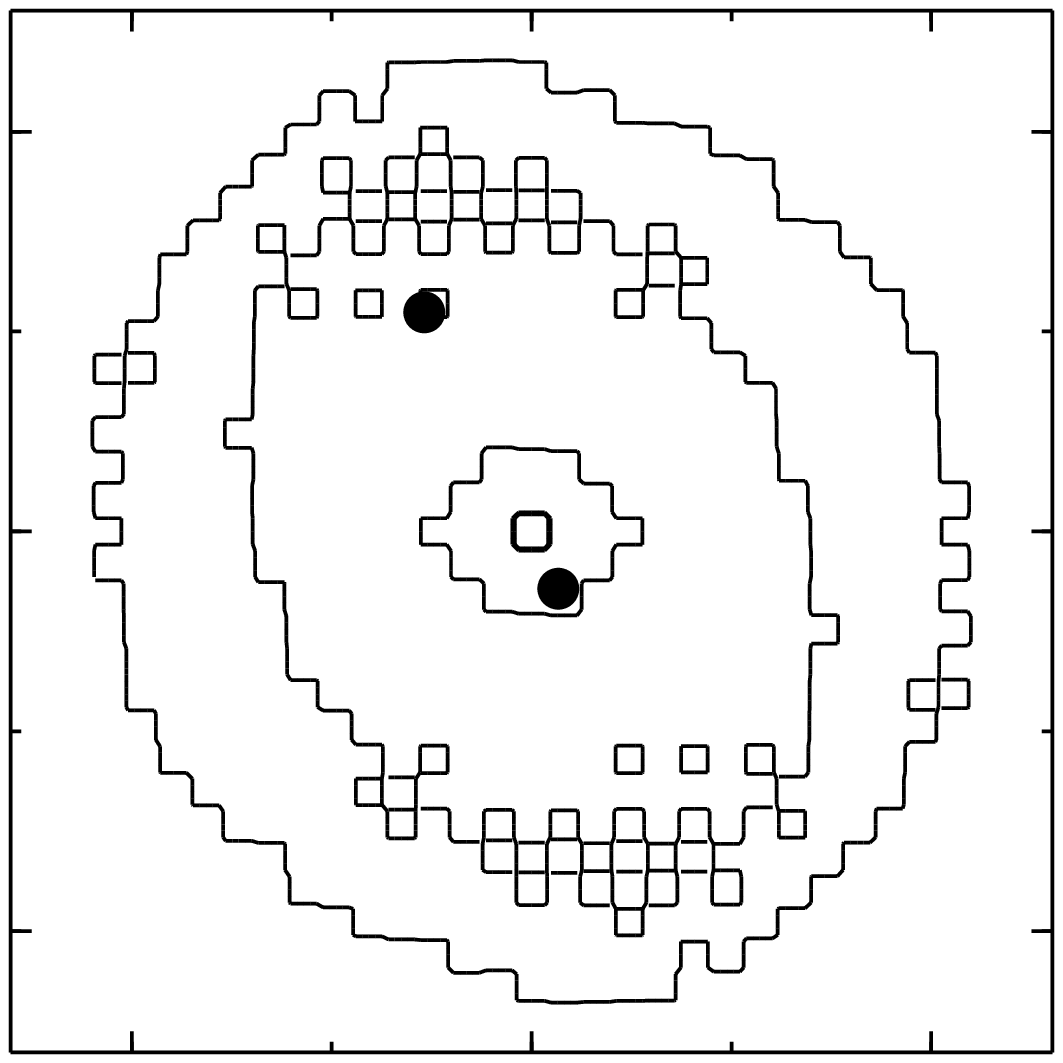}\\
\end{minipage}
\vspace{+0.1cm}

\hspace{1cm}\begin{minipage}[l]{0.15\textwidth}
\small
\begin{tiny}
\begin{verbatim}
object MG0414+053
redshifts 0.960 2.640
shear -30
- 3 quads -
\end{verbatim}
\end{tiny}\end{minipage}
\begin{minipage}[r]{0.15\textwidth}
\includegraphics[width=49pt, bb = 0 35 320 320]{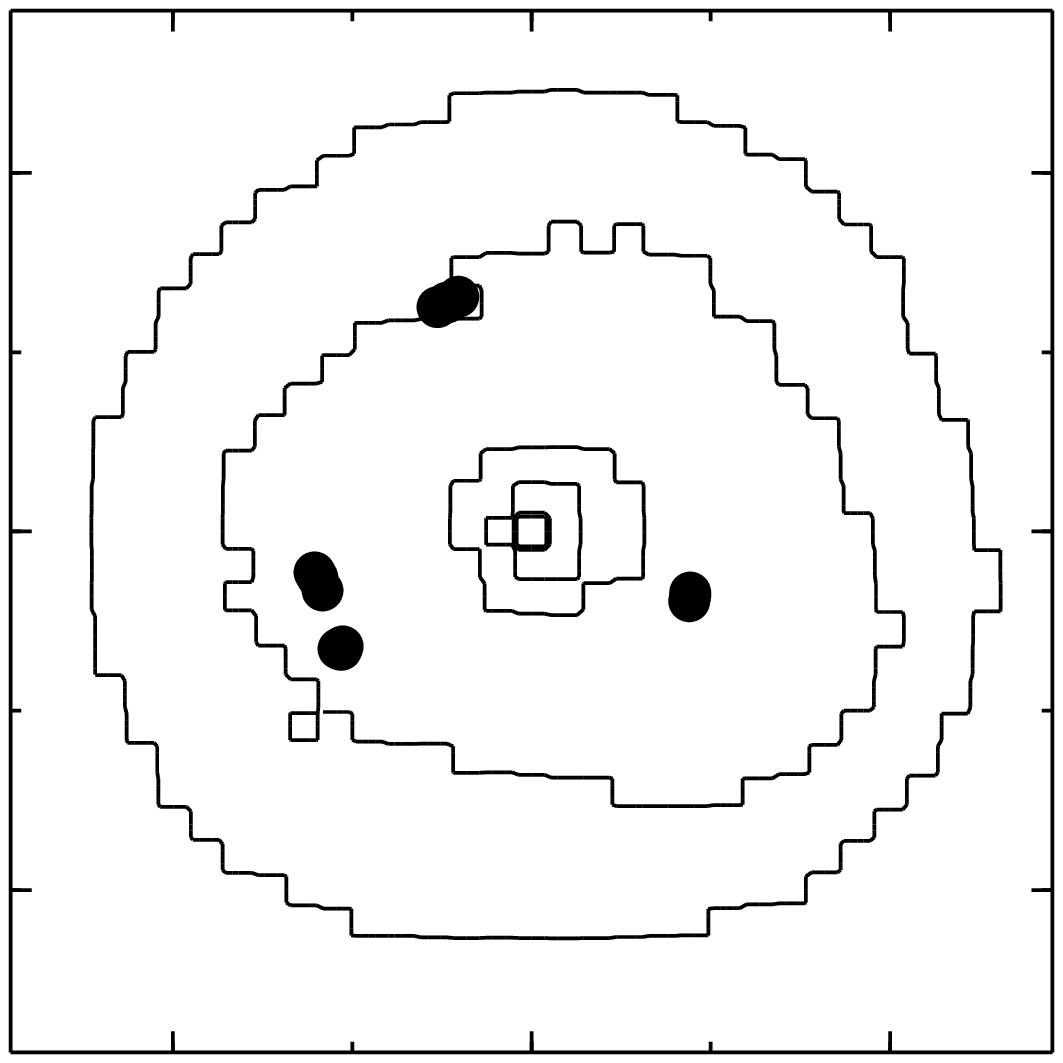}\\
\end{minipage}\begin{minipage}[l]{0.15\textwidth}
\small
\begin{tiny}
\begin{verbatim}
object MG2016+112
zlens 1.01
shear +45
- 3 ptmasses -
multi 3 3.27
-1.735  1.778 1
 0.317 -1.470 1
 1.269  0.274 2
\end{verbatim}
\end{tiny}\end{minipage}
\begin{minipage}[r]{0.15\textwidth}
\includegraphics[width=49pt, bb = 0 36 320 320]{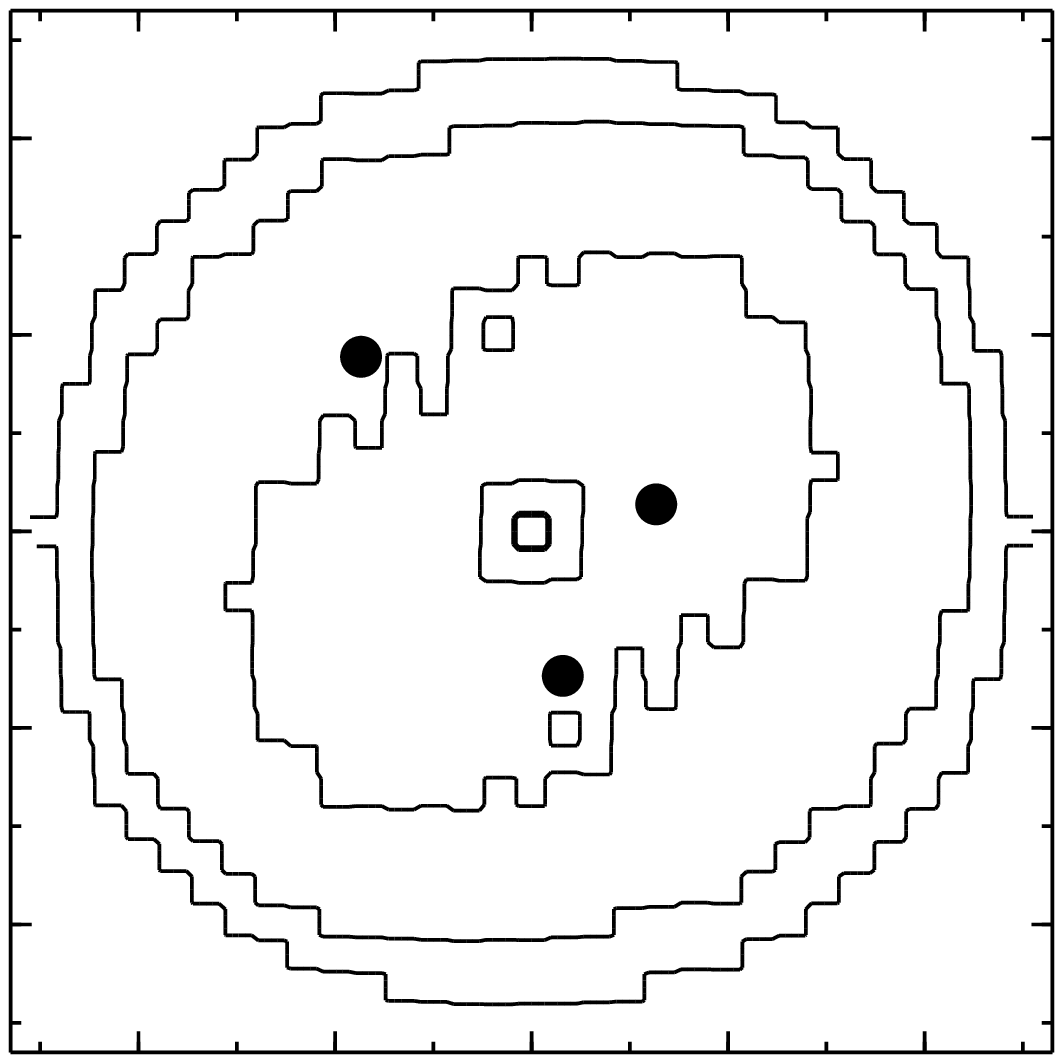}\\
\end{minipage}\begin{minipage}[l]{0.15\textwidth}
\small
\begin{tiny}
\begin{verbatim}
object PG1115+080
redshifts 0.311 1.722
shear -30
quad
 0.3550  1.3220
-0.9090 -0.7140 13.3
-1.0930 -0.2600 0
 0.7170 -0.6270 11.7
\end{verbatim}
\end{tiny}\end{minipage}
\begin{minipage}[r]{0.15\textwidth}
\includegraphics[width=49pt, bb = 0 35 320 320]{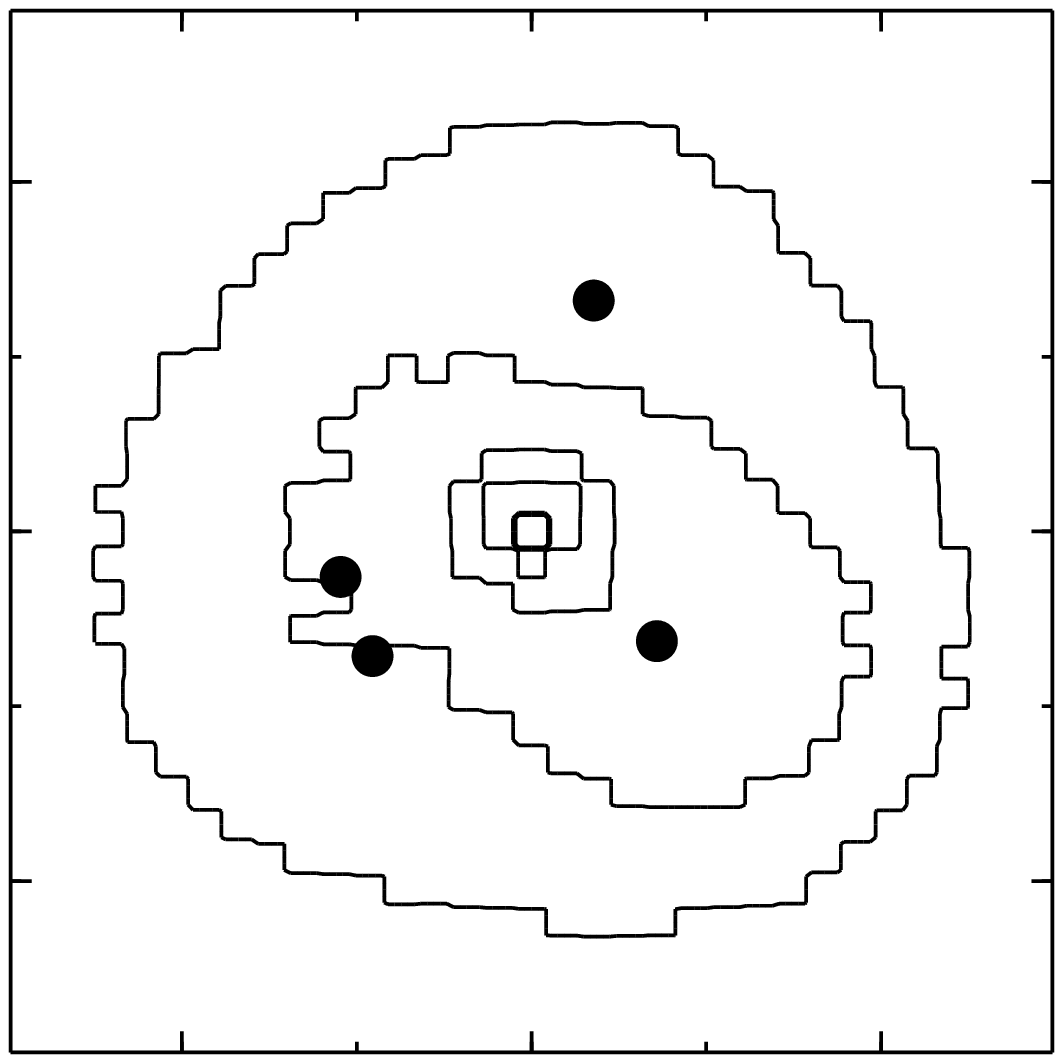}\\
\end{minipage}
\vspace{+0.1cm}

\hspace{1cm}\begin{minipage}[l]{0.15\textwidth}
\small
\begin{tiny}
\begin{verbatim}
object Q0047-281
redshifts 0.485 3.6
quad
 1.270  0.105
-0.630 -0.995  0
 0.520 -1.045  0
-0.730  0.705  0
\end{verbatim}
\end{tiny}\end{minipage}
\begin{minipage}[r]{0.15\textwidth}
\includegraphics[width=49pt, bb = 0 35 320 320]{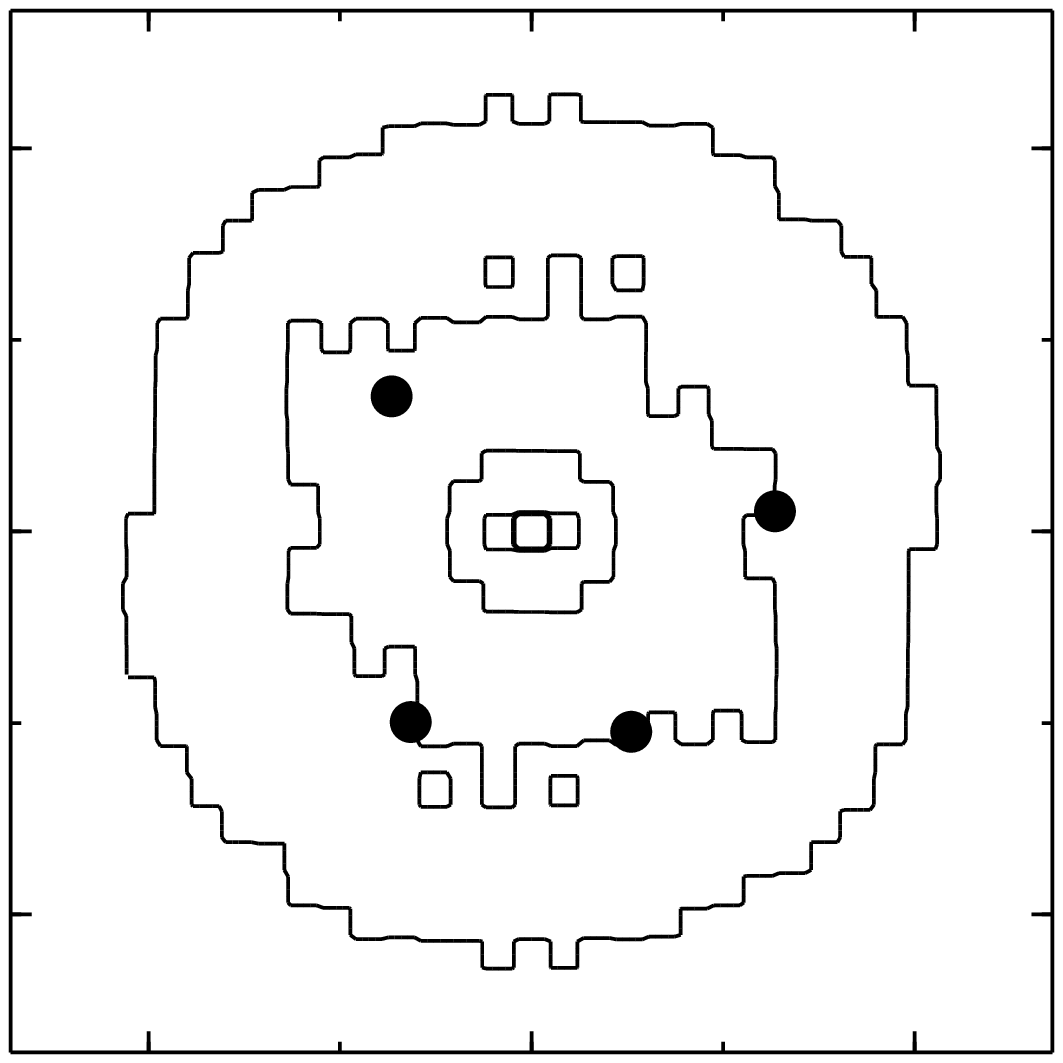}\\
\end{minipage}\begin{minipage}[l]{0.15\textwidth}
\small
\begin{tiny}
\begin{verbatim}
object Q0142-100
redshifts 0.490 2.720
shear 0
double
 1.764  0.574
-0.381 -0.039 0
\end{verbatim}
\end{tiny}\end{minipage}
\begin{minipage}[r]{0.15\textwidth}
\includegraphics[width=49pt, bb = 0 36 320 320]{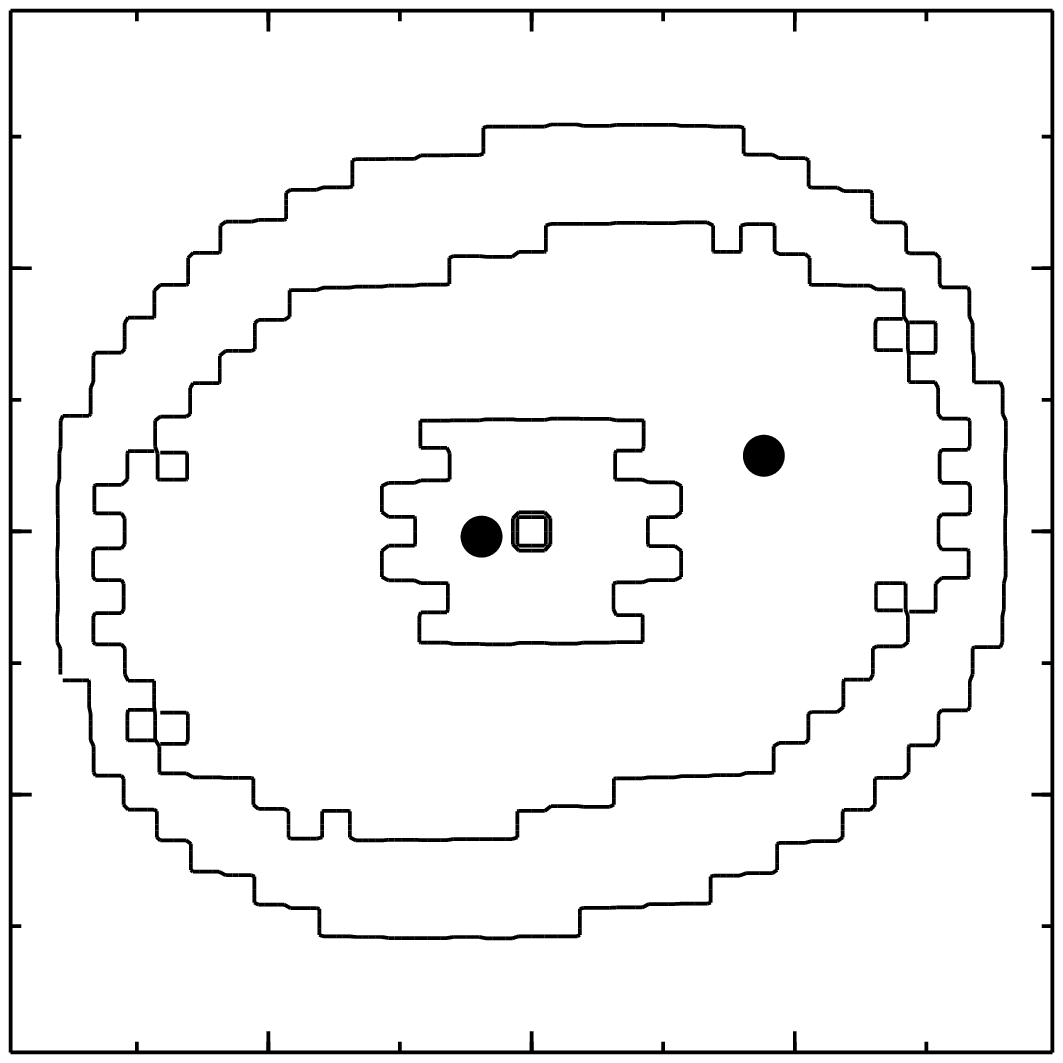}\\
\end{minipage}\begin{minipage}[l]{0.15\textwidth}
\small
\begin{tiny}
\begin{verbatim}
object Q0957+561
redshifts 0.356  1.41
shear 0
double
 1.408  5.034
 0.182 -1.018 423
double
 2.860  3.470
-1.540 -0.050 0
\end{verbatim}
\end{tiny}\end{minipage}
\begin{minipage}[r]{0.15\textwidth}
\includegraphics[width=49pt, bb = 0 35 320 320]{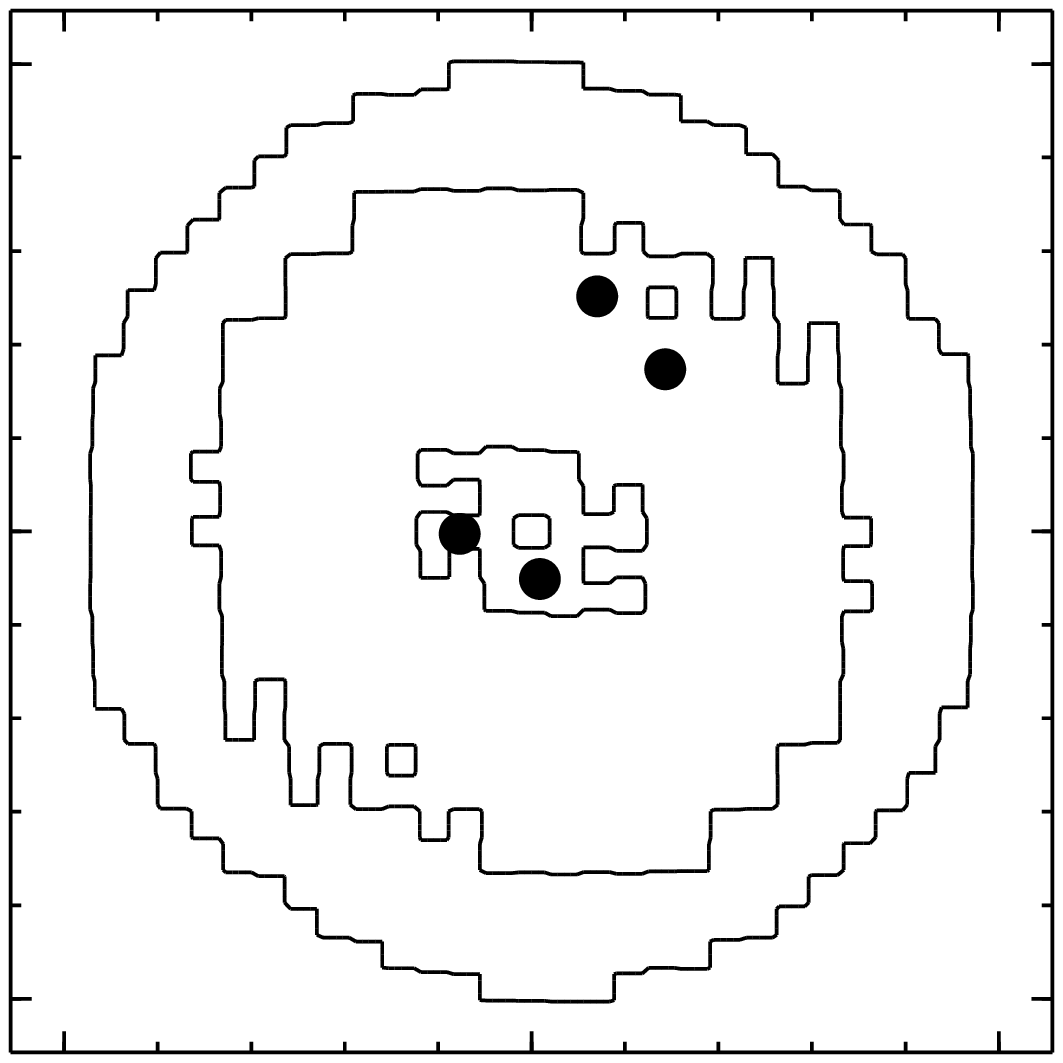}\\
\end{minipage}
\vspace{+0.1cm}

\hspace{1cm}\begin{minipage}[l]{0.15\textwidth}
\small
\begin{tiny}
\begin{verbatim}
object Q2237+030
redshifts 0.039 1.69
shear -45
maxsteep 3
quad 
 0.598  0.758
-0.075 -0.939 0 
 0.791 -0.411 0
-0.710  0.271 0
\end{verbatim}
\end{tiny}\end{minipage}
\begin{minipage}[r]{0.15\textwidth}
\includegraphics[width=49pt, bb = 0 35 320 320]{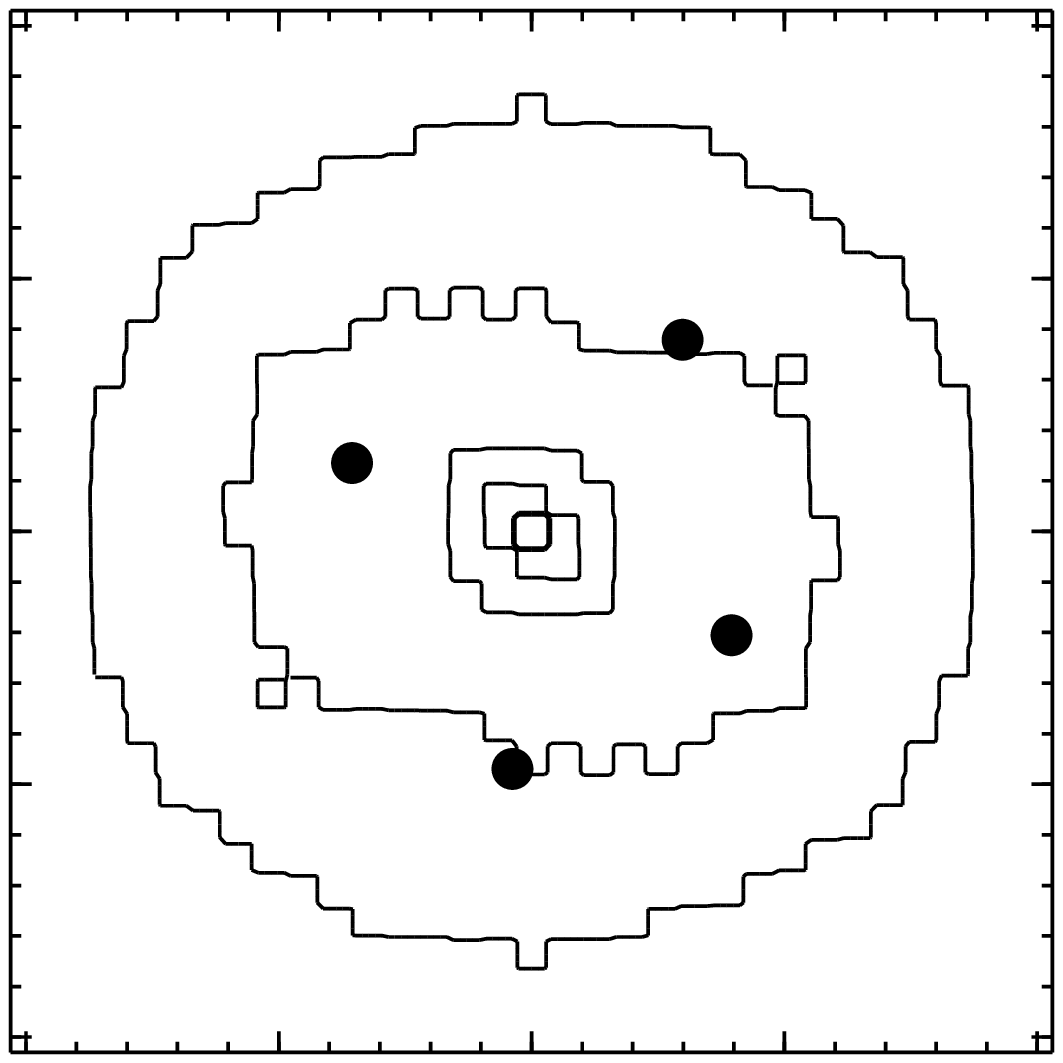}\\
\end{minipage}\begin{minipage}[l]{0.15\textwidth}
\small
\begin{tiny}
\begin{verbatim}
object RXJ0911+055
redshifts 0.769  2.80
shear +90
quad
 2.226  0.278
-0.968 -0.105  146
-0.709 -0.507    0
-0.696  0.439    0
\end{verbatim}
\end{tiny}\end{minipage}
\begin{minipage}[r]{0.15\textwidth}
\includegraphics[width=49pt, bb = 0 36 320 320]{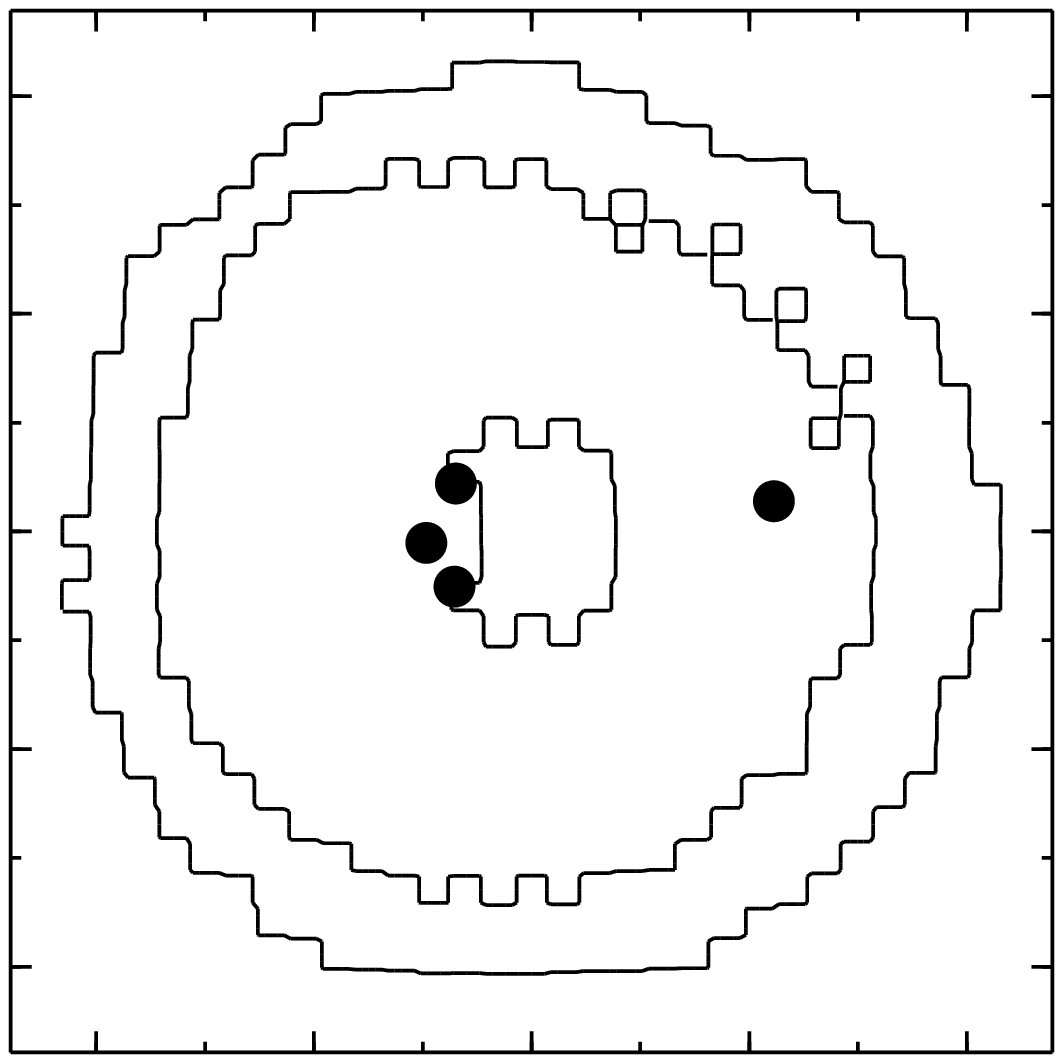}\\
\end{minipage}\begin{minipage}[l]{0.15\textwidth}
\small
\begin{tiny}
\begin{verbatim}
object SBS1520+530
redshifts 0.71 1.855
shear +45
double
 1.141   0.395
-0.288  -0.257 130
\end{verbatim}
\end{tiny}\end{minipage}
\begin{minipage}[r]{0.15\textwidth}
\includegraphics[width=49pt, bb = 0 35 320 320]{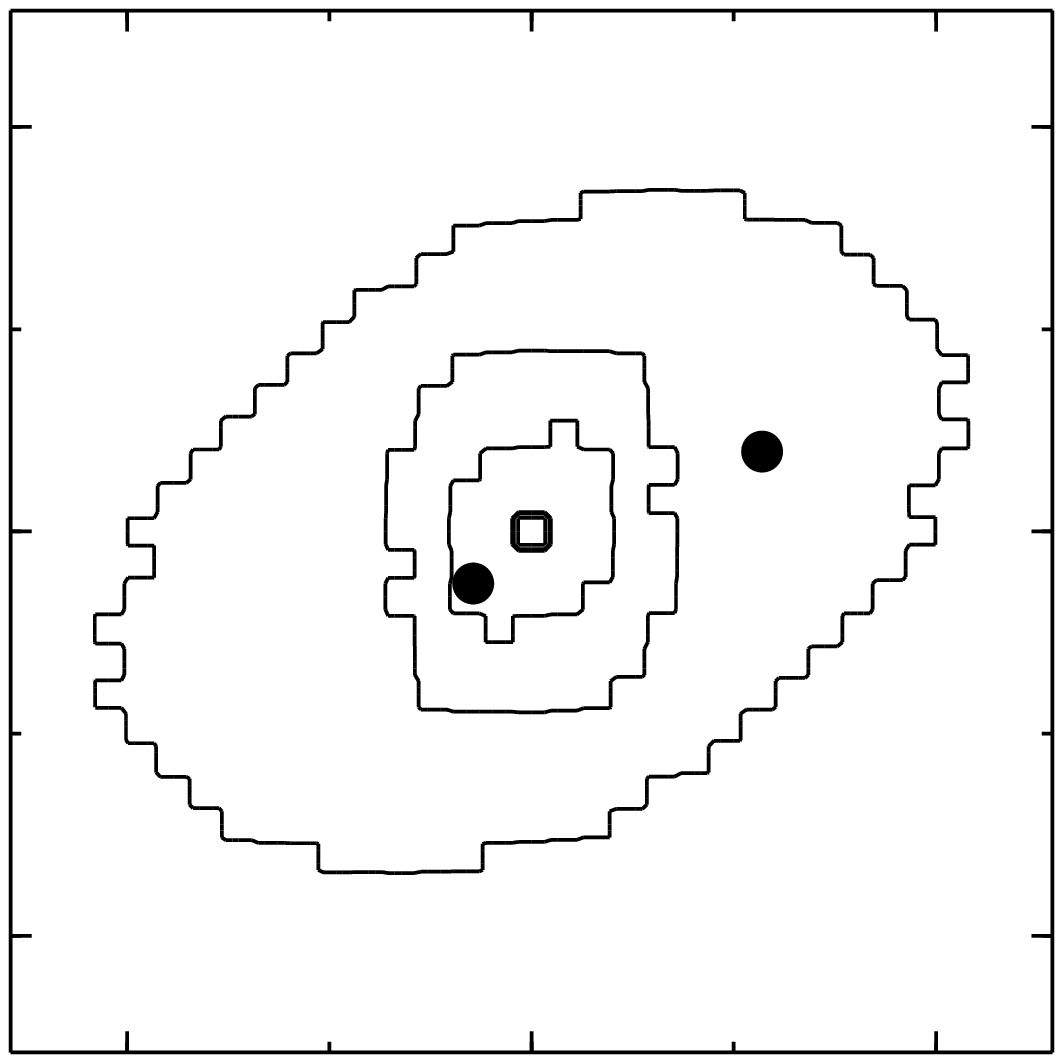}\\
\end{minipage}
\vspace{+0.1cm}

\caption{PIXELENS input data and projected mass distribution for the lens sample. The black dots mark the multiple images. All maps have a radius of 15 pixel. All mass maps have a radius of 2 $\rl$, which corresponds roughly to 2 $\rein$.  All lens properties as well as respective length specifications are in Table \ref{tab1}.  \label{fig16}}
\end{figure*}
\end{center}

\section{Dust reddening}
\label{appa}

\begin{figure}
\figurenum{16}
\center
\includegraphics[scale=0.7]{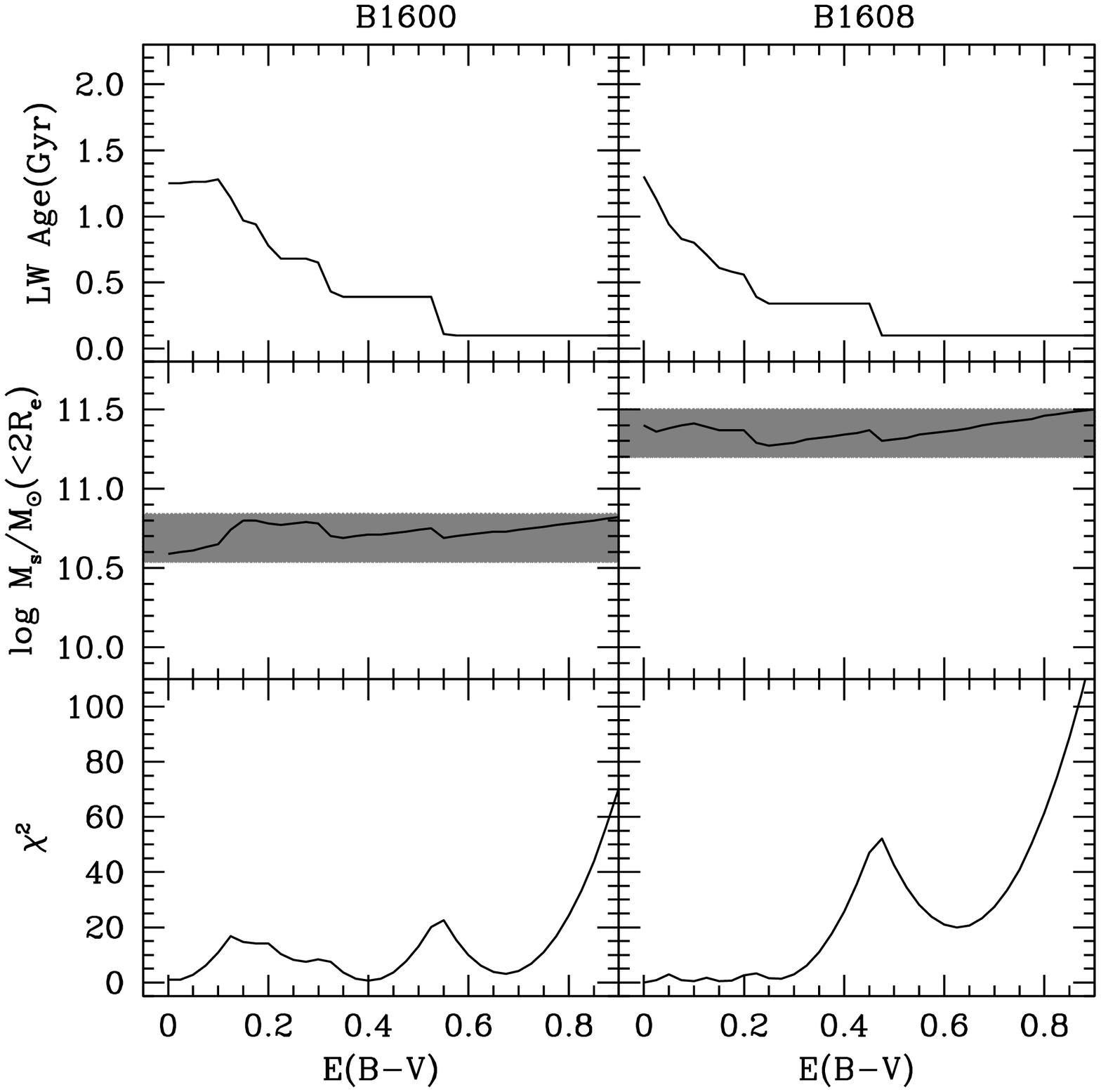}
\caption{From top to bottom: Age, stellar mass and minimum $\chi^2$ versus
  reddening for the lenses B1600 and B1608.\label{figA}}
\end{figure}

In this section we study the impact of dust reddening. Fig.~\ref{figA} shows
what happens when dust is included in the analysis of two lenses where
the contribution of dust could be important: B1600 and B1608. For this
exercise we take a simpler set of models, but the underlying effect
from dust is similar to the more complex case involving exponentially
decaying star formation histories. We run a set of simple stellar
populations with solar metallicity, changing the age and the dust
content.  The models are reddened according to a single parameter
$E(B-V)$ -- that follows a Galactic reddening curve \cite[e.g.][]{fitz99},
assuming R$\equiv A_V/E(B-V)=3.1$ (other reddening laws will
not introduce significant differences). The observed V--i and i--H
colours constrain the stellar M/L in the observed H band, which is
used to determine the stellar mass content, following the standard
methodology of this paper. The panels, from top to bottom, show the
best SSP-based luminosity-weighted age, stellar mass, and $\chi^2$ as
a function of the reddening parameter $E(B-V)$.
We find that dust "conspires" with age such that an increase in dust is compensated by a
younger age to give the same colours, yielding a small variation of
the estimated stellar mass with respect to dust reddening. Most
importantly, the value of $\chi^2$ worsens for high amounts of
reddening. Hence we can safely say that even in the case of B1600 and
B1608, the systematics on the stellar mass cannot be any larger than
about 0.3 dex in $\log(M_s)$, shown as a shaded grey region in the middle
panels of the figure. The other lens from our sample that could be
affected by dust, Q2237 (i.e. the bulge of a late-type galaxy) is at a
very low redshift ($z=0.039$), so that stellar masses are determined
from {\sl rest-frame} H-band, which is even less sensitive to dust
\citep{fe10}.

\label{lastpage}

\end{document}